# IMPROVED V II log(gf) VALUES, HYPERFINE STRUCTURE CONSTANTS, AND ABUNDANCE DETERMINATIONS IN THE PHOTOSPHERES OF THE SUN AND METAL-POOR STAR HD 84937

M. P. Wood[1], J. E. Lawler[1], E. A. Den Hartog[1], C. Sneden[2], and J. J. Cowan[3]

[1]Department of Physics, University of Wisconsin, Madison, WI 53706; mpwood@wisc.edu, jelawler@wisc.edu, eadenhar@wisc.edu

[2]Department of Astronomy and McDonald Observatory, University of Texas, Austin, TX 78712; chris@verdi.as.utexas.edu

[3]Homer L. Dodge Department of Physics and Astronomy, University of Oklahoma, Norman, OK 73019; cowan@nhn.ou.edu

ABSTRACT

New experimental absolute atomic transition probabilities are reported for 203 lines of V II. Branching fractions are measured from spectra recorded using a Fourier transform spectrometer and an echelle spectrometer. The branching fractions are normalized with radiative lifetime measurements to determine the new transition probabilities. Generally good agreement is found between this work and previously reported V II transition probabilities. Use of two spectrometers, independent radiometric calibration methods, and independent data analysis routines enables a reduction in systematic uncertainties, in particular those due to optical depth errors. In addition, new hyperfine structure constants are measured for selected levels by least squares fitting line profiles in the FTS spectra. The new V II data are applied to high resolution visible and UV spectra of the Sun and metal-poor star HD 84937 to determine new, more accurate V abundances. Lines covering a range of wavelength and excitation potential are used to search for non-LTE effects. Very good agreement is found between our new solar photospheric V abundance, log $\varepsilon$(V) = 3.95 from 15 V II lines, and the solar-system meteoritic value. In HD 84937, we derive [V/H] = -2.08 from 68 lines, leading to a value of [V/Fe] = 0.24.

1. INTRODUCTION

Stellar abundances, particularly their trends as a function of metallicity, are tests for models of nucelosynthesis and provide valuable information regarding the evolution of chemical elements in the Galaxy. Abundances in old, metal-poor stars are especially valuable since they represent the "fossil-record" of nucleosynthesis in the earliest generations of stars. Studies of metal-poor stars have found large and unexpected iron (Fe)-group abundance trends (McWilliam et al. 1995a, 1995b, McWilliam 1997, Westin et al. 2000, Cowan et al. 2002, Sneden et al. 2003, Cayrel et al. 2004, Barklem et al. 2005, Lai et al. 2008, Bonifacio et al. 2009, Roederer 2009, Suda et al. 2011, Yong et al. 2013) which have thus far refused explanation using current models of early Galactic supernova yields. These relative Fe-group abundance trends can cover ±1 dex for metallicities ranging from solar ([Fe/H] ≡ 0) to -4 (e.g. Figure 12 of McWilliam 1997) [1].

It may be that these unexpected trends represent a failure for models for nucleosynthesis in the early Galaxy. However, these trends may also indicate the breakdown of standard abundance derivation techniques in low metallicity stars, or they may result from inaccurate laboratory atomic data. Fairly comprehensive databases of atomic transition probabilities exist (e.g. NIST Atomic Spectra Database[2] and Vienna Atomic Line Database[3]). These are of utmost importance for stellar abundance determinations. To obtain the most accurate abundances it is crucial to use lines that are unsaturated in the photosphere of the star being investigated. For studies covering a wide metallicity range, this requirement necessitates the use of many lines covering a range of excitation potential (E.P.) and log($gf$) values, which introduces the possibility that inaccurate laboratory atomic data are affecting the measured abundances. In the photospheres of stars of interest, neutral atoms (first spectra) are a minor ionization stage while singly ionized atoms (second spectra) represent the dominant ionization stage. Therefore, in order to avoid saturation in higher metallicity stars, one makes use of weak first spectra lines that arise from high E.P. levels. In lower metallicity stars, however, one must change to stronger lines with lower E.P. values, and assuming suitable second spectra lines exist in the observed spectra, singly ionized lines become important as well. The strength of these high and low E.P. lines can vary by orders of magnitude, making it rather difficult to measure both with small uncertainties in the same laboratory spectra. If the laboratory atomic data are not at fault, the unexpected Fe-group abundance trends might result from the failure of 1D/LTE (one-dimensional/local thermodynamic equilibrium) approximations traditionally incorporated into photospheric models used for abundance determinations in metal-poor stars of interest (e.g. Asplund 2005). Giant stars are favored in studies of metal-poor stars to provide large photon fluxes for high signal-to-noise (S/N), high-resolution spectra. The combination of low-density atmospheres and reduced electron pressure from a lower metal content leads to lower collision rates in metal-poor giant stars, which may result in departures from LTE. The two possible explanations described above for the unexpected trends can be investigated with expanded and more accurate experimental Fe-group transition probabilities. One approach to determine if the trends result from 3D/non-LTE effects is to search for anomalous abundance measurements from various lines covering a range of E.P., log(*gf*), and wavelength for a wide range of stellar types. If improved atomic data and targeted searches for 3D/non-LTE effects fail to eliminate the observed Fe-group abundance trends, it would provide evidence that nucleosynthetic models for the early Galaxy are incomplete and need to be reexamined.

---

[1] We adopt standard spectroscopic notations. For elements X and Y, the relative abundances are written [X/Y] = $\log_{10}(N_X/N_Y)_{star} - \log_{10}(N_X/N_Y)_\odot$. For element X, the "absolute" abundance is written log ε(X) = $\log_{10}(N_X/N_H) + 12$. Metallicity is defined as the [Fe/H] value.
[2] Available at http://www.nist.gov/pml/data/asd.cfm.
[3] Available at http://www.astro.uu.se/~vald/php/vald.php.

Our group has an effort underway to expand sets of transition probabilities and reduce transition probability uncertainties for Fe-group lines. Den Hartog et al. (2011) focuses on selected multiplets in Mn I and Mn II that cover small wavelength ranges and/or are Russell Saunders (LS) multiplets. Given these benefits, it is possible to reduce the Mn transition probability uncertainties to 0.02 dex with $2\sigma$ confidence. Such small uncertainties are practical only under favorable conditions and in general and difficult to achieve. A broader approach is taken in the recent work on Ti I (Lawler et al. 2013), Ti II (Wood et al. 2013), and Ni I (Wood et al. 2014) by attempting measurements on every possible line connecting to upper levels with previously reported radiative lifetimes. This results in a larger set of transition probability measurements, though often with higher uncertainties (0.02 to ~0.10 dex). However, small (~0.02 dex) uncertainties, such as those reported by Den Hartog et al. (2011), are not necessary for the detection of non-LTE effects in metal-poor stars. Mn I resonance lines connected to the ground level show non-LTE effects of 0.5 to 1 dex (Sobeck et al. 2014, in preparation).

The V II study reported herein follows the broad approach previously used for Ti I, Ti II, and Ni I by attempting transition probability measurements for all lines connected to the 31 odd-parity upper levels with laser induced fluorescence (LIF) lifetime measurements by Den Hartog et al. (2014), plus an additional odd-parity upper level with a previously reported lifetime (Biémont et al. 1989, Xu et al 2006). The result is a set of 203 transition probabilities covering a wide range of E.P., log(*gf*), and wavelength. Uncertainties range from 0.02 dex for dominant branches, primarily determined by the LIF lifetime uncertainties, to ~0.12 dex for weak branches widely separated in wavelength from the dominant branch(es). The uncertainties on weak branches, which are often the most important for accurate abundance measurements, primarily result from S/N effects, uncertainty in the radiometric calibration, or both. The use of both a

Fourier transform spectrometer (FTS) and an echelle spectrometer, with independent radiometric calibrations, serves to reduce systematic uncertainties on weak line measurements.

In Section 2 and Section 3 we describe the spectra recorded using the FTS and echelle spectrometer, and in Section 4 we discuss the determination of V II branching fractions from these spectra. In Section 5 we present new absolute transition probabilities with comparisons to previous results in the literature. In Section 6 new hyperfine structure (HFS) constants and component patterns derived from our FTS spectra are introduce. We apply the new V II data to determine the photospheric V abundance of the Sun in Section 7 and metal-poor star HD 84937 in Section 8.

## 2. FOURIER TRANSFORM SPECTROMETER DATA

This V II transition probability study makes use of archived FTS data from the 1.0 m FTS previously at the National Solar Observatory (NSO) on Kitt Peak. The NSO 1.0 m FTS has a large etendue (like all interferometric spectrometers), a resolution limit as small as 0.01 cm$^{-1}$, wavenumber accuracy to 1 part in 10$^8$, broad spectral coverage from the near ultraviolet (UV) to infrared (IR), and a high data collection rate (Brault 1976). Unfortunately the NSO FTS has been dismantled, and while there are plans to restore it to full operational status at a university laboratory, it is currently unavailable to guest observers. Table 1 lists the 13 FTS spectra used in this V II branching fraction study. All spectra, raw interferograms, and header files are available in the NSO electronic archives.[4]

Multiple FTS spectra are needed to determine high-quality branching fractions. Optimum sensitivity is achieved for different spectral ranges using various beam splitter, filter, and detector combinations. Although branching fractions are determined mainly from spectra obtained with lamps having an Ar gas fill, additional spectra are needed from lamps with a Ne buffer gas to allow for the correction of blends between V and Ar lines. In addition, spectra are needed with the lamp operating over a range of current. Overlapping visible-UV and IR spectra of the lamps operating at high current are needed for high S/N measurements on very weak branches to all known lower levels. Conversely, one also needs visible-UV spectra of the lamps operating at low currents in which the dominant branches are optically thin. Optical depth errors can still be present even for the lowest current FTS spectra used in this study, and these concerns are addressed using the echelle spectrometer as described in Section 3.

---
[4] Available at http://nsokp.nso.edu/.

A relative radiometric calibration of the FTS spectra is essential for the measurement of accurate emission branching fractions. As in our past branching fraction studies we make use of the Ar I and Ar II line calibration technique. Sets of well-known branching ratios for Ar I and Ar II lines have been established for this purpose in the 4300 - 35000 cm$^{-1}$ range by Adams & Whaling (1981), Danzmann & Kock (1982), Hashiguchi & Hasikuni (1985), and Whaling et al. (1993). Intensities for these lines are measured and compared to known branching ratios in order to construct a relative radiometric calibration. This technique is internal to the HCD lamp and captures the wavelength-dependent response of the detectors, spectrometer optics, lamp windows, and any reflections which contribute to the measured signal. As described in the next Section, the use of an echelle spectrometer allows us to extend the branching fraction measurements beyond 35000 cm$^{-1}$.

## 3. ECHELLE SPECTROMETER DATA

As mentioned in the previous section, one motivation for the construction of an echelle spectrometer at the University of Wisconsin is the closure of the NSO 1.0 m FTS. A further motivation is the need to reduce optical depth errors as a source of systematic uncertainty in branching fraction measurements, especially on weak lines. FTS instruments suffer from inherent multiplex noise, in which quantum statistical (Poisson) noise from all spectral features, particularly strong visible and near-IR branches, is smoothly redistributed throughout the entire spectrum. This can be a hindrance when measuring the weak UV transitions that are typically the most important for accurate Fe-group abundance determinations. Often, as the lamp current is reduced, the weak lines become comparable to the multiplex noise before the dominant branch(es) from the common upper level are optically thin. The dispersive echelle spectrometer is free from multiplex noise and provides adequate S/N on these astrophysically important weak lines even at very low lamp current, reducing the possibility of optical depth errors.

The new spectrometer incorporates a 3.0 m focal length grating spectrograph with a large (128 × 254 mm ruled area), coarse (23.2 grooves mm$^{-1}$) echelle grating with a 63.5° blaze angle. Attached to the spectrograph is a custom 0.5 m focal length orthogonal order separator, which generates and images a 2D spectrum onto a large UV-sensitive CCD array while also serving to compensate aberrations in the spectrograph. The echelle spectrometer has a resolving power of 250,000, broad wavelength coverage, and superb UV sensitivity. While the echelle spectrometer has lower resolution and wavenumber precision compared to a FTS, it has the main advantage of being free from multiplex noise. This allows us to record spectra of commercial sealed hollow cathode discharge (HCD) lamps operating at very low currents, in order to eliminate optical depth errors, while still being able to detect weak lines with adequate S/N. Two HCD lamps are used, one each with Ar and Ne buffer gases, to check for and eliminate possible blends. A more detailed description of the echelle spectrometer, including a thorough aberration analysis, is presented by Wood & Lawler (2012).

In addition to the 13 FTS spectra listed in Table 1, the 75 CCD frames of echelle spectrometer spectra listed in Table 2 are also part of this V II transition probability study. These spectra are radiometrically calibrated using the UV continuum from a NIST-traceable $D_2$ standard lamp. This lamp is periodically checked against a NIST-calibrated Ar mini-arc to ensure an accurate UV calibration. The use of standard lamps to calibrate a FTS is often difficult due to ghosts, and we instead rely upon the calibration method described in Section 2, but it is our preferred method for calibrating the echelle spectrometer. The use of a $D_2$ standard lamp also enables branching fraction measurements to wavelengths shorter than that allowed using the Ar I and Ar II branching ratio technique since the standard lamp is calibrated to 2000 Å.

## 4. V II BRANCHING FRACTIONS

All possible transitions wavenumbers between known energy levels of V II from Thorne et al. (2013) that satisfy both the parity change and $|\Delta J| \leq 1$ selection rules are computed and used during this branching fraction analysis. Transitions which violate these two selection rules are suppressed by a factor of ~$10^6$ and are unimportant for stellar abundance studies. These two selection rules are obeyed throughout the periodic table whereas many important Fe-group transitions violate the $\Delta S = 0$ and $|\Delta L| \leq 1$ selection rules of LS coupling. We can make measurements for branching fractions as weak as 0.0001, and therefore systematic errors from missing branches are negligible in this study.

Branching fraction measurements are completed for all 31 odd-parity upper levels with LIF lifetime measurements by Den Hartog et al. (2014). In addition, branching fractions are measured for the odd-parity upper level at 39403.787 cm$^{-1}$, which has previously reported lifetime measurements from Biémont et al. (1989) and Xu et al. (2006). As in our previous work, thousands of possible spectral line observations are analyzed in both FTS and echelle spectra to calculate the branching fractions. Integration limits and non-zero baselines are set "interactively" during data analysis. Non-zero baselines are necessary for the echelle spectra, which are not background corrected, and are occasionally needed for the FTS spectra when a line falls on the wing of a dominant feature. A simple numerical integration technique is used to determine un-calibrated V II line intensities. This method is preferred since the majority of V II lines have unresolved hyperfine structure that leads to variations in the observed line width. For consistency, this method is also applied to lines with partially resolved hyperfine structure. This same integration technique is also used on selected Ar I and Ar II lines to establish a relative radiometric calibration of the FTS spectra.

Branching fraction uncertainties depend on the S/N of the data, the line strengths, and the wavelength separation of lines from a common upper level. Branching fraction uncertainty always migrates to the weakest lines because branching fractions sum to 1.0 by definition. Uncertainties on weak lines near the dominant branch(es) from the common upper level tend to be limited by S/N. For lines that are widely separated in wavelength from the dominant branch(es), systematic errors in the radiometric calibration tend to be the dominant source of uncertainty. The systematic uncertainty in the calibration is estimated using the product of 0.001%/cm$^{-1}$ and the wavenumber difference between the line of interest and the dominant branch from the common upper level, as presented and tested by Wickliffe et al. (2000). The calibration uncertainty is combined with the standard deviation of measurements from multiple spectra to determine the total branching fraction uncertainty. The final uncertainty, especially for lines widely separated from the dominant branch(es) from a common upper level, is primarily systematic and it is therefore impractical to state whether it represents 1σ or 2σ error bars. The combination of data from both the FTS and echelle spectrometer, which make use of independent radiometric calibration methods, is important in assessing and controlling systematic uncertainties.

## 5. V II TRANSITION PROBABILITIES AND COMPARISON TO EARLIER MEASUREMENTS

Absolute transition probability measurements are given for 203 lines of V II in Table 3. Branching fraction measurements from a combination of FTS and echelle data are normalized with published LIF radiative lifetimes (Den Hartog et al. 2014) to determine the transition probabilities. Air wavelengths in the table are computed from V II energy levels (Thorne et al. 2013) using the standard index of air (Peck & Reeder 1972).

Often lines must be omitted if they are too weak to have reliable S/N, have uncertain classifications, or are too seriously blended to be separated. The effect of these problem lines can be seen by summing all transition probabilities for a given upper level in Table 3 and comparing the sum to the inverse upper level lifetime (Den Hartog et al. 2014). The sum is typically > 90% of the inverse level lifetime. While these problem lines have large fractional uncertainty in their branching fractions, this does not have a significant effect on the uncertainties of the lines kept in Table 3. The transition probability uncertainties quoted in Table 3 are found by combining branching fraction uncertainties and radiative lifetime uncertainties in quadrature.

Figures 1-3 compare our new V II transition probability data to the NIST Atomic Spectra Database as of 2014 June 11 (Kramida et al. 2013; see footnote 2). The figures are distinguished by the accuracy ratings assigned to each value in the database. Figure 1 is a comparison of 62

log(*gf*) values in common to this work and to NIST database values with a "B" (≤10%) accuracy rating, plotted as a function of wavelength in the upper panel (a) and the log(*gf*) value reported in this study in the lower panel (b). Individual error bars on the log(*gf*) differences represent the uncertainties on measurements from this work. The central solid line represents perfect agreement at a logarithmic difference of zero, while the grey dotted lines represent ±10% differences in *f*-values. Figure 2 is a comparison of 19 lines in common between this work and the NIST database having accuracy grades of "C+" (≤18%) or "C" (≤25%), again plotted as a function of wavelength in the upper panel (a) and the log(*gf*) value reported in this study in the lower panel (b). The error bars and central solid line have the same meaning as in Figure 1, while the grey dotted lines represent ±25% differences in *f*-values. Karamatskos et al. (1986) are cited as the source of the "B", "C+", and "C" rated data in the NIST database. Similarly to this work, those authors used a combination of LIF lifetimes and emission branching fractions to determine their transition probabilities. Aside from a few outliers, the majority of lines agree within combined uncertainties. The slight dip in Figure 1a for lines near 3550 Å was observed previously by Biémont et al. (1989), who suggested it results from an incorrect calibration of the FTS spectra used by Karamatskos et al.

A comparison of 15 log(*gf*) values in common to this work and the NIST database with accuracy grades of "D" (≤50%) is plotted as a function of wavelength in the upper panel (a) of Figure 3 and the log(*gf*) value measured in this work in the lower panel (b). The error bars and central solid line have the same meaning as in Figure 1, whereas the grey dotted lines represent ±50% differences in *f*-values. Both the work of Wujec & Musielok (1986) and the earlier work of Roberts et al. (1973) are cited as the sources of the "D" rated values in the NIST database. Roberts et al. determined their log(*gf*) values from a combination of lifetimes measured using the beam-foil technique with branching fractions measured from an arc discharge. Wujec & Musielok (1986) also performed branching fraction measurements using an arc discharge, with some log(*gf*) values determined using the Roberts et al. lifetimes, while other values were determined using a Boltzmann analysis to set relative log(*gf*) values on an absolute scale. The agreement between this work and previous measurements is not as good as in Figures 1-2. The overall trend of higher log(*gf*) values in this work can be explained by our use of the new LIF lifetime measurements of Den Hartog et al. (2014), which in many cases are lower than the lifetimes of Roberts et al. Biémont et al. (1989) noted that the beam-foil lifetimes of Roberts et al. are long by as much as a factor of two, which they attribute to the possibility of cascading transitions in the beam-foil excitation. In addition, the tendency for stronger lines to be enhanced compared to weaker lines, as seen in Figure 3b, is evidence for optical depth errors in the earlier measurements. This current work makes use of a new echelle spectrometer which was specifically developed to address optical depth concerns in transition probability measurements.

Figure 4 is a comparison of log(*gf*) values for 137 lines in common to this work and the work of Biémont et al. (1989), plotted as a function of wavelength in the upper panel (a) and the log(*gf*) value measured in this work in the lower panel (b). The solid line represents perfect agreement at a logarithmic difference of 0, while the error bars represent the uncertainty reported by Biémont et al. (1989) and the uncertainty reported herein combined in quadrature. Similarly to this work, Biémont et al. used a combination of LIF radiative lifetimes and emission branching fractions from FTS spectra to determine the majority of their transition probabilities. Transition probabilities for two additional levels were measured by interpolating upper level populations in an inductively-coupled plasma source. While six of the LIF lifetimes were new measurements (including the lifetime value utilized in this study for the 39403.787 cm$^{-1}$ upper level), Biémont et al. also utilized earlier lifetime measurements from Karamatskos et al. (1986). As in this work, spectra were recording using the NSO 1.0 m FTS, and as such there is some overlap in the FTS spectra used by Biémont et al. and in this work (e.g., Index #6 in Table 1). Over 70% of the lines plotted in Figure 4 agree within combined uncertainties. However, it is likely that the agreement is actually better than this. A significant number of the transition probabilities in Table 2 of the Biémont et al. work have quoted uncertainties less than the uncertainties on the corresponding level lifetimes used to derive the transition probability values. As stated in Biémont et al., the transition probability uncertainties contain contributions from the lifetime uncertainties, S/N effects, and uncertainty in the FTS calibration. For these transition probability values, we substitute the corresponding lifetime uncertainty in place of the quoted transition probability uncertainty to determine the error bars, since the lifetime uncertainty represents a lower limit on the transition probability uncertainty. It is likely the true uncertainties for these lines are larger than this, which would increase the size of the combined error bars and bring more lines into agreement.

6. MEASUREMENTS OF V II HYPERFINE STRUCUTRE CONSTANTS

Vanadium is essentially monoisotopic, with $^{51}$V being the only naturally occurring stable[5] isotope, and therefore isotope shifts are unimportant for this study. However, since $^{51}$V has a non-zero nuclear spin (I = 7/2), hyperfine structure (HFS) leads to a broadening of many V II transitions observed in this study. Several V II levels have previously reported experimental HFS $A$ constants. Arvidsson (2003) measured 26 $A$ constants using least squares fitting of HFS patterns in two FTS spectra, one recorded using the NSO 1.0 m FTS (likely #7 in our Table 1) and an additional FTS spectrum, recorded at Lund University, to capture the deep UV. More recently, Armstrong et al. (2011) published a set of 55 high accuracy HFS $A$ constants from LIF measurements made using a single-frequency laser on a beam of V II atoms.

---

[5] $^{50}$V is nearly stable, with a half-life of ~$10^{17}$ years. However, its solar-system abundance is only 0.25%, entirely negligible for this study.

Similarly to Arvidsson (2003), in this work we use least squares fitting of V II line profiles in order to determine new HFS $A$ constants. We make use of the Casimir formula, as presented in the text by Woodgate (1980),

$$\Delta E = \frac{A}{2}K + \frac{B}{8}\frac{3K(K+1) - 4I(I+1)J(J+1)}{I(2I-1)J(2J-1)},$$

where $\Delta E$ is the wavenumber shift of an HFS sublevel ($F,J$) from the center of gravity of the fine-structure level ($J$),

$$K = F(F+1) - J(J+1) - I(I+1),$$

$F$ is the total atomic angular momentum, $J$ is the total electronic angular momentum, and $I$ is the nuclear spin. Unfortunately the FTS spectra we utilize do not have adequate resolution or S/N, or both, in order to determine any HFS $B$ constants, and therefore we neglect the electric quadrupole interaction term in determining the energy shifts. For this same reason, rather than taking a broad approach and measuring as many new HFS $A$ constants as possible, we instead target transitions which are used in either the abundance analyses of the Sun (Section 7) or HD 84937 (Section 8). We choose to focus on lines that are broadened and/or show HFS in our FTS spectra and cause a non-negligible amount of broadening in the solar and stellar spectra, as we are able to obtain the most reliable results for these lines. For these transitions, we start by using the LIF measurements of Armstrong et al. (2011) to fix either the upper or lower level HFS $A$ constant, and then nonlinear least-squares fit the observed HFS pattern in order to determine the HFS $A$ constant for the other level. By fixing these newly measured HFS $A$ constants, we can then proceed to fit HFS patterns for transitions which have neither an upper or lower LIF HFS constant measurement. In addition to the HFS *A* constants of the upper or lower level, the fitting parameters include the center of gravity wavenumber, the total intensity of the line, and one line-profile parameter which represents the convolution of the instrumental sinc function with a variable-temperature Doppler-broadened Gaussian function.

Table 4 lists 21 new magnetic dipole HFS *A* constants measured in this study. These values are determined from a S/N weighted mean of HFS pattern fits in four FTS spectra (#1, 2, 3, and 13 in Table 1), while error bars represent the standard deviation of the measurements. In general there is good agreement with the earlier work of Arvidsson (2003), with a few exceptions. This work benefits from newly measured level energies and classifications (Thorne et al. 2013), which help identify possible blends, as well as being tied to new and accurate LIF HFS constants from Armstrong et al. (2011), so we are confident in our reported values. Please note that while the HFS *A* constants from Armstrong et al. are not listed in Table 4, they did serve as reference values in our study due to the high spectral resolution and S/N of their single frequency laser measurements. For the level at 2605.040 cm$^{-1}$, the agreement is significantly worse, with our new result having almost equal magnitude but opposite sign to that reported by Arvidsson (2003). However, the value presented here is reinforced by new theoretical relativistic configuration interaction calculations from Beck & Abdalmoneam (2014). Using the LIF HFS measurements of Armstrong et al. (2011) in combination with the new HFS *A* constants from Table 4, HFS component patterns for selected lines used in the solar and/or HD 84973 V II abundance determinations are listed in Table 5. Several patterns listed in Table 5 connect to the ground term, even though these levels have no previously reported HFS *A* constants and are not included in Table 4. While we are unable to reliably measure the HFS constants for these levels, we conclude from the FTS spectra that they are small, and for the purposes of Table 5 the HFS *A* constants for levels in the ground term have been set to zero. Individual energy shifts are calculated using the Casimir formula quoted above and the component strengths are normalized to sum to unity. Given the relatively large error bars on some of the newly measured HFS *A* constants listed in Table 4, we choose not to attempt HFS pattern determinations on lines which are not needed in the abundance analyses presented in Section 7 and Section 8. Therefore we caution that Table 5 is not meant to be an exhaustive list of lines with measurable HFS in V II, and there may be other lines with detectable HFS broadening.

## 7. THE VANADIUM ABUNDANCE IN THE SOLAR PHOTOSPHERE

We apply our new V II transition probability and HFS data to produce a new V abundance for the solar photosphere. We follow the techniques described in our previous studies of Fe-group species: Ti I (Lawler et al. 2013), Ti II (Wood et al. 2013), and Ni I (Wood et al. 2014). Since we employ synthetic spectrum analyses for each feature, substitution of full sets of HFS components for individual lines is straightforward; see Lawler et al. (2001a, 2001b) for previous examples of this procedure. In our Ni I study (Wood et al. 2014) we included the effects of isotopic wavelength shifts in our solar abundance determinations, but with only one naturally-occurring stable isotope ($^{51}$V), isotopic shifts are irrelevant here. However, whereas HFS was unimportant in our Ni I study, it must be incorporated into our abundance analysis here.

To begin, as in our previous papers, we estimate approximate V II absorption transition strengths with the simple formula,

$$\mathrm{STR} \equiv \log(gf) - \theta\chi$$

with the log(*gf*) values given in Table 3, excitation energies $\chi$ (eV), and inverse temperature $\theta = 5040/T$ (we assume $\theta = 1.0$ for this rough calculation). The STR values are plotted as a function of wavelength in Figure 5. Red circles call attention to those lines that we use in the solar abundance computations (see below). These relative strengths apply only to V II because they include neither Saha ionization factors (which could allow comparison to other vanadium species) nor the vanadium solar abundance (which could allow comparison to other elements). They do, however, indicate the relative line strengths for V II lines. In Figure 5, the horizontal line at STR = –4.1 indicates the strengths of V II lines that lie at the approximate weak-line limit of features that are useful in a solar abundance analysis. Expressing the line equivalent width $EW$ as log of the reduced width $\log(RW) = \log(EW/\lambda)$, the weak-line limit is approximately $\log(RW) \sim -6$. For Figure 5 the corresponding weak-line strength of –4.1 is determined empirically, by measuring the $EW$s of the weakest V II lines of Table 3. All of the lines in our study are stronger than $\log(RW) \sim -6$ in the solar center-of-disk spectrum (Delbouille et al. 1973)[6]. Thus if unblended they are potentially useful photospheric vanadium abundance indicators.

Table 3 lists nearly 110 V II lines with wavelengths longer than the atmospheric cutoff (3000 Å, indicated with the blue vertical line in Figure 5). However, almost all of these lines arise at wavelengths $\lambda < 4100$ Å, in the crowded near-UV spectral region. Therefore the main impediment to their use is contamination by transitions of other atomic and molecular species. We follow the procedures of our previous papers to determine which V II lines can be employed. We inspect each line in the Delbouille et al. (1973) photospheric spectrum, and then use the Moore et al. (1966) solar line identification compendium and the Kurucz (2011)[7] atomic/molecular line database to identify those V II lines that are too blended to yield trustworthy vanadium abundances. Unfortunately, this procedure eliminates the vast majority of V II lines in Table 3, and we are left with only 25 V II lines meriting further investigation.

---

[6] http://bass2000.obspm.fr/solar_spect.php
[7] http://kurucz.harvard.edu/linelists.html

We compute synthetic spectra for these surviving transitions with the current version of the LTE line analysis code MOOG[8] (Sneden 1973). Line list assembly is described in detail by Lawler et al. (2013). Briefly, we begin with the Kurucz (2011) line database, gathering atomic and CN, CH, NH, and OH molecular lines in a typically 4 Å interval centered on each V II line, but modify transition probabilities and account for isotopic/hyperfine substructure as needed from recent lab studies on these species: second spectra rare earth atoms (Sneden et al. 2009 and references therein), Cr I (Sobeck et al. 2007), Ti I (Lawler et al. 2013), Ti II (Wood et al. 2014), and Ni I (Wood et al. 2014). To be consistent with our previous work beginning with Lawler et al. (2001), we adopt the Holweger & Müller (1974) empirical model photosphere. The line lists and solar model are used as inputs in MOOG, and the output raw synthetic spectra are then convolved with Gaussian smoothing functions to empirically match the broadening effects of the spectrograph instrument profile (a negligible effect for the Delbouille et al. 1973 solar spectral atlas) and solar macroturbulence. For V II lines and for the lines with lab transition data named above, the lab data are accepted without alteration. For other contaminants (the majority of features in most near-UV spectral windows), adjustments are made to their log(*gf*) values to best reproduce the observed solar photospheric spectrum.

The comparisons of observed and synthetic solar spectra result in the elimination of more V II lines, due either to unacceptably large contamination by other species or because they are simply too strong to be sensitive to abundance changes. In the end we are left with only 15 lines that are appropriate for a solar vanadium abundance determination. In Figure 6 we show observed and synthetic spectra for representative V II lines at 3530.78 and 4564.58 Å. This

---

[8] Available at http://www.as.utexas.edu/~chris/moog.html

figure also shows the positions and fractional strengths of the HFS components for each line. Inclusion of HFS always serves to broaden and desaturate a transition, resulting in a derived abundance which, compared to single-line assumptions, ranges from slightly smaller (a few percent) for weak lines to factors approaching five times smaller for strong (saturated) lines.

The abundances from individual lines are listed in Table 6, in which we also include columns for line wavelengths, excitation energies, oscillator strengths, and whether HFS is included in the synthetic spectrum computations. While the majority of lines listed with a "no" in Table 6 have negligible HFS, there are some for which HFS patterns could not be determined. This may be due to a lack of resolved structure, a lack of S/N, or both in the available FTS spectra. However, the majority of lines for which HFS has a detectable effect on derived abundances have HFS patterns reported in Table 5. Inclusion of the missing HFS patterns for lines listed "no" in Table 6 would likely have a negligibly small effect on the abundance determinations. The line abundances are plotted as functions of wavelengths in the top panel of Figure 7. With this small data set we do not find any obvious trends of abundance with line wavelength, excitation energy, transition probability, or overall line strength. From these 15 lines we derive a new solar photospheric vanadium abundance: $\langle \log \varepsilon(V) \rangle = 3.95 \pm 0.01$ with $\sigma = 0.05$.

Our solar analysis unfortunately has to exclude many promising V II lines in the 3100–3300 Å spectral range, those with relative strengths STR $\geq -1$ in Figure 5. These lines include ones for which we have produced syntheses but in the end must neglect, because they are so strong that their absorptions depend more on microturbulent velocity, damping, and outer-atmosphere line formation physics than they do on V abundance. Here are four examples, giving the $\log(RW)$ values from Moore et al. (1966) in parentheses after their wavelengths: 3121.15 Å

(–4.38); 3126.22 Å (–4.32); 3188.71 Å (–4.45), and 3276.14 Å (–4.49). All these lines, and many others in the near-UV spectral region, are strong enough to appear on the "flat" or "damping" parts of the solar photospheric curve-of-growth. Proper accounting of HFS does not remove enough of the overall line saturation of these transitions to make them reliable abundance indicators. However, our trial syntheses of these, and some other lines that are too strong for solar V abundance estimation, suggest that their values are compatible with the abundances derived from weaker lines. Their abundance uncertainties are simply too large to be of use here, but their great strengths will render them as prime V abundance indicators in very metal-poor stars.

The standard deviation of the mean solar abundance determined from the 15 lines of our study is only ±0.01, and so the dominant source of total uncertainty is external, through choice of model solar atmosphere and analytical technique. Scott et al. (2014) have recently completed new analyses of the solar photospheric abundances of Fe-group elements, using EWs and a variety of solar models and line formation methodologies. They concur with previous studies that the Fe-group elements exist almost exclusively in the singly ionized species, whose absorption features can be analyzed to adequate accuracy with an assumption of LTE. For V II, Scott et al. find only five transitions amenable to EW measurement, and from them derive <log ε(V)> = 4.03 with the Holweger & Müller (1974) solar model. But the application of other models yields essentially the same result, and the mean V abundance varies only in the range 3.98–4.04; see Scott et al. for extended discussion of these abundance exercises. They urge caution in interpretation of their V II result due to the few lines that they trusted for this species.

Our mean photospheric abundance of <log ε(V)> = 3.95 is 0.08 smaller than that of Scott et al. (2014) with the same solar model assumption. If we repeat the Scott et al. *EW* analysis with their log(*gf*) values we recover their mean abundance. However, for their lines our transition probabilities are 0.02 dex larger on average than they used, and thus their revised abundance would become 4.01 with application of our log(*gf*) values. Our synthetic spectrum analysis of the Scott et al. lines yields <log ε(V)> = 3.97, another decrease of 0.04 dex. We suspect that the synthetic spectrum approach may account better for blending transitions in the crowded near-UV spectral region where the V II lines arise. Finally, we note that Scott et al. (2014) recommend a solar-system meteoritic abundance of log ε(V) = 3.96 ± 0.02, which is in excellent accord with our photospheric V abundance.

## 8. THE VANADIUM ABUNDANCE OF METAL-POOR STAR HD 84937

Determining vanadium abundances in low metallicity stars from new V II lab data continues our efforts to apply improved basic lab data to Fe-group elements whose nucleosynthesis and Galactic evolution can be predicted theoretically (e.g. Kobayashi et al. 2006 and references therein). Concerns exist about the reliability of Fe-group abundances at low metallicity, including the need to gather spectra of high quality UV and near-UV wavelengths, and the need to explore the limitations of standard analytical assumptions (e.g. LTE, plane-parallel geometry) used in deriving abundances. Here we see what can be understood from application of traditional abundance techniques to this new large set of accurate V II line data for one well-studied star.

HD 84937 is a metal-poor main-sequence turnoff star ($T_{eff}$ = 6300 K, log g = 4.0, [Fe/H] = –2.15, and $v_t$ = 1.5 km s$^{-1}$). Inspection of its near-UV spectrum reveals many promising V II transitions, since contamination from other species decreases greatly and many V II lines that are completely saturated in the solar photosphere weaken enough to be useful abundance indicators.

We derive the V abundance in HD 84937 in similar fashion to our previous Fe-group abundance studies.

Our data set for HD 84937 consists of an optical *ESO VLT UVES* spectrum and an *HST/STIS* UV high-resolution spectrum[9]. Lawler et al. (2013) describe these spectra in detail. The availability of a UV spectrum for this star causes us to repeat the transition candidate selection process from the beginning, which results in nearly 75 possible V II lines for analysis. Line-by-line abundance derivation via spectrum syntheses is done as described in Lawler et al. (2013). This yields a mean abundance in HD 84937 of $\langle\log\varepsilon(V)\rangle = 1.871 \pm 0.009$, with $\sigma = 0.075$ from 68 lines. Individual abundances derived from V II lines in HD 84937 are listed in Table 7 and displayed in the lower panel of Figure 7. Then with the mean solar abundance derived in §6, we compute a relative V abundance of $\langle[V/H]\rangle = 1.87 - 3.95 = -2.08$. There are six V II lines in common between our analyses of the solar photosphere (Table 6) and HD 84937 (Table 7). From just those lines, we derive $\langle[V/H]\rangle = -2.08 \pm 0.03$ ($\sigma = 0.08$), in excellent agreement with the value computed from the complete line sets for both stars.

Following the discussion of Lawler et al. (2013), we estimate that internal line-to-line scatter uncertainties are $\leq 0.04$ dex. For external error estimates, we derive abundances of 14 typical V II lines with model atmosphere parameters varied in accord with the HD 84937 uncertainties. If $T_{eff}$ is increased by 150 K, then on average $\Delta(\log\varepsilon) \approx +0.09$; if log g is increased by 0.3, then $\Delta(\log\varepsilon) \approx +0.08$; if the metallicity is decreased to [Fe/H] = $-2.45$, then $\Delta(\log\varepsilon) \approx 0.00$ (unchanged); and if $v_t$ is decreased to 1.25 km s$^{-1}$ then $\Delta(\log\varepsilon) \approx +0.00$ to $+0.08$, depending on the strength of the measured transition. However, we note that the response of V II transitions to changes in model atmospheric parameters is very similar to other ions of the Fe-group, which have similar Saha ionization properties. Thus [V/Ti] or [V/Fe] relative abundance ratios are mostly insensitive to model parameter variations. We reserve comment on the statistical equilibrium of vanadium in HD 84937 until completion of our study of V I transitions in the Sun and this star (Lawler et al. 2014).

---
[9] Obtained from the HST archive; the spectrum was gathered originally under proposal #7402 (P.I., R. C. Peterson).

## 9. IMPLICATIONS FOR Fe-GROUP NUCLEOSYNTHESIS

Our group has recently been examining the Fe-peak abundances in both the Sun and selected metal-poor stars. The Fe-peak elements were synthesized in Type II supernovae (SNe) early in the history of the Galaxy. As described in the previous section, our new mean photospheric abundance for HD 84937 is [V/H] = 2.08 (based upon log $\varepsilon$(V) = 1.87 for HD 84937 and log $\varepsilon$(V) = 3.95 for the Sun). Lawler et al. (2013) had previously found that [Fe/H] = -2.32 in HD 84937, leading to a newly determined value of [V/Fe] = 0.24 for this metal-poor star. Vanadium is an odd-Z element, and as such it has a lower photospheric abundance than the nearby (even-Z) Fe-peak element Ti, even though it is made in a similar manner (see discussion in Lawler et al. 2013). For HD 84937, we now have new precise values for [Ti/Fe] = 0.47 (Lawler et al. 2013, Wood et al. 2013), [Ni/Fe] = -0.07 (Wood et al. 2014) and [V/Fe] = 0.24 (this paper). These new abundance ratios can provide constraints on inputs for SNe models. The Fe-peak elements are synthesized in either complete (e.g., Ni) or incomplete (e.g., Ti and V) silicon burning during the (core collapse) SN phase. Thus, the abundances are a direct indication of such parameters as the SN energy and where the mass cut, the boundary above which matter is ejected in the explosion, is located (see Nakamura et al. 1999). Previously, Lawler et al. (2013) suggested that the precise value of the Ti abundance in HD84937 was higher than various SNe model predictions. One can employ the new V abundance value reported herein to determine [V/Ti] = -0.23 in this one metal-poor star. The only major stable isotope of vanadium is $^{51}$V (see footnote 5), and therefore model predictions (see e.g., Thielemann et al. 1996) would need to compare mass production of isotopic abundances 51 and 48 (the main isotope for Ti) in, for example, a typical 15 $M_\odot$ model.

There have not been extensive listings of data for V, and in particular V II, in the literature. However, previous studies (e.g., Lai et al. 2008, Henry et al. 2010) have suggested that [V/Fe] is solar (i.e., remains flat) over a wide range of metallicity. Our new value for HD 84937 of [V/Fe] = 0.24 at [Fe/H] = -2.32 is slightly elevated, and suggests that [V/Fe] might rise at lower metallicity. We caution that there is some uncertainty in the Fe abundance, which could then lead to uncertainty in the relative [V/Fe] abundance ratio. McWilliam et al. (1995b) had found a few stars with very high [V/Fe] values at low metallicities (see also Johnson 2002). Recently, Roederer (2014, private communication) has assembled a large data base of elements in metal-poor stars, and in Figure 8 we plot that new data for [V/Fe] along with that of McWilliam et al. (1995b) and Gratton & Sneden (1991). There does indeed appear to be a rise in [V/Fe], with large scatter, at very low metallicities. Our new value for HD 84937, indicated by a filled red circle, is consistent with this increasing scatter in [V/Fe] that starts to occur below a metallicity of approximately [Fe/H] = -1. Clearly, additional studies - abundance analyses of additional stars are planned in the near future - will be required to identify any V abundance trends at low metallicities.

## 10. SUMMARY

We report 203 new experimental transition probabilities in V II from a combination of branching fractions measured using FTS and echelle spectra and new LIF radiative lifetimes. Generally good agreement is found with previously reported V II transition probabilities. The use of two spectrometers with independent radiometric calibration methods leads to a reduction in systematic uncertainties and allows for a more thorough examination of optical depth effects. The FTS spectra also yield new measurements for V II HFS constants of selected levels. The new V II transition probabilities and HFS data are used to re-determine the vanadium abundance of the Sun and metal-poor star HD 84937 using lines covering a range of wavelength, E.P., and log(*gf*) values to search for non-LTE effects. Our new solar photospheric vanadium abundance, log ε(V) = 3.95, is slightly lower than previous results, but shows excellent agreement with the solar-system meteoritic value. In HD 84937, we derive [V/H] = -2.08, yielding a value of [V/Fe] = 0.24 for this star, which is consistent with a rise in [V/Fe] at low metallicity subject to uncertainty in the Fe abundance.

The authors acknowledge the contribution of N. Brewer on data analysis for this project. This work is supported in part by NASA grant NNX10AN93G (J.E.L.) and NSF grant AST-1211585 (C.S.). The authors thank Ian Roederer for making available his abundance data on metal-poor stars.

FIGURE CAPTIONS

Figure 1. Comparison of 62 log(*gf*) values with accuracy rank "B" (≤10%) from the NIST ASD by Kramida et al. (2013) to results of this work. The log(*gf*) differences are plotted as a function of wavelength in the upper panel (a) and as a function of the log(*gf*) value reported in this study in the lower panel (b). The solid central line represents perfect agreement, the grey dashed lines indicate ±10% differences in *f*-values, and the error bars are from this work only.

Figure 2. Comparison of 19 log(*gf*) values with accuracy rank "C" (≤18%) or "C+" (≤25%) from the NIST ASD by Kramida et al. (2013) to results of this work. The log(*gf*) differences are plotted as a function of wavelength in the upper panel (a) and as a function of the log(*gf*) value reported in this study in the lower panel (b). The solid central line and error bars have the same meaning as in Figure 1, while the grey dotted lines represent ±25% differences in *f*-values.

Figure 3. Comparison of 15 log(*gf*) values with accuracy rank "D" (≤50%) from the NIST ASD by Kramida et al. (2013) to results of this work. The log(*gf*) differences are plotted as a function of wavelength in the upper panel (a) and as a function of the log(*gf*) value reported in this study in the lower panel (b). The solid central line and error bars have the same meaning as in Figure 1, while the grey dotted lines represent ±50% differences in *f*-values.

Figure 4. Comparison of 137 log(*gf*) values in common between this work and the work of Biémont et al. (1989) plotted as a function of wavelength in the upper panel (a) and as a function of the log(*gf*) value reported in this work in the lower panel (b). The solid line represents perfect

agreement, and the error bars represent the uncertainties reported by Biémont et al. (1989) and the uncertainties reported in this work combined in quadrature.

Figure 5. Relative strengths of V II lines as defined by the vertical axis label; see text for further discussion. The vertical blue line indicates the atmospheric cutoff for ground-based spectroscopy. The horizontal blue line indicates the STR values of barely detectable lines (reduced widths $\log(RW) = -6$). Red circles show the 14 lines used in our solar analysis.

Figure 6. Synthetic and observed photospheric spectra of two V II lines. For each transition, the vertical lines indicate the wavelengths and relative strengths of the hyperfine components. These lines are placed at the component wavelengths, with vertical extents equal to their fractional contribution to the total transition probability of the total V II feature. There are 9 such components for the 3530.7 Å line and 15 for the 4564.4 Å line. The green open circles represent every 4$^{th}$ point from the Delbouille et al. (1973) solar center-of-disk spectrum.

Figure 7. Vanadium abundances in the solar photosphere (top panel) and HD 84937 (bottom panel) derived from V II lines, plotted as functions of wavelength. In each panel, the solid horizontal line represents the mean abundance, and the two dotted lines are placed ±1σ from the mean. The abundance statistics are given in the panels.

Figure 8. [V/Fe] values as a function of metallicity, showing a slight rise and large scatter for [Fe/H] < -1. As indicated in the figure legend, the data sources are McWilliam et al. (1995b) (green diamonds), Gratton & Sneden (1991) (blue squares), and Roederer (2014, private

communication) (cyan triangles), with the HD 84937 result presented in the paper indicated by the red circle.

REFERENCES

Adams, D. L., & Whaling, W. 1981, JOSA, 71, 1036

Asplund, M. 2005, ARA&A, 43, 481

Armstrong, N. M. R., Rosner, S. D., & Holt, R. A. 2011, Phys Scr, 84, 055301

Arvidsson, K. 2003, Master's Thesis (Lund Observatory, Lund University, Sweden)

Barklem, P. S., Christlieb, N., Beers, T. C., et al. 2005, A&A, 439, 129

Beck, D., & Abdalmoneam, M. 2014, in Bulletin of the American Physical Society, 59, 8
    (http://meetings.aps.org/link/BAPS.2014.DAMOP.D1.40)

Biémont, E., Grevesse, N., Faires, L. M., et al. 1989, A&A, 209, 391

Bonifacio, P., Spite, M., Cayrel, R., et al. 2009, A&A, 501, 519

Brault, J. W. 1976, JOSA, 66, 1081

Cayrel, R., Depange, E., Spite, M., et al. 2004, A&A, 416, 1117

Cowan, J. J., Sneden, C., Spite, M., et al. 2002, ApJ, 572, 861

Danzmann, K., & Kock, M. 1982, JOSA, 72, 1556

Delbouille, L., Roland, G., & Neven, L. 1973, Photometric Atlas of the Solar Spectrum from
    λ3000 to λ10000 (Liège: Inst. d'Ap., Univ. de Liège)

Den Hartog, E. A., Lawler, J. E., Sobeck, J. S., Sneden, C., & Cowan, J. J. 2011, ApJS, 194, 35

Den Hartog, E. A., et al. 2014, in preparation

Gratton, R. G., & Sneden, C. 1991, A&A, 541, 501

Hashiguchi, S., & Hasikuni, M. 1985, JPSJ, 54, 1290

Henry, R. B. C., Cowan, J. J., & Sobeck, J. 2010, ApJ, 709, 715

Holweger, H., & Müller, E. A. 1974, SoPh, 39, 19

Johnson, J. A. 2002, ApJS, 139, 219

Table 1. Fourier transform spectra of a custom water-cooled V hollow cathode discharge (HCD) lamp. All spectra were recorded using the 1 m FTS on the McMath telescope at the National Solar Observatory, Kitt Peak, AZ.

| Index | Date | Serial Number | Buffer Gas | Lamp Current (mA) | Wavenumber Range (cm$^{-1}$) | Limit of Resolution (cm$^{-1}$) | Coadds | Beam Splitter | Filter | Detector[a] |
|---|---|---|---|---|---|---|---|---|---|---|
| 1 | 1984 Dec. 9 | 3 | Ar-Ne | 600 | 7764 - 49105 | 0.057 | 12 | UV | | Mid Range Si Diode |
| 2 | 1984 Dec. 9 | 4 | Ar-Ne | 300 | 7764 - 49105 | 0.057 | 8 | UV | | Mid Range Si Diode |
| 3 | 1984 Dec. 9 | 5 | Ar-Ne | 150 | 7764 - 49105 | 0.057 | 8 | UV | | Mid Range Si Diode |
| 4 | 1986 July 30 | 9 | Ar | 500 | 14924 - 37018 | 0.048 | 4 | UV | CuSO$_4$ | Large UV Si Diode |
| 5 | 1986 July 30 | 10 | Ar | 500 | 14924 - 37018 | 0.048 | 4 | UV | CuSO$_4$ | Large UV Si Diode |
| 6 | 1981 June 16 | 7 | Ar | 332 | 6924 - 37564 | 0.043 | 8 | UV | WG295 | UV Mid Range Si Diode |
| 7 | 1981 June 15 | 3 | Ar | 250 | 14878 - 36533 | 0.043 | 8 | UV | CuSO$_4$ WG295 | UV Mid Range Si Diode |
| 8 | 1979 Dec. 12 | 9 | Ar | 300 | 12422 - 31054 | 0.042 | 10 | UV | TC+ 4-97 | Mid Range Si Diode |
| | | | | | | | | | WG345 | |
| 9 | 1979 Dec. 12 | 8 | Ar | 300 | 7716 - 22421 | 0.030 | 8 | UV | GG945 | Super Blue Si Diode |
| 10 | 1980 Sept. 4 | 1 | Ar | 110 | 0 - 17837 | 0.023 | 5 | UV | RG610 | InSb |
| 11 | 1983 Nov. 30 | 3 | Ar | 460 | 2799 - 9518 | 0.011 | 17 | $CaF_2$ | Si | InSb |
| 12 | 1983 Apr. 17 | 4 | Ne-Ar | 370 | 1534 - 5769 | 0.011 | 80 | $CaF_2$ | Ge | InSb |
| 13 | 1984 July 25 | 5 | Ne | 1000 | 12948 - 45407 | 0.054 | 8 | UV | $CuSO_4$ | R166 photomultiplier |
| | | | | | | | | | | Mid Range Si Diode |

[a]Detector types include the Super Blue silicon (Si) photodiode, Large UV Si photodiode, Mid Range Si photodiode, UV Mid Range Si photodiode, a solar blind R166 photomulitplier, and InSb detectors for the IR. The UV beam splitter is fused silica.

Table 2. Echelle spectra of commercial V HCD lamps.

| Index | Date | Serial Numbers[a] | Buffer Gas | Lamp Current (mA) | Wavelength Range (Å) | Resolving Power | Coadds | Exposure Time (s) |
|---|---|---|---|---|---|---|---|---|
| 47-51 | 2013 May 24 | 1, 3, 5, 7, 9 | Ne | 3 | 2200-3900 | 250,000 | 60 | 90 |
| 52-56 | 2013 May 21 | 1, 3, 5, 7, 9 | Ne | 5 | 2200-3900 | 250,000 | 90 | 60 |
| 57-61 | 2013 May 22 | 1, 3, 5, 7, 9 | Ne | 10 | 2200-3900 | 250,000 | 90 | 60 |
| 62-66 | 2013 May 23 | 1, 3, 5, 7, 9 | Ne | 15 | 2200-3900 | 250,000 | 60 | 90 |
| 67-71 | 2013 May 20 | 1, 3, 5, 7, 9 | Ar | 3 | 2200-3900 | 250,000 | 6 | 900 |
| 72-76 | 2013 May 15 | 1, 3, 5, 7, 9 | Ar | 5 | 2200-3900 | 250,000 | 18 | 300 |
| 77-81 | 2013 May 16 | 1, 3, 5, 7, 9 | Ar | 10 | 2200-3900 | 250,000 | 88 | 60 |
| 82-86 | 2013 May 17 | 1, 3, 5, 7, 9 | Ar | 12 | 2200-3900 | 250,000 | 60 | 90 |
| 87-91 | 2014 Jan. 30 | 1, 3, 5, 7, 9 | Ne | 5 | 2000-2800 | 250,000 | 18 | 300 |
| 92-96 | 2014 Jan. 31 | 1, 3, 5, 7, 9 | Ne | 10 | 2000-2800 | 250,000 | 36 | 150 |
| 97-101 | 2014 Feb. 3 | 1, 3, 5, 7, 9 | Ne | 15 | 2000-2800 | 250,000 | 72 | 75 |
| 102-106 | 2014 Feb. 5 | 1, 3, 5, 7, 9 | Ar | 15 | 2000-2800 | 250,000 | 45 | 120 |
| 127-131 | 2014 May 13 | 1, 3, 5, 7, 9 | Ar | 10 | 2100-3200 | 250,000 | 40 | 180 |
| 132-136 | 2014 May 9 | 1, 3, 5, 7, 9 | Ar | 15 | 2100-3200 | 250,000 | 90 | 60 |
| 137-141 | 2014 May 14 | 1, 3, 5, 7, 9 | Ne | 15 | 2100-3200 | 250,000 | 120 | 45 |

[a]At least 3 CCD frames are needed to capture a complete echelle grating order in the UV. In the above data 5 CCD frames are used to provide redundancy and a check for lamp drift.

Table 3. Experimental atomic transition probabilities for 203 lines of V II from upper odd parity levels organized by increasing wavelength in air.

| Wavelength in air[a] (Å) | Upper Level[b] Energy (cm$^{-1}$) | J | Lower Level[b] Energy (cm$^{-1}$) | J | Transition Probability (10$^6$ s$^{-1}$) | log$_{10}$(gf) |
|---|---|---|---|---|---|---|
| 2123.3231 | 47420.230 | 4 | 339.125 | 4 | 5.8 ± 1.0 | -1.45 |
| 2126.9227 | 47108.079 | 1 | 106.643 | 1 | 5.8 ± 0.9 | -1.93 |
| 2129.4687 | 47051.889 | 3 | 106.643 | 1 | 5.9 ± 1.1 | -1.55 |
| 2131.8351 | 47101.932 | 2 | 208.790 | 3 | 9.1 ± 1.3 | -1.51 |
| 2134.0805 | 46879.911 | 2 | 36.102 | 1 | 13.3 ± 2.1 | -1.34 |

Note – Table 3 is available in its entirety via the link to the machine-readable version online.

[a]Wavelength values computed from energy levels using the standard index of air from Peck & Reeder (1972).

[b]Energy levels, parities, and J values are from Thorne et al. (2013).

```
Title: Improved V II log(gf) Values, Hyperfine Structure Constants, and
       Abundance Determinations in the Photospheres of the Sun and
       Metal-poor Star HD 84937
Authors: Wood M.P., Lawler J.E., Den Hartog E.A., Sneden C., & Cowan J.J.
Table: Experimental atomic transition probabilities for 203 lines of V II
       from upper odd-parity levels organized by increasing wavelength in
       air.
================================================================================
Byte-by-byte Description of file: Table3mr.txt
--------------------------------------------------------------------------------
   Bytes Format Units     Label     Explanations
--------------------------------------------------------------------------------
   1-  9 F9.4   0.1nm     WaveAir   Wavelength in air; Angstroms (1)
  11- 19 F9.3   cm-1      UpLev     Upper level (2)
  21     I1     ---       UpJ       Upper level J value (2)
  23- 31 F9.3   cm-1      LowLev    Lower level (2)
  33     I1     ---       LowJ      Lower level J value (2)
  35- 41 F7.3   10+6/s    TranP     Transition probability
  43- 48 F6.3   10+6/s    e_TranP   Total uncertainty in TranP
  50- 54 F5.2   ---       log(gf)   Log of degeneracy times oscillator
                                     strength
--------------------------------------------------------------------------------
Note (1): Computed from energy levels using the standard index of air
          from Peck & Reeder (1972).
Note (2): From Thorne er al. (2013).
--------------------------------------------------------------------------------

2123.3231 47420.230 4   339.125 4   5.8    1.0   -1.45
2126.9227 47108.079 1   106.643 2   5.8    0.9   -1.93
2129.4687 47051.889 3   106.643 2   5.9    1.1   -1.55
2131.8351 47101.932 2   208.790 3   9.1    1.3   -1.51
2134.0805 46879.911 2    36.102 1  13.3    2.1   -1.34
2134.1128 47051.889 3   208.790 3  30.     5.    -0.85
2137.2994 46879.911 2   106.643 2  44.     6.    -0.82
2138.1559 46754.533 1     0.000 0  25.     4.    -1.29
2139.8084 46754.533 1    36.102 1  46.     7.    -1.02
2140.0680 47051.889 3   339.125 4  86.    12.    -0.38
2141.9777 46879.911 2   208.790 3  46.     7.    -0.80
2143.0445 46754.533 1   106.643 2  34.     5.    -1.16
2145.9909 46690.495 1   106.643 2   8.7    1.4   -1.74
2148.4186 46740.008 2   208.790 3   7.8    1.2   -1.57
2672.0004 37520.665 3   106.643 2  21.7    1.4   -0.79
2677.7959 37369.154 2    36.102 1  32.2    2.2   -0.76
2678.5644 37531.132 4   208.790 3  12.2    0.9   -0.93
2679.3159 37520.665 3   208.790 3  31.4    2.0   -0.63
2682.8655 37369.154 2   106.643 2  17.8    1.4   -1.02
2683.0803 37259.529 1     0.000 0  32.9    2.5   -0.97
2685.6826 37259.529 1    36.102 1   6.2    0.5   -1.70
2687.9517 37531.132 4   339.125 4  85.     6.    -0.08
2688.7084 37520.665 3   339.125 4  13.9    0.9   -0.98
2689.8735 37201.538 0    36.102 1  97.     6.    -0.98
2690.2406 37369.154 2   208.790 3  32.2    2.1   -0.76
2690.7822 37259.529 1   106.643 2  52.     4.    -0.77
2694.7359 37205.021 3   106.643 2   0.188  0.022 -2.84
```

```
2700.9275 37352.464 5    339.125 4   35.4    2.0    -0.37
2702.1765 37205.021 3    208.790 3   26.8    1.6    -0.69
2705.2145 36954.686 1      0.000 0    4.5    0.4    -1.83
2706.1564 37150.615 4    208.790 3   33.2    1.9    -0.48
2706.6904 37041.179 2    106.643 2   16.8    1.1    -1.03
2707.8600 36954.686 1     36.102 1   12.3    0.8    -1.39
2711.7303 37205.021 3    339.125 4    8.5    0.7    -1.18
2713.0442 36954.686 1    106.643 2    6.4    0.4    -1.67
2714.1973 37041.179 2    208.790 3    7.6    0.5    -1.38
2715.6547 36919.266 3    106.643 2   30.3    1.8    -0.63
2715.7383 37150.615 4    339.125 4    2.42   0.18   -1.62
2723.2115 36919.266 3    208.790 3    2.33   0.16   -1.74
2728.6373 36673.584 2     36.102 1   20.1    1.2    -0.95
2733.9014 36673.584 2    106.643 2    2.40   0.16   -1.87
2742.4220 36489.437 1     36.102 1    7.9    0.8    -1.57
2742.6731 39612.964 5   3162.966 5    1.92   0.17   -1.62
2743.7721 39403.787 4   2968.389 4    0.84   0.08   -2.07
2808.6905 47108.079 1  11514.784 1    3.20   0.25   -1.94
2824.4280 46690.495 1  11295.513 0    2.91   0.27   -1.98
2840.0887 47108.079 1  11908.261 2    6.0    0.4    -1.67
2840.5848 47101.932 2  11908.261 2    4.7    0.4    -1.55
2869.9608 37520.665 3   2687.208 2    1.89   0.13   -1.79
2875.6857 37369.154 2   2605.040 1    3.30   0.22   -1.69
2879.1594 37531.132 4   2808.959 3    4.3    0.3    -1.32
2880.0276 37520.665 3   2808.959 3   26.1    1.6    -0.64
2882.4990 37369.154 2   2687.208 2   45.5    2.8    -0.55
2884.7830 37259.529 1   2605.040 1   61.     4.     -0.64
2889.6187 37201.538 0   2605.040 1  216.    14.     -0.57
2891.6396 37259.529 1   2687.208 2  154.    10.     -0.24
2892.4409 37531.132 4   2968.389 4   41.4    2.8    -0.33
2892.6542 37369.154 2   2808.959 3  147.     9.     -0.03
2893.3172 37520.665 3   2968.389 4  123.     7.     +0.03
2896.2060 37205.021 3   2687.208 2   21.7    1.2    -0.72
2903.0754 37041.179 2   2605.040 1   31.6    1.7    -0.70
2906.4581 37205.021 3   2808.959 3   80.     4.     -0.15
2907.4714 37352.464 5   2968.389 4   25.4    1.5    -0.45
2908.8174 37531.132 4   3162.966 5  177.    12.     +0.31
2910.0193 37041.179 2   2687.208 2   95.     5.     -0.22
2910.3858 36954.686 1   2605.040 1  119.     6.     -0.34
2911.0629 37150.615 4   2808.959 3   38.7    2.1    -0.35
2917.3647 36954.686 1   2687.208 2   18.4    1.2    -1.15
2919.9933 37205.021 3   2968.389 4   13.6    0.8    -0.92
2920.3835 36919.266 3   2687.208 2   33.1    1.8    -0.53
2924.0190 37352.464 5   3162.966 5  188.     9.     +0.42
2924.6411 37150.615 4   2968.389 4  130.     7.     +0.18
2930.8078 36919.266 3   2808.959 3   61.     3.     -0.26
2934.4007 36673.584 2   2605.040 1   17.4    1.0    -0.95
2941.3852 37150.615 4   3162.966 5   41.7    2.2    -0.31
2941.4954 36673.584 2   2687.208 2   29.7    1.7    -0.71
2944.5712 36919.266 3   2968.389 4   82.     4.     -0.13
2950.3486 36489.437 1   2605.040 1   41.     3.     -0.80
2952.0712 36673.584 2   2808.959 3   72.     4.     -0.33
2957.5207 36489.437 1   2687.208 2   51.     4.     -0.70
2968.3787 47420.230 4  13741.640 3  212.    11.     +0.40
```

```
2975.6514 47108.079 1 13511.799 1  93.    5.    -0.43
2976.1960 47101.932 2 13511.799 1  97.    5.    -0.19
2983.5619 47101.932 2 13594.723 2  53.7   2.9   -0.45
2988.0247 47051.889 3 13594.723 2  40.    3.    -0.43
2995.9995 46879.911 2 13511.799 1  48.    4.    -0.49
3001.2041 47051.889 3 13741.640 3 223.   19.    +0.32
3003.4639 46879.911 2 13594.723 2  69.    6.    -0.33
3007.2997 46754.533 1 13511.799 1  15.8   1.7   -1.19
3008.6143 46740.008 2 13511.799 1  31.9   1.8   -0.66
3013.1043 46690.495 1 13511.799 1 105.    6.    -0.37
3014.8205 46754.533 1 13594.723 2 237.   18.    -0.01
3016.1417 46740.008 2 13594.723 2  23.1   1.4   -0.80
3016.7801 46879.911 2 13741.640 3 130.   10.    -0.05
3093.1002 35483.606 6  3162.966 5 200.   10.    +0.57
3102.3005 35193.182 5  2968.389 4 178.    9.    +0.45
3110.7101 34946.637 4  2808.959 3 157.    8.    +0.31
3118.3816 34745.828 3  2687.208 2 147.    7.    +0.18
3121.1470 35193.182 5  3162.966 5  21.8   1.1   -0.46
3125.2856 34592.843 2  2605.040 1 149.    7.    +0.04
3126.2194 34946.637 4  2968.389 4  37.1   1.9   -0.31
3130.2701 34745.828 3  2808.959 3  46.6   2.3   -0.32
3133.3346 34592.843 2  2687.208 2  43.1   2.3   -0.50
3145.3375 34592.843 2  2808.959 3   3.09  0.22  -1.64
3145.3586 34946.637 4  3162.966 5   1.18  0.19  -1.80
3145.9755 34745.828 3  2968.389 4   2.60  0.20  -1.57
3164.8395 40430.087 4  8842.050 3   4.42  0.25  -1.22
3168.1325 40195.567 3  8640.362 2   7.2   0.8   -1.12
3187.7122 40001.754 2  8640.362 2 112.    6.    -0.07
3188.5129 40195.567 3  8842.050 3 108.    6.    +0.06
3190.6825 40430.087 4  9097.889 4 125.    6.    +0.24
3208.3461 40001.754 2  8842.050 3  18.2   1.0   -0.85
3214.7456 40195.567 3  9097.889 4  14.3   0.8   -0.81
3267.7022 39234.086 3  8640.362 2 157.    8.    +0.25
3271.1225 39403.787 4  8842.050 3 160.    8.    +0.36
3276.1247 39612.964 5  9097.889 4 170.    9.    +0.48
3289.3882 39234.086 3  8842.050 3  10.5   0.6   -0.92
3298.7378 39403.787 4  9097.889 4   7.4   0.4   -0.97
3469.5166 47108.079 1 18293.871 2   9.0   1.0   -1.31
3477.4946 47101.932 2 18353.827 3   5.6   0.8   -1.30
3479.8324 37369.154 2  8640.362 2   2.50  0.26  -1.64
3485.9210 37520.665 3  8842.050 3   4.0   0.4   -1.29
3493.1622 37259.529 1  8640.362 2   7.0   0.8   -1.42
3499.8282 37205.021 3  8640.362 2   0.49  0.07  -2.20
3504.4357 37369.154 2  8842.050 3  18.6   1.8   -0.77
3514.4108 46740.008 2 18293.871 2   4.2   0.7   -1.41
3517.2994 37520.665 3  9097.889 4  44.    4.    -0.24
3517.5215 46690.495 1 18269.514 1   3.7   0.6   -1.69
3520.0190 37041.179 2  8640.362 2   8.1   0.8   -1.13
3520.5388 46690.495 1 18293.871 2   7.9   0.9   -1.36
3521.8340 46740.008 2 18353.827 3  14.3   1.1   -0.88
3524.7160 37205.021 3  8842.050 3   7.9   0.7   -0.99
3530.7720 36954.686 1  8640.362 2  53.    4.    -0.53
3531.4904 37150.615 4  8842.050 3   0.36  0.04  -2.21
3538.2387 37352.464 5  9097.889 4   1.13  0.11  -1.63
```

```
3545.1959 37041.179 2  8842.050 3  51.     4.     -0.32
3556.7999 37205.021 3  9097.889 4  64.     5.     -0.07
3560.5897 36919.266 3  8842.050 3   3.0    0.3    -1.40
3566.1777 36673.584 2  8640.362 2   8.8    0.8    -1.08
3574.3485 47101.932 2 19132.791 2  12.3    1.4    -0.93
3589.7591 36489.437 1  8640.362 2  78.     4.     -0.35
3592.0216 36673.584 2  8842.050 3  52.     4.     -0.30
3593.3330 36919.266 3  9097.889 4  20.9    1.7    -0.55
3621.2087 46740.008 2 19132.791 2  25.9    1.9    -0.59
3625.6113 46740.008 2 19166.314 1   6.1    0.6    -1.22
3627.7151 46690.495 1 19132.791 2  11.0    1.3    -1.19
3632.1336 46690.495 1 19166.314 1   4.2    0.8    -1.61
3674.6849 47108.079 1 19902.608 0   8.9    0.8    -1.27
3700.1245 47108.079 1 20089.650 1  12.7    1.2    -1.11
3700.9665 47101.932 2 20089.650 1   4.5    0.5    -1.33
3703.8190 39612.964 5 12621.485 5   0.54   0.06   -1.91
3709.3258 47108.079 1 20156.670 0   6.1    0.7    -1.42
3715.4638 39612.964 5 12706.078 6  26.3    1.7    -0.22
3718.1508 40430.087 4 13542.645 3   2.36   0.18   -1.36
3722.1315 39403.787 4 12545.100 4   0.58   0.08   -1.97
3727.3412 40430.087 4 13608.939 4  31.7    2.2    -0.23
3731.9693 46690.495 1 19902.608 0   9.2    0.9    -1.24
3732.7476 39403.787 4 12621.485 5  25.7    1.6    -0.32
3735.1560 47108.079 1 20343.046 2  12.9    1.2    -1.09
3736.0141 47101.932 2 20343.046 2   9.2    0.8    -1.01
3743.5972 40195.567 3 13490.883 2   2.77   0.27   -1.39
3745.7992 39234.086 3 12545.100 4  26.6    1.8    -0.41
3750.8677 40195.567 3 13542.645 3  26.8    2.0    -0.40
3760.2208 40195.567 3 13608.939 4   4.8    0.4    -1.15
3767.7039 46690.495 1 20156.670 0   7.0    0.7    -1.35
3770.9662 40001.754 2 13490.883 2  31.5    2.5    -0.47
3774.6699 47101.932 2 20617.073 2   2.7    0.6    -1.55
3778.3435 40001.754 2 13542.645 3   6.3    0.5    -1.17
3787.2392 46740.008 2 20343.046 2  24.1    2.0    -0.59
3794.3565 46690.495 1 20343.046 2  11.2    1.1    -1.14
3826.9679 46740.008 2 20617.073 2   7.5    0.8    -1.08
3863.7854 40430.087 4 14556.068 4   3.15   0.27   -1.20
3865.7093 39403.787 4 13542.645 3   0.30   0.04   -2.21
3866.7219 37369.154 2 11514.784 1   2.8    0.3    -1.50
3875.6446 39403.787 4 13608.939 4   0.62   0.12   -1.90
3878.7074 40430.087 4 14655.607 5  13.4    1.1    -0.57
3884.8361 40195.567 3 14461.748 3   2.64   0.23   -1.38
3896.1379 36954.686 1 11295.513 0   5.6    0.6    -1.42
3899.1276 40195.567 3 14556.068 4  10.6    0.9    -0.77
3903.2525 37520.665 3 11908.261 2   7.8    0.9    -0.91
3916.4045 37041.179 2 11514.784 1   7.3    0.8    -1.07
3926.4802 37369.154 2 11908.261 2   0.94   0.13   -1.96
3929.7201 36954.686 1 11514.784 1   3.7    0.5    -1.59
3951.9570 37205.021 3 11908.261 2  11.3    1.2    -0.73
3968.0884 36489.437 1 11295.513 0   7.2    0.5    -1.29
3973.6283 36673.584 2 11514.784 1   7.4    0.8    -1.06
3977.7204 37041.179 2 11908.261 2   2.30   0.26   -1.56
3989.7890 39612.964 5 14556.068 4   0.47   0.06   -1.90
3997.1098 36919.266 3 11908.261 2   3.7    0.4    -1.20
```

```
4002.9279 36489.437 1 11514.784 1   5.0    0.4    -1.44
4005.7021 39612.964 5 14655.607 5  13.5    1.1    -0.45
4008.1622 39403.787 4 14461.748 3   0.67   0.08   -1.83
4023.3772 39403.787 4 14556.068 4  11.1    0.9    -0.61
4035.6204 39234.086 3 14461.748 3  11.0    0.9    -0.72
4036.7636 36673.584 2 11908.261 2   2.21   0.24   -1.57
4051.0450 39234.086 3 14556.068 4   0.62   0.10   -1.97
4183.4285 40430.087 4 16532.983 5   3.7    0.4    -1.06
4202.3526 37531.132 4 13741.640 3   1.7    0.3    -1.38
4205.0842 40195.567 3 16421.528 4   3.5    0.4    -1.19
4225.2146 40001.754 2 16340.981 3   3.6    0.4    -1.32
4528.4829 40430.087 4 18353.827 3   3.2    0.3    -1.06
4564.5771 40195.567 3 18293.871 2   2.8    0.4    -1.22
4600.1697 40001.754 2 18269.514 1   2.7    0.3    -1.37
```

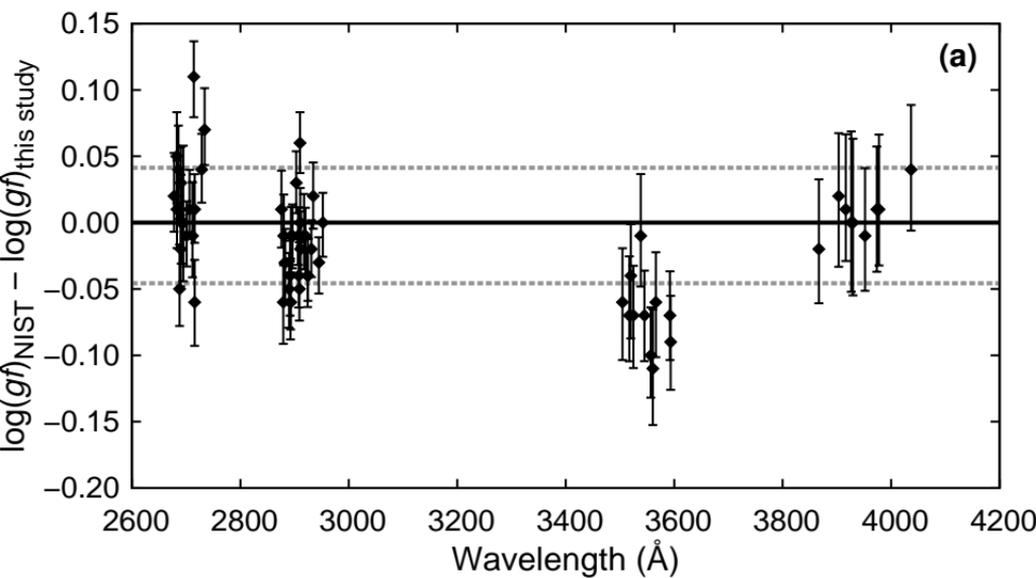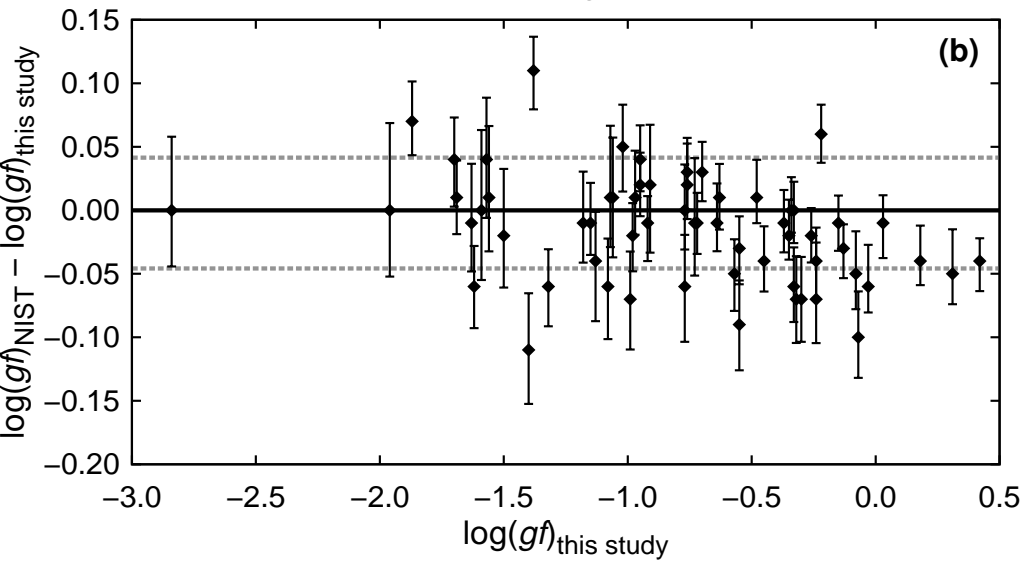

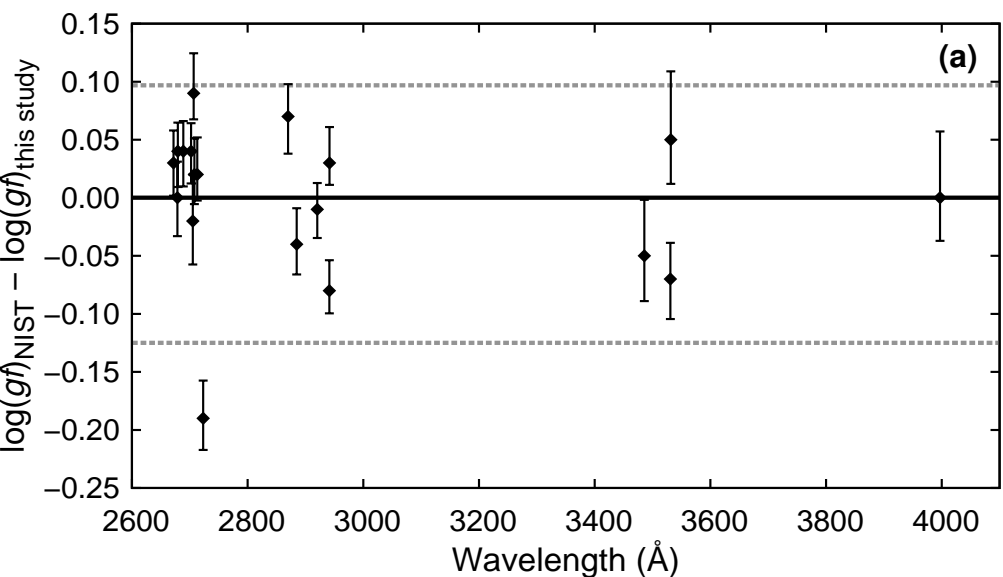
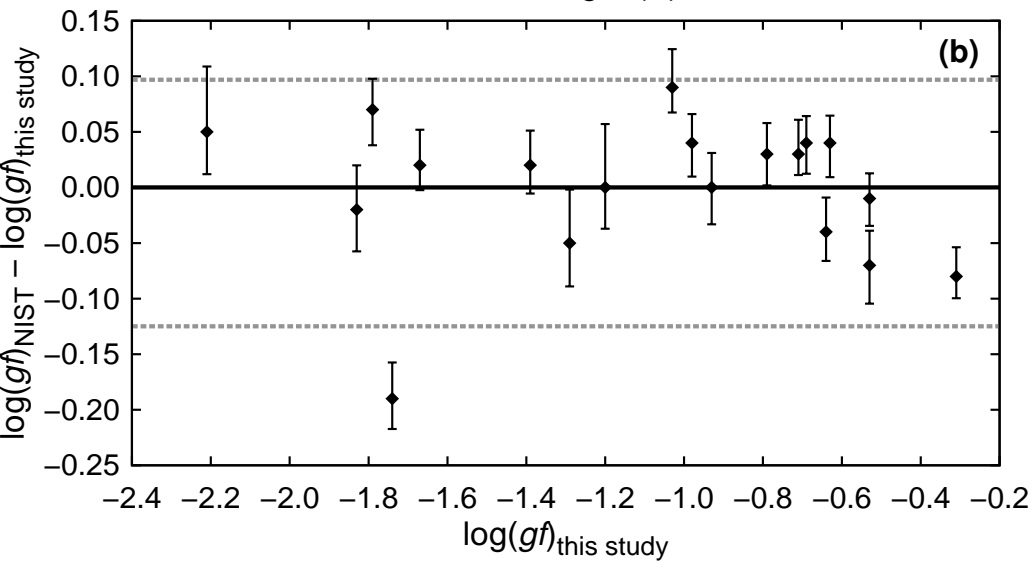

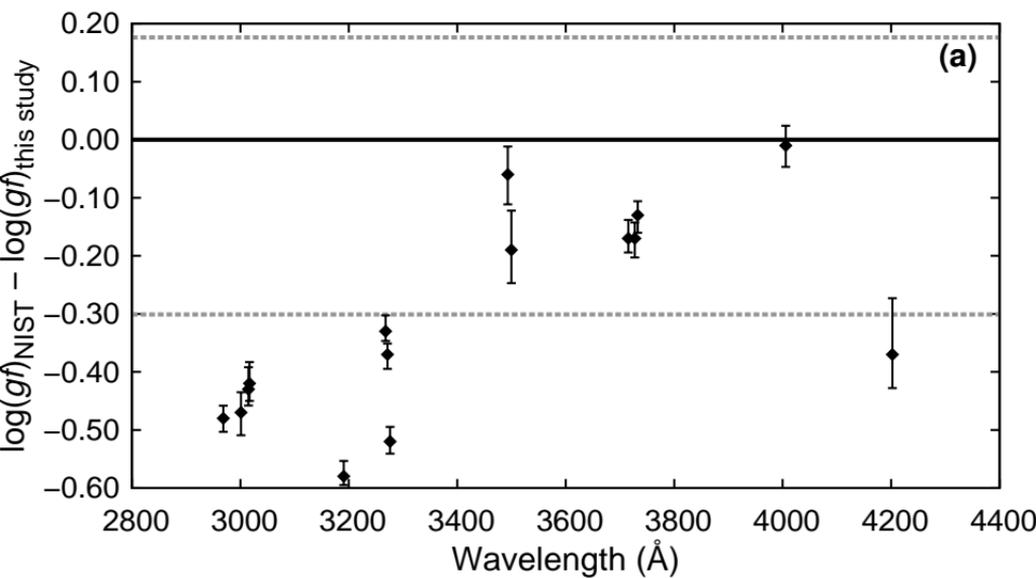
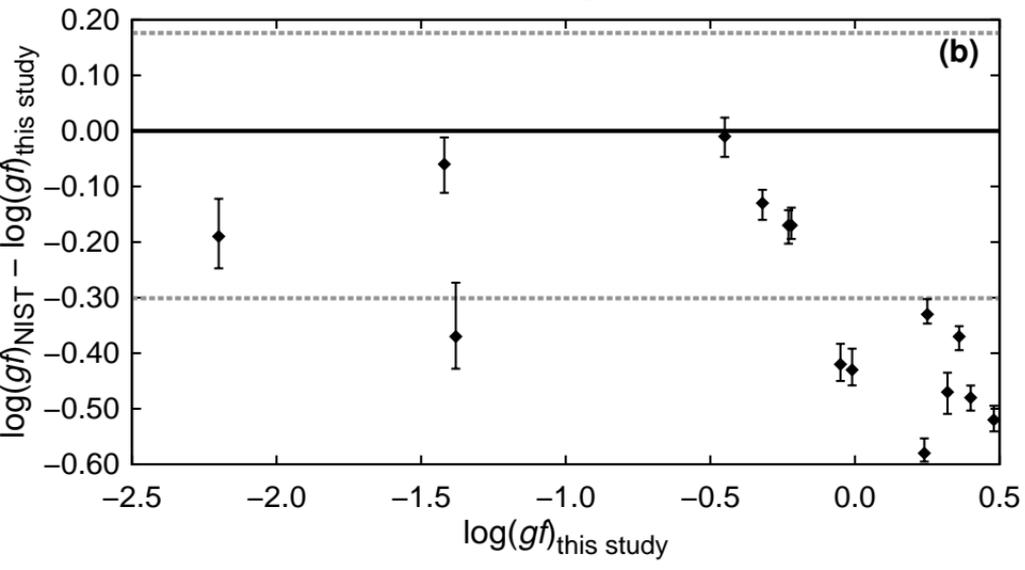

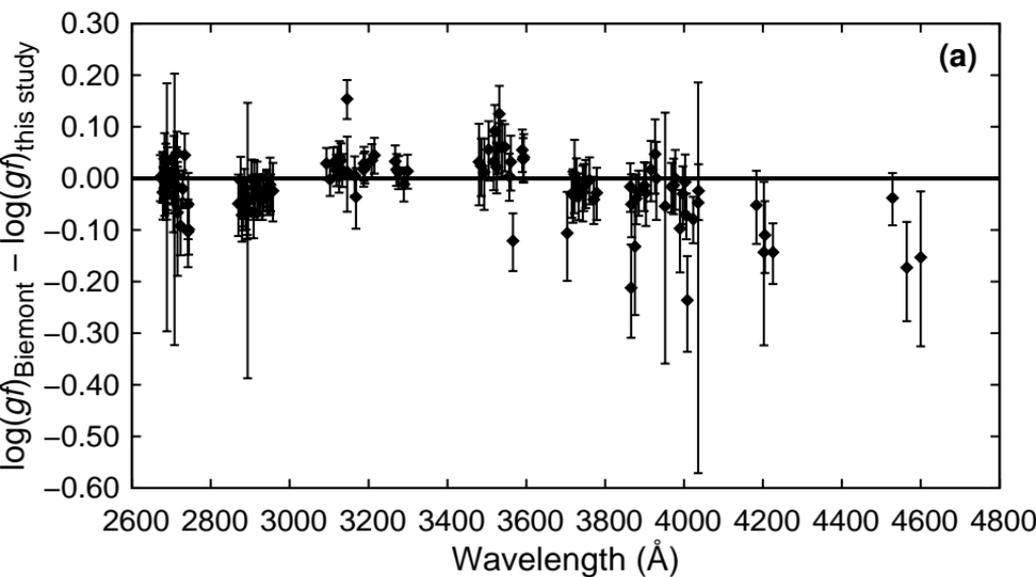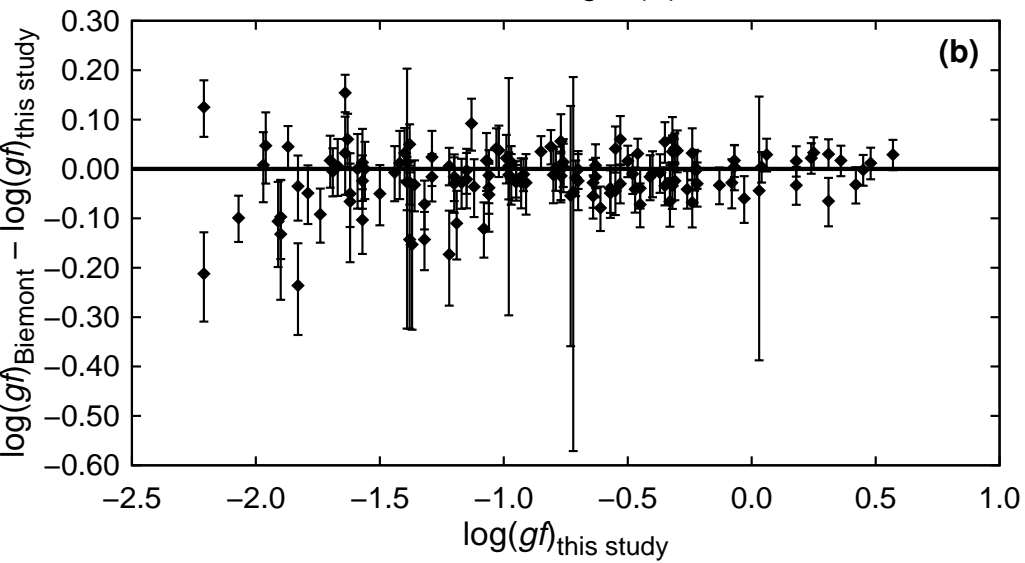

Table 4. New V II magnetic dipole hyperfine structure constants from FTS spectra.

| Configuration[a] | Term[a] | J[a] | Parity[a] | Level Energy[a] (cm$^{-1}$) | Hyperfine Structure Constant $A$ ($10^{-3}$ cm$^{-1}$) | |
|---|---|---|---|---|---|---|
| | | | | | This Work | Arvidsson[b] |
| $3d^3(^4F)4s$ | a$^5$F | 1 | ev | 2605.040 | −13.7 ± 1.0 | 17.0 ± 5.7 |
| $3d^3(^4F)4s$ | a$^5$F | 2 | ev | 2687.208 | 13.5 ± 3.0 | 11.1 ± 1.9 |
| $3d^3(^4F)4s$ | a$^5$F | 3 | ev | 2808.959 | 18.4 ± 1.0 | 19.8 ± 2.0 |
| $3d^3(^4F)4s$ | a$^5$F | 4 | ev | 2968.389 | 20.5 ± 1.0 | 27.6 ± 1.9 |
| $3d^3(^4F)4s$ | a$^5$F | 5 | ev | 3162.966 | 21.8 ± 1.0 | 24.4 ± 1.5 |
| $3d^3(^4F)4s$ | a$^3$F | 2 | ev | 8640.362 | 30.5 ± 1.0 | 31.3 ± 0.2 |
| $3d^3(^4F)4s$ | a$^3$F | 3 | ev | 8842.050 | 6.0 ± 1.0 | |
| $3d^3(^4F)4s$ | a$^3$F | 4 | ev | 9097.889 | −2.9 ± 1.0 | |
| $3d^4$ | a$^3$P | 1 | ev | 11514.784 | 0.0 ± 2.0 | |
| $3d^4$ | a$^3$P | 2 | ev | 11908.261 | 0.0 ± 2.0 | |
| $3d^4$ | a$^3$G | 3 | ev | 14461.748 | 5.0 ± 1.0 | |

| | | | | | |
|---|---|---|---|---|---|
| $3d^3(^4F)4p$ | $z^5G$ | 2 | od | 34592.843 | 26.0 ± 1.5 |
| $3d^3(^4F)4p$ | $z^5G$ | 3 | od | 34745.828 | 14.4 ± 1.5 |
| $3d^3(^4F)4p$ | $z^5G$ | 4 | od | 34946.637 | 9.0 ± 1.5 |
| $3d^3(^4F)4p$ | $z^5G$ | 5 | od | 35193.182 | 6.7 ± 1.5 |
| $3d^3(^4F)4p$ | $z^5G$ | 6 | od | 35483.606 | 5.2 ± 2.0 |
| $3d^3(^4F)4p$ | $z^5F$ | 3 | od | 36919.266 | 4.9 ± 1.5 |
| $3d^3(^4F)4p$ | $z^5F$ | 4 | od | 37150.615 | 4.9 ± 1.5 |
| $3d^3(^4F)4p$ | $z^5D$ | 3 | od | 37520.665 | 2.5 ± 1.5 |
| $3d^3(^4F)4p$ | $z^5P$ | 2 | od | 46879.911 | -4.2 ± 1.5 |
| $3d^3(^4F)4p$ | $z^5P$ | 3 | od | 47051.889 | 1.1 ± 1.5 |

[a]Level energies and classifications are from Thorne et al. (2013).

[b]Arvidsson (2003) master's thesis at Lund University, Lund, Sweden.

Table 5. Hyperfine structure line component patterns for V II.

| Center of gravity wavenumber[a] (cm$^{-1}$) | Center of gravity air wavelength[b] (Å) | $F_{upper}$[a] | $F_{lower}$[a] | Component offset from center of gravity wavenumber (cm$^{-1}$) | Component offset from center of gravity wavelength (Å) | Component strength[c] |
|---|---|---|---|---|---|---|
| 37414.022 | 2672.0004 | 6.5 | 5.5 | 0.02625 | -0.001875 | 0.25000 |
| 37414.022 | 2672.0004 | 5.5 | 5.5 | 0.01000 | -0.000714 | 0.04545 |
| 37414.022 | 2672.0004 | 5.5 | 4.5 | 0.01000 | -0.000714 | 0.16883 |
| 37414.022 | 2672.0004 | 4.5 | 5.5 | -0.00375 | 0.000268 | 0.00455 |
| 37414.022 | 2672.0004 | 4.5 | 4.5 | -0.00375 | 0.000268 | 0.06926 |

Note –Table 5 is available in its entirety via the link to the machine-readable version online.

[a]Energy levels and total angular momentum F values are from Thorne et al. (2013).

[b]Wavelength values computed from energy levels using the standard index of air from Peck & Reeder (1972).

[c]Component strengths are normalized to sum to unity.

```
Title: Improved V II log(gf) Values, Hyperfine Structure Constants, and
       Abundance Determinations in the Photospheres of the Sun and
       Metal-poor Star HD 84937
Authors: Wood M.P., Lawler J.E., Den Hartog E.A., Sneden C., & Cowan J.J.
Table: Hyperfine structure line component patterns for V II.
================================================================================
Byte-by-byte Description of file: Table5mr.txt
--------------------------------------------------------------------------------
   Bytes Format Units   Label    Explanations
--------------------------------------------------------------------------------
   1-  9 F9.3   cm-1    WaveNum  Center-of-Gravity Wavenumber
  11- 19 F9.4   0.1nm   WaveAir  Center-of-Gravity  Air Wavelength in
                                 Angstroms
  21- 23 F3.1   ---     Fupp     Component upper level F or total angular
                                 momentum
  25- 27 F3.1   ---     Flow     Component lower level F or total angular
                                 momentum
  29- 36 F8.5   cm-1    CWaveR   Component offset Wavenumber with respect to
                                 Center-of-gravity Wavenumber
  38- 46 F9.6   0.1nm   CWave    Component offset Wavelength with respect to
                                 Center-of-gravity Wavelength
  48- 54 F7.5   ---     Str      Component strength (1)
--------------------------------------------------------------------------------
Note (1): Normalized to sum to one.
--------------------------------------------------------------------------------

37414.022 2672.0004 6.5 5.5  0.02625 -0.001875 0.25000
37414.022 2672.0004 5.5 5.5  0.01000 -0.000714 0.04545
37414.022 2672.0004 5.5 4.5  0.01000 -0.000714 0.16883
37414.022 2672.0004 4.5 5.5 -0.00375  0.000268 0.00455
37414.022 2672.0004 4.5 4.5 -0.00375  0.000268 0.06926
37414.022 2672.0004 4.5 3.5 -0.00375  0.000268 0.10476
37414.022 2672.0004 3.5 4.5 -0.01500  0.001071 0.01190
37414.022 2672.0004 3.5 3.5 -0.01500  0.001071 0.07483
37414.022 2672.0004 3.5 2.5 -0.01500  0.001071 0.05612
37414.022 2672.0004 2.5 3.5 -0.02375  0.001696 0.02041
37414.022 2672.0004 2.5 2.5 -0.02375  0.001696 0.06531
37414.022 2672.0004 2.5 1.5 -0.02375  0.001696 0.02143
37414.022 2672.0004 1.5 2.5 -0.03000  0.002143 0.02857
37414.022 2672.0004 1.5 1.5 -0.03000  0.002143 0.04286
37414.022 2672.0004 0.5 1.5 -0.03375  0.002410 0.03571
37311.875 2679.3159 6.5 6.5  0.02625 -0.001885 0.21635
37311.875 2679.3159 6.5 5.5  0.02625 -0.001885 0.03365
37311.875 2679.3159 5.5 6.5  0.01000 -0.000718 0.03365
37311.875 2679.3159 5.5 5.5  0.01000 -0.000718 0.12787
37311.875 2679.3159 5.5 4.5  0.01000 -0.000718 0.05276
37311.875 2679.3159 4.5 5.5 -0.00375  0.000269 0.05276
37311.875 2679.3159 4.5 4.5 -0.00375  0.000269 0.06629
37311.875 2679.3159 4.5 3.5 -0.00375  0.000269 0.05952
37311.875 2679.3159 3.5 4.5 -0.01500  0.001077 0.05952
37311.875 2679.3159 3.5 3.5 -0.01500  0.001077 0.02721
37311.875 2679.3159 3.5 2.5 -0.01500  0.001077 0.05612
37311.875 2679.3159 2.5 3.5 -0.02375  0.001706 0.05612
37311.875 2679.3159 2.5 2.5 -0.02375  0.001706 0.00638
```

```
37311.875 2679.3159 2.5 1.5 -0.02375  0.001706 0.04464
37311.875 2679.3159 1.5 2.5 -0.03000  0.002154 0.04464
37311.875 2679.3159 1.5 1.5 -0.03000  0.002154 0.00000
37311.875 2679.3159 1.5 0.5 -0.03000  0.002154 0.02679
37311.875 2679.3159 0.5 1.5 -0.03375  0.002424 0.02679
37311.875 2679.3159 0.5 0.5 -0.03375  0.002424 0.00893
37192.007 2687.9517 7.5 7.5  0.03997 -0.002889 0.20148
37192.007 2687.9517 7.5 6.5  0.03997 -0.002889 0.02074
37192.007 2687.9517 6.5 7.5  0.01856 -0.001341 0.02074
37192.007 2687.9517 6.5 6.5  0.01856 -0.001341 0.14005
37192.007 2687.9517 6.5 5.5  0.01856 -0.001341 0.03365
37192.007 2687.9517 5.5 6.5  0.00000  0.000000 0.03365
37192.007 2687.9517 5.5 5.5  0.00000  0.000000 0.09324
37192.007 2687.9517 5.5 4.5  0.00000  0.000000 0.03977
37192.007 2687.9517 4.5 5.5 -0.01570  0.001135 0.03977
37192.007 2687.9517 4.5 4.5 -0.01570  0.001135 0.05899
37192.007 2687.9517 4.5 3.5 -0.01570  0.001135 0.04012
37192.007 2687.9517 3.5 4.5 -0.02855  0.002064 0.04012
37192.007 2687.9517 3.5 3.5 -0.02855  0.002064 0.03527
37192.007 2687.9517 3.5 2.5 -0.02855  0.002064 0.03571
37192.007 2687.9517 2.5 3.5 -0.03855  0.002786 0.03571
37192.007 2687.9517 2.5 2.5 -0.03855  0.002786 0.02012
37192.007 2687.9517 2.5 1.5 -0.03855  0.002786 0.02750
37192.007 2687.9517 1.5 2.5 -0.04568  0.003302 0.02750
37192.007 2687.9517 1.5 1.5 -0.04568  0.003302 0.01185
37192.007 2687.9517 1.5 0.5 -0.04568  0.003302 0.01620
37192.007 2687.9517 0.5 1.5 -0.04997  0.003612 0.01620
37192.007 2687.9517 0.5 0.5 -0.04997  0.003612 0.01157
37181.540 2688.7084 6.5 7.5  0.02625 -0.001898 0.22222
37181.540 2688.7084 6.5 6.5  0.02625 -0.001898 0.02618
37181.540 2688.7084 6.5 5.5  0.02625 -0.001898 0.00160
37181.540 2688.7084 5.5 6.5  0.01000 -0.000723 0.16827
37181.540 2688.7084 5.5 5.5  0.01000 -0.000723 0.04196
37181.540 2688.7084 5.5 4.5  0.01000 -0.000723 0.00406
37181.540 2688.7084 4.5 5.5 -0.00375  0.000271 0.12311
37181.540 2688.7084 4.5 4.5 -0.00375  0.000271 0.04885
37181.540 2688.7084 4.5 3.5 -0.00375  0.000271 0.00661
37181.540 2688.7084 3.5 4.5 -0.01500  0.001085 0.08598
37181.540 2688.7084 3.5 3.5 -0.01500  0.001085 0.04837
37181.540 2688.7084 3.5 2.5 -0.01500  0.001085 0.00850
37181.540 2688.7084 2.5 3.5 -0.02375  0.001718 0.05612
37181.540 2688.7084 2.5 2.5 -0.02375  0.001718 0.04209
37181.540 2688.7084 2.5 1.5 -0.02375  0.001718 0.00893
37181.540 2688.7084 1.5 2.5 -0.03000  0.002170 0.03274
37181.540 2688.7084 1.5 1.5 -0.03000  0.002170 0.03175
37181.540 2688.7084 1.5 0.5 -0.03000  0.002170 0.00694
37181.540 2688.7084 0.5 1.5 -0.03375  0.002441 0.01488
37181.540 2688.7084 0.5 0.5 -0.03375  0.002441 0.02083
37013.339 2700.9275 8.5 7.5  0.09399 -0.006859 0.20455
37013.339 2700.9275 7.5 7.5  0.04834 -0.003528 0.01697
37013.339 2700.9275 7.5 6.5  0.04834 -0.003528 0.16485
37013.339 2700.9275 6.5 7.5  0.00806 -0.000588 0.00071
37013.339 2700.9275 6.5 6.5  0.00806 -0.000588 0.02785
37013.339 2700.9275 6.5 5.5  0.00806 -0.000588 0.13054
```

```
37013.339  2700.9275  5.5  6.5  -0.02686   0.001960  0.00175
37013.339  2700.9275  5.5  5.5  -0.02686   0.001960  0.03338
37013.339  2700.9275  5.5  4.5  -0.02686   0.001960  0.10124
37013.339  2700.9275  4.5  5.5  -0.05640   0.004116  0.00275
37013.339  2700.9275  4.5  4.5  -0.05640   0.004116  0.03428
37013.339  2700.9275  4.5  3.5  -0.05640   0.004116  0.07660
37013.339  2700.9275  3.5  4.5  -0.08057   0.005879  0.00337
37013.339  2700.9275  3.5  3.5  -0.08057   0.005879  0.03127
37013.339  2700.9275  3.5  2.5  -0.08057   0.005879  0.05628
37013.339  2700.9275  2.5  3.5  -0.09936   0.007251  0.00325
37013.339  2700.9275  2.5  2.5  -0.09936   0.007251  0.02494
37013.339  2700.9275  2.5  1.5  -0.09936   0.007251  0.04000
37013.339  2700.9275  1.5  2.5  -0.11279   0.008231  0.00212
37013.339  2700.9275  1.5  1.5  -0.11279   0.008231  0.01556
37013.339  2700.9275  1.5  0.5  -0.11279   0.008231  0.02778
36996.231  2702.1765  6.5  6.5   0.05612  -0.004099  0.21635
36996.231  2702.1765  6.5  5.5   0.05612  -0.004099  0.03365
36996.231  2702.1765  5.5  6.5   0.02138  -0.001561  0.03365
36996.231  2702.1765  5.5  5.5   0.02138  -0.001561  0.12787
36996.231  2702.1765  5.5  4.5   0.02138  -0.001561  0.05276
36996.231  2702.1765  4.5  5.5  -0.00802   0.000586  0.05276
36996.231  2702.1765  4.5  4.5  -0.00802   0.000586  0.06629
36996.231  2702.1765  4.5  3.5  -0.00802   0.000586  0.05952
36996.231  2702.1765  3.5  4.5  -0.03207   0.002342  0.05952
36996.231  2702.1765  3.5  3.5  -0.03207   0.002342  0.02721
36996.231  2702.1765  3.5  2.5  -0.03207   0.002342  0.05612
36996.231  2702.1765  2.5  3.5  -0.05077   0.003709  0.05612
36996.231  2702.1765  2.5  2.5  -0.05077   0.003709  0.00638
36996.231  2702.1765  2.5  1.5  -0.05077   0.003709  0.04464
36996.231  2702.1765  1.5  2.5  -0.06413   0.004684  0.04464
36996.231  2702.1765  1.5  1.5  -0.06413   0.004684  0.00000
36996.231  2702.1765  1.5  0.5  -0.06413   0.004684  0.02679
36996.231  2702.1765  0.5  1.5  -0.07215   0.005270  0.02679
36996.231  2702.1765  0.5  0.5  -0.07215   0.005270  0.00893
36918.584  2707.8600  4.5  4.5   0.09660  -0.007085  0.25463
36918.584  2707.8600  4.5  3.5   0.09660  -0.007085  0.16204
36918.584  2707.8600  3.5  4.5  -0.02760   0.002024  0.16204
36918.584  2707.8600  3.5  3.5  -0.02760   0.002024  0.01058
36918.584  2707.8600  3.5  2.5  -0.02760   0.002024  0.16071
36918.584  2707.8600  2.5  3.5  -0.12419   0.009110  0.16071
36918.584  2707.8600  2.5  2.5  -0.12419   0.009110  0.08929
36848.043  2713.0442  4.5  5.5   0.09660  -0.007113  0.30000
36848.043  2713.0442  4.5  4.5   0.09660  -0.007113  0.09722
36848.043  2713.0442  4.5  3.5   0.09660  -0.007113  0.01944
36848.043  2713.0442  3.5  4.5  -0.02760   0.002032  0.15278
36848.043  2713.0442  3.5  3.5  -0.02760   0.002032  0.12698
36848.043  2713.0442  3.5  2.5  -0.02760   0.002032  0.05357
36848.043  2713.0442  2.5  3.5  -0.12419   0.009145  0.05357
36848.043  2713.0442  2.5  2.5  -0.12419   0.009145  0.09643
36848.043  2713.0442  2.5  1.5  -0.12419   0.009145  0.10000
36812.623  2715.6547  6.5  5.5   0.05145  -0.003796  0.25000
36812.623  2715.6547  5.5  5.5   0.01960  -0.001446  0.04545
36812.623  2715.6547  5.5  4.5   0.01960  -0.001446  0.16883
36812.623  2715.6547  4.5  5.5  -0.00735   0.000542  0.00455
```

```
36812.623  2715.6547  4.5  4.5  -0.00735   0.000542  0.06926
36812.623  2715.6547  4.5  3.5  -0.00735   0.000542  0.10476
36812.623  2715.6547  3.5  4.5  -0.02940   0.002169  0.01190
36812.623  2715.6547  3.5  3.5  -0.02940   0.002169  0.07483
36812.623  2715.6547  3.5  2.5  -0.02940   0.002169  0.05612
36812.623  2715.6547  2.5  3.5  -0.04655   0.003434  0.02041
36812.623  2715.6547  2.5  2.5  -0.04655   0.003434  0.06531
36812.623  2715.6547  2.5  1.5  -0.04655   0.003434  0.02143
36812.623  2715.6547  1.5  2.5  -0.05880   0.004338  0.02857
36812.623  2715.6547  1.5  1.5  -0.05880   0.004338  0.04286
36812.623  2715.6547  0.5  1.5  -0.06615   0.004880  0.03571
36811.490  2715.7383  7.5  7.5   0.06860  -0.005061  0.20148
36811.490  2715.7383  7.5  6.5   0.06860  -0.005061  0.02074
36811.490  2715.7383  6.5  7.5   0.03185  -0.002350  0.02074
36811.490  2715.7383  6.5  6.5   0.03185  -0.002350  0.14005
36811.490  2715.7383  6.5  5.5   0.03185  -0.002350  0.03365
36811.490  2715.7383  5.5  6.5   0.00000   0.000000  0.03365
36811.490  2715.7383  5.5  5.5   0.00000   0.000000  0.09324
36811.490  2715.7383  5.5  4.5   0.00000   0.000000  0.03977
36811.490  2715.7383  4.5  5.5  -0.02695   0.001988  0.03977
36811.490  2715.7383  4.5  4.5  -0.02695   0.001988  0.05899
36811.490  2715.7383  4.5  3.5  -0.02695   0.001988  0.04012
36811.490  2715.7383  3.5  4.5  -0.04900   0.003615  0.04012
36811.490  2715.7383  3.5  3.5  -0.04900   0.003615  0.03527
36811.490  2715.7383  3.5  2.5  -0.04900   0.003615  0.03571
36811.490  2715.7383  2.5  3.5  -0.06615   0.004880  0.03571
36811.490  2715.7383  2.5  2.5  -0.06615   0.004880  0.02012
36811.490  2715.7383  2.5  1.5  -0.06615   0.004880  0.02750
36811.490  2715.7383  1.5  2.5  -0.07840   0.005784  0.02750
36811.490  2715.7383  1.5  1.5  -0.07840   0.005784  0.01185
36811.490  2715.7383  1.5  0.5  -0.07840   0.005784  0.01620
36811.490  2715.7383  0.5  1.5  -0.08575   0.006326  0.01620
36811.490  2715.7383  0.5  0.5  -0.08575   0.006326  0.01157
36637.482  2728.6373  5.5  4.5   0.05593  -0.004165  0.30000
36637.482  2728.6373  4.5  4.5   0.01198  -0.000893  0.09722
36637.482  2728.6373  4.5  3.5   0.01198  -0.000893  0.15278
36637.482  2728.6373  3.5  4.5  -0.02397   0.001785  0.01944
36637.482  2728.6373  3.5  3.5  -0.02397   0.001785  0.12698
36637.482  2728.6373  3.5  2.5  -0.02397   0.001785  0.05357
36637.482  2728.6373  2.5  3.5  -0.05193   0.003868  0.05357
36637.482  2728.6373  2.5  2.5  -0.05193   0.003868  0.09643
36637.482  2728.6373  1.5  2.5  -0.07191   0.005356  0.10000
34711.706  2880.0276  6.5  6.5  -0.16695   0.013853  0.21635
34711.706  2880.0276  6.5  5.5  -0.04735   0.003929  0.03365
34711.706  2880.0276  5.5  6.5  -0.18320   0.015201  0.03365
34711.706  2880.0276  5.5  5.5  -0.06360   0.005277  0.12787
34711.706  2880.0276  5.5  4.5   0.03760  -0.003120  0.05276
34711.706  2880.0276  4.5  5.5  -0.07735   0.006418  0.05276
34711.706  2880.0276  4.5  4.5   0.02385  -0.001979  0.06629
34711.706  2880.0276  4.5  3.5   0.10665  -0.008849  0.05952
34711.706  2880.0276  3.5  4.5   0.01260  -0.001045  0.05952
34711.706  2880.0276  3.5  3.5   0.09540  -0.007916  0.02721
34711.706  2880.0276  3.5  2.5   0.15980  -0.013259  0.05612
34711.706  2880.0276  2.5  3.5   0.08665  -0.007190  0.05612
```

```
34711.706 2880.0276 2.5 2.5   0.15105 -0.012533 0.00638
34711.706 2880.0276 2.5 1.5   0.19705 -0.016350 0.04464
34711.706 2880.0276 1.5 2.5   0.14480 -0.012015 0.04464
34711.706 2880.0276 1.5 1.5   0.19080 -0.015831 0.00000
34711.706 2880.0276 1.5 0.5   0.21840 -0.018121 0.02679
34711.706 2880.0276 0.5 1.5   0.18705 -0.015520 0.02679
34711.706 2880.0276 0.5 0.5   0.21465 -0.017810 0.00893
34562.743 2892.4409 7.5 7.5  -0.24703  0.020674 0.20148
34562.743 2892.4409 7.5 6.5  -0.09328  0.007806 0.02074
34562.743 2892.4409 6.5 7.5  -0.26844  0.022466 0.02074
34562.743 2892.4409 6.5 6.5  -0.11469  0.009599 0.14005
34562.743 2892.4409 6.5 5.5   0.01856 -0.001553 0.03365
34562.743 2892.4409 5.5 6.5  -0.13325  0.011152 0.03365
34562.743 2892.4409 5.5 5.5   0.00000  0.000000 0.09324
34562.743 2892.4409 5.5 4.5   0.11275 -0.009436 0.03977
34562.743 2892.4409 4.5 5.5  -0.01570  0.001314 0.03977
34562.743 2892.4409 4.5 4.5   0.09705 -0.008122 0.05899
34562.743 2892.4409 4.5 3.5   0.18930 -0.015842 0.04012
34562.743 2892.4409 3.5 4.5   0.08420 -0.007046 0.04012
34562.743 2892.4409 3.5 3.5   0.17645 -0.014767 0.03527
34562.743 2892.4409 3.5 2.5   0.24820 -0.020772 0.03571
34562.743 2892.4409 2.5 3.5   0.16645 -0.013931 0.03571
34562.743 2892.4409 2.5 2.5   0.23820 -0.019935 0.02012
34562.743 2892.4409 2.5 1.5   0.28945 -0.024224 0.02750
34562.743 2892.4409 1.5 2.5   0.23107 -0.019338 0.02750
34562.743 2892.4409 1.5 1.5   0.28232 -0.023627 0.01185
34562.743 2892.4409 1.5 0.5   0.31307 -0.026200 0.01620
34562.743 2892.4409 0.5 1.5   0.27803 -0.023268 0.01620
34562.743 2892.4409 0.5 0.5   0.30878 -0.025842 0.01157
34552.276 2893.3172 6.5 7.5  -0.26075  0.021836 0.22222
34552.276 2893.3172 6.5 6.5  -0.10700  0.008960 0.02618
34552.276 2893.3172 6.5 5.5   0.02625 -0.002198 0.00160
34552.276 2893.3172 5.5 6.5  -0.12325  0.010321 0.16827
34552.276 2893.3172 5.5 5.5   0.01000 -0.000837 0.04196
34552.276 2893.3172 5.5 4.5   0.12275 -0.010279 0.00406
34552.276 2893.3172 4.5 5.5  -0.00375  0.000314 0.12311
34552.276 2893.3172 4.5 4.5   0.10900 -0.009128 0.04885
34552.276 2893.3172 4.5 3.5   0.20125 -0.016853 0.00661
34552.276 2893.3172 3.5 4.5   0.09775 -0.008186 0.08598
34552.276 2893.3172 3.5 3.5   0.19000 -0.015911 0.04837
34552.276 2893.3172 3.5 2.5   0.26175 -0.021919 0.00850
34552.276 2893.3172 2.5 3.5   0.18125 -0.015178 0.05612
34552.276 2893.3172 2.5 2.5   0.25300 -0.021186 0.04209
34552.276 2893.3172 2.5 1.5   0.30425 -0.025478 0.00893
34552.276 2893.3172 1.5 2.5   0.24675 -0.020663 0.03274
34552.276 2893.3172 1.5 1.5   0.29800 -0.024955 0.03175
34552.276 2893.3172 1.5 0.5   0.32875 -0.027530 0.00694
34552.276 2893.3172 0.5 1.5   0.29425 -0.024641 0.01488
34552.276 2893.3172 0.5 0.5   0.32500 -0.027216 0.02083
34517.813 2896.2060 6.5 5.5  -0.03838  0.003221 0.25000
34517.813 2896.2060 5.5 5.5  -0.07312  0.006136 0.04545
34517.813 2896.2060 5.5 4.5   0.00113 -0.000095 0.16883
34517.813 2896.2060 4.5 5.5  -0.10252  0.008602 0.00455
34517.813 2896.2060 4.5 4.5  -0.02827  0.002372 0.06926
```

```
34517.813 2896.2060 4.5 3.5   0.03248 -0.002726 0.10476
34517.813 2896.2060 3.5 4.5  -0.05232  0.004390 0.01190
34517.813 2896.2060 3.5 3.5   0.00843 -0.000708 0.07483
34517.813 2896.2060 3.5 2.5   0.05568 -0.004672 0.05612
34517.813 2896.2060 2.5 3.5  -0.01027  0.000862 0.02041
34517.813 2896.2060 2.5 2.5   0.03698 -0.003103 0.06531
34517.813 2896.2060 2.5 1.5   0.07073 -0.005935 0.02143
34517.813 2896.2060 1.5 2.5   0.02362 -0.001982 0.02857
34517.813 2896.2060 1.5 1.5   0.05737 -0.004814 0.04286
34517.813 2896.2060 0.5 1.5   0.04935 -0.004141 0.03571
34436.139 2903.0754 5.5 4.5   0.11808 -0.009955 0.30000
34436.139 2903.0754 4.5 4.5   0.06298 -0.005310 0.09722
34436.139 2903.0754 4.5 3.5   0.00133 -0.000112 0.15278
34436.139 2903.0754 3.5 4.5   0.01789 -0.001508 0.01944
34436.139 2903.0754 3.5 3.5  -0.04376  0.003689 0.12698
34436.139 2903.0754 3.5 2.5  -0.09171  0.007732 0.05357
34436.139 2903.0754 2.5 3.5  -0.07883  0.006646 0.05357
34436.139 2903.0754 2.5 2.5  -0.12678  0.010688 0.09643
34436.139 2903.0754 1.5 2.5  -0.15182  0.012800 0.10000
34396.062 2906.4581 6.5 6.5  -0.13708  0.011584 0.21635
34396.062 2906.4581 6.5 5.5  -0.01748  0.001477 0.03365
34396.062 2906.4581 5.5 6.5  -0.17182  0.014520 0.03365
34396.062 2906.4581 5.5 5.5  -0.05222  0.004413 0.12787
34396.062 2906.4581 5.5 4.5   0.04898 -0.004139 0.05276
34396.062 2906.4581 4.5 5.5  -0.08162  0.006897 0.05276
34396.062 2906.4581 4.5 4.5   0.01958 -0.001655 0.06629
34396.062 2906.4581 4.5 3.5   0.10238 -0.008652 0.05952
34396.062 2906.4581 3.5 4.5  -0.00447  0.000377 0.05952
34396.062 2906.4581 3.5 3.5   0.07833 -0.006619 0.02721
34396.062 2906.4581 3.5 2.5   0.14273 -0.012061 0.05612
34396.062 2906.4581 2.5 3.5   0.05963 -0.005039 0.05612
34396.062 2906.4581 2.5 2.5   0.12403 -0.010481 0.00638
34396.062 2906.4581 2.5 1.5   0.17003 -0.014368 0.04464
34396.062 2906.4581 1.5 2.5   0.11067 -0.009352 0.04464
34396.062 2906.4581 1.5 1.5   0.15667 -0.013239 0.00000
34396.062 2906.4581 1.5 0.5   0.18427 -0.015571 0.02679
34396.062 2906.4581 0.5 1.5   0.14865 -0.012562 0.02679
34396.062 2906.4581 0.5 0.5   0.17625 -0.014894 0.00893
34384.075 2907.4714 8.5 7.5  -0.19301  0.016321 0.20455
34384.075 2907.4714 7.5 7.5  -0.23866  0.020182 0.01697
34384.075 2907.4714 7.5 6.5  -0.08491  0.007180 0.16485
34384.075 2907.4714 6.5 7.5  -0.27894  0.023588 0.00071
34384.075 2907.4714 6.5 6.5  -0.12519  0.010587 0.02785
34384.075 2907.4714 6.5 5.5   0.00806 -0.000681 0.13054
34384.075 2907.4714 5.5 6.5  -0.16011  0.013539 0.00175
34384.075 2907.4714 5.5 5.5  -0.02686  0.002271 0.03338
34384.075 2907.4714 5.5 4.5   0.08589 -0.007263 0.10124
34384.075 2907.4714 4.5 5.5  -0.05640  0.004769 0.00275
34384.075 2907.4714 4.5 4.5   0.05635 -0.004765 0.03428
34384.075 2907.4714 4.5 3.5   0.14860 -0.012566 0.07660
34384.075 2907.4714 3.5 4.5   0.03218 -0.002722 0.00337
34384.075 2907.4714 3.5 3.5   0.12443 -0.010522 0.03127
34384.075 2907.4714 3.5 2.5   0.19618 -0.016590 0.05628
34384.075 2907.4714 2.5 3.5   0.10564 -0.008933 0.00325
```

```
34384.075 2907.4714 2.5 2.5   0.17739 -0.015000 0.02494
34384.075 2907.4714 2.5 1.5   0.22864 -0.019334 0.04000
34384.075 2907.4714 1.5 2.5   0.16396 -0.013865 0.00212
34384.075 2907.4714 1.5 1.5   0.21521 -0.018198 0.01556
34384.075 2907.4714 1.5 0.5   0.24596 -0.020799 0.02778
34368.166 2908.8174 7.5 8.5  -0.34153  0.028907 0.20455
34368.166 2908.8174 7.5 7.5  -0.15623  0.013223 0.01697
34368.166 2908.8174 7.5 6.5   0.00727 -0.000616 0.00071
34368.166 2908.8174 6.5 7.5  -0.17764  0.015036 0.16485
34368.166 2908.8174 6.5 6.5  -0.01414  0.001197 0.02785
34368.166 2908.8174 6.5 5.5   0.12756 -0.010797 0.00175
34368.166 2908.8174 5.5 6.5  -0.03270  0.002768 0.13054
34368.166 2908.8174 5.5 5.5   0.10900 -0.009226 0.03338
34368.166 2908.8174 5.5 4.5   0.22890 -0.019374 0.00275
34368.166 2908.8174 4.5 5.5   0.09330 -0.007897 0.10124
34368.166 2908.8174 4.5 4.5   0.21320 -0.018045 0.03428
34368.166 2908.8174 4.5 3.5   0.31130 -0.026348 0.00337
34368.166 2908.8174 3.5 4.5   0.20035 -0.016957 0.07660
34368.166 2908.8174 3.5 3.5   0.29845 -0.025261 0.03127
34368.166 2908.8174 3.5 2.5   0.37475 -0.031719 0.00325
34368.166 2908.8174 2.5 3.5   0.28845 -0.024415 0.05628
34368.166 2908.8174 2.5 2.5   0.36475 -0.030873 0.02494
34368.166 2908.8174 2.5 1.5   0.41925 -0.035486 0.00212
34368.166 2908.8174 1.5 2.5   0.35762 -0.030269 0.04000
34368.166 2908.8174 1.5 1.5   0.41212 -0.034881 0.01556
34368.166 2908.8174 0.5 1.5   0.40783 -0.034519 0.02778
34353.971 2910.0193 5.5 5.5  -0.02437  0.002064 0.23636
34353.971 2910.0193 5.5 4.5   0.04988 -0.004226 0.06364
34353.971 2910.0193 4.5 5.5  -0.07947  0.006732 0.06364
34353.971 2910.0193 4.5 4.5  -0.00522  0.000442 0.09470
34353.971 2910.0193 4.5 3.5   0.05553 -0.004704 0.09167
34353.971 2910.0193 3.5 4.5  -0.05031  0.004262 0.09167
34353.971 2910.0193 3.5 3.5   0.01044 -0.000885 0.01905
34353.971 2910.0193 3.5 2.5   0.05769 -0.004887 0.08929
34353.971 2910.0193 2.5 3.5  -0.02463  0.002086 0.08929
34353.971 2910.0193 2.5 2.5   0.02262 -0.001917 0.00071
34353.971 2910.0193 2.5 1.5   0.05637 -0.004776 0.06000
34353.971 2910.0193 1.5 2.5  -0.00242  0.000205 0.06000
34353.971 2910.0193 1.5 1.5   0.03133 -0.002654 0.04000
34349.646 2910.3858 4.5 4.5   0.14455 -0.012248 0.25463
34349.646 2910.3858 4.5 3.5   0.08290 -0.007024 0.16204
34349.646 2910.3858 3.5 4.5   0.02035 -0.001724 0.16204
34349.646 2910.3858 3.5 3.5  -0.04130  0.003499 0.01058
34349.646 2910.3858 3.5 2.5  -0.08925  0.007562 0.16071
34349.646 2910.3858 2.5 3.5  -0.13789  0.011684 0.16071
34349.646 2910.3858 2.5 2.5  -0.18584  0.015747 0.08929
34341.656 2911.0629 7.5 6.5  -0.12460  0.010563 0.22222
34341.656 2911.0629 6.5 6.5  -0.16135  0.013678 0.02618
34341.656 2911.0629 6.5 5.5  -0.04175  0.003539 0.16827
34341.656 2911.0629 5.5 6.5  -0.19320  0.016378 0.00160
34341.656 2911.0629 5.5 5.5  -0.07360  0.006239 0.04196
34341.656 2911.0629 5.5 4.5   0.02760 -0.002340 0.12311
34341.656 2911.0629 4.5 5.5  -0.10055  0.008524 0.00406
34341.656 2911.0629 4.5 4.5   0.00065 -0.000055 0.04885
```

```
34341.656 2911.0629 4.5 3.5   0.08345 -0.007074 0.08598
34341.656 2911.0629 3.5 4.5  -0.02140  0.001814 0.00661
34341.656 2911.0629 3.5 3.5   0.06140 -0.005205 0.04837
34341.656 2911.0629 3.5 2.5   0.12580 -0.010664 0.05612
34341.656 2911.0629 2.5 3.5   0.04425 -0.003751 0.00850
34341.656 2911.0629 2.5 2.5   0.10865 -0.009210 0.04209
34341.656 2911.0629 2.5 1.5   0.15465 -0.013110 0.03274
34341.656 2911.0629 1.5 2.5   0.09640 -0.008172 0.00893
34341.656 2911.0629 1.5 1.5   0.14240 -0.012071 0.03175
34341.656 2911.0629 1.5 0.5   0.17000 -0.014411 0.01488
34341.656 2911.0629 0.5 1.5   0.13505 -0.011448 0.00694
34341.656 2911.0629 0.5 0.5   0.16265 -0.013788 0.02083
34189.498 2924.0190 8.5 8.5  -0.28751  0.024590 0.19051
34189.498 2924.0190 8.5 7.5  -0.10221  0.008742 0.01404
34189.498 2924.0190 7.5 8.5  -0.33316  0.028495 0.01404
34189.498 2924.0190 7.5 7.5  -0.14786  0.012646 0.14460
34189.498 2924.0190 7.5 6.5   0.01564 -0.001338 0.02318
34189.498 2924.0190 6.5 7.5  -0.18814  0.016092 0.02318
34189.498 2924.0190 6.5 6.5  -0.02464  0.002108 0.10794
34189.498 2924.0190 6.5 5.5   0.11706 -0.010012 0.02797
34189.498 2924.0190 5.5 6.5  -0.05956  0.005094 0.02797
34189.498 2924.0190 5.5 5.5   0.08214 -0.007026 0.07947
34189.498 2924.0190 5.5 4.5   0.20204 -0.017280 0.02893
34189.498 2924.0190 4.5 5.5   0.05260 -0.004499 0.02893
34189.498 2924.0190 4.5 4.5   0.17250 -0.014754 0.05820
34189.498 2924.0190 4.5 3.5   0.27060 -0.023144 0.02652
34189.498 2924.0190 3.5 4.5   0.14833 -0.012687 0.02652
34189.498 2924.0190 3.5 3.5   0.24643 -0.021077 0.04329
34189.498 2924.0190 3.5 2.5   0.32273 -0.027602 0.02110
34189.498 2924.0190 2.5 3.5   0.22764 -0.019469 0.02110
34189.498 2924.0190 2.5 2.5   0.30394 -0.025995 0.03435
34189.498 2924.0190 2.5 1.5   0.35844 -0.030656 0.01273
34189.498 2924.0190 1.5 2.5   0.29051 -0.024846 0.01273
34189.498 2924.0190 1.5 1.5   0.34501 -0.029507 0.03273
34182.226 2924.6411 7.5 7.5  -0.21840  0.018687 0.20148
34182.226 2924.6411 7.5 6.5  -0.06465  0.005532 0.02074
34182.226 2924.6411 6.5 7.5  -0.25515  0.021832 0.02074
34182.226 2924.6411 6.5 6.5  -0.10140  0.008676 0.14005
34182.226 2924.6411 6.5 5.5   0.03185 -0.002725 0.03365
34182.226 2924.6411 5.5 6.5  -0.13325  0.011401 0.03365
34182.226 2924.6411 5.5 5.5   0.00000  0.000000 0.09324
34182.226 2924.6411 5.5 4.5   0.11275 -0.009647 0.03977
34182.226 2924.6411 4.5 5.5  -0.02695  0.002306 0.03977
34182.226 2924.6411 4.5 4.5   0.08580 -0.007341 0.05899
34182.226 2924.6411 4.5 3.5   0.17805 -0.015235 0.04012
34182.226 2924.6411 3.5 4.5   0.06375 -0.005455 0.04012
34182.226 2924.6411 3.5 3.5   0.15600 -0.013348 0.03527
34182.226 2924.6411 3.5 2.5   0.22775 -0.019487 0.03571
34182.226 2924.6411 2.5 3.5   0.13885 -0.011881 0.03571
34182.226 2924.6411 2.5 2.5   0.21060 -0.018020 0.02012
34182.226 2924.6411 2.5 1.5   0.26185 -0.022405 0.02750
34182.226 2924.6411 1.5 2.5   0.19835 -0.016972 0.02750
34182.226 2924.6411 1.5 1.5   0.24960 -0.021357 0.01185
34182.226 2924.6411 1.5 0.5   0.28035 -0.023988 0.01620
```

```
34182.226 2924.6411 0.5 1.5  0.24225 -0.020728 0.01620
34182.226 2924.6411 0.5 0.5  0.27300 -0.023359 0.01157
34068.544 2934.4007 5.5 4.5  0.10388 -0.008948 0.30000
34068.544 2934.4007 4.5 4.5  0.05993 -0.005162 0.09722
34068.544 2934.4007 4.5 3.5 -0.00172  0.000148 0.15278
34068.544 2934.4007 3.5 4.5  0.02398 -0.002066 0.01944
34068.544 2934.4007 3.5 3.5 -0.03767  0.003245 0.12698
34068.544 2934.4007 3.5 2.5 -0.08562  0.007375 0.05357
34068.544 2934.4007 2.5 3.5 -0.06563  0.005653 0.05357
34068.544 2934.4007 2.5 2.5 -0.11358  0.009784 0.09643
34068.544 2934.4007 1.5 2.5 -0.13356  0.011504 0.10000
33950.877 2944.5712 6.5 7.5 -0.23555  0.020430 0.22222
33950.877 2944.5712 6.5 6.5 -0.08180  0.007095 0.02618
33950.877 2944.5712 6.5 5.5  0.05145 -0.004462 0.00160
33950.877 2944.5712 5.5 6.5 -0.11365  0.009857 0.16827
33950.877 2944.5712 5.5 5.5  0.01960 -0.001700 0.04196
33950.877 2944.5712 5.5 4.5  0.13235 -0.011479 0.00406
33950.877 2944.5712 4.5 5.5 -0.00735  0.000637 0.12311
33950.877 2944.5712 4.5 4.5  0.10540 -0.009142 0.04885
33950.877 2944.5712 4.5 3.5  0.19765 -0.017143 0.00661
33950.877 2944.5712 3.5 4.5  0.08335 -0.007229 0.08598
33950.877 2944.5712 3.5 3.5  0.17560 -0.015230 0.04837
33950.877 2944.5712 3.5 2.5  0.24735 -0.021454 0.00850
33950.877 2944.5712 2.5 3.5  0.15845 -0.013743 0.05612
33950.877 2944.5712 2.5 2.5  0.23020 -0.019966 0.04209
33950.877 2944.5712 2.5 1.5  0.28145 -0.024411 0.00893
33950.877 2944.5712 1.5 2.5  0.21795 -0.018904 0.03274
33950.877 2944.5712 1.5 1.5  0.26920 -0.023349 0.03175
33950.877 2944.5712 1.5 0.5  0.29995 -0.026016 0.00694
33950.877 2944.5712 0.5 1.5  0.26185 -0.022711 0.01488
33950.877 2944.5712 0.5 0.5  0.29260 -0.025378 0.02083
33884.397 2950.3486 4.5 4.5  0.13152 -0.011452 0.25463
33884.397 2950.3486 4.5 3.5  0.06987 -0.006084 0.16204
33884.397 2950.3486 3.5 4.5  0.02407 -0.002096 0.16204
33884.397 2950.3486 3.5 3.5 -0.03758  0.003272 0.01058
33884.397 2950.3486 3.5 2.5 -0.08553  0.007447 0.16071
33884.397 2950.3486 2.5 3.5 -0.12115  0.010549 0.16071
33884.397 2950.3486 2.5 2.5 -0.16910  0.014724 0.08929
33864.625 2952.0712 5.5 6.5 -0.13727  0.011967 0.25000
33864.625 2952.0712 5.5 5.5 -0.01767  0.001541 0.04545
33864.625 2952.0712 5.5 4.5  0.08353 -0.007282 0.00455
33864.625 2952.0712 4.5 5.5 -0.06162  0.005371 0.16883
33864.625 2952.0712 4.5 4.5  0.03958 -0.003451 0.06926
33864.625 2952.0712 4.5 3.5  0.12238 -0.010669 0.01190
33864.625 2952.0712 3.5 4.5  0.00363 -0.000317 0.10476
33864.625 2952.0712 3.5 3.5  0.08643 -0.007535 0.07483
33864.625 2952.0712 3.5 2.5  0.15083 -0.013149 0.02041
33864.625 2952.0712 2.5 3.5  0.05847 -0.005097 0.05612
33864.625 2952.0712 2.5 2.5  0.12287 -0.010711 0.06531
33864.625 2952.0712 2.5 1.5  0.16887 -0.014721 0.02857
33864.625 2952.0712 1.5 2.5  0.10289 -0.008970 0.02143
33864.625 2952.0712 1.5 1.5  0.14889 -0.012980 0.04286
33864.625 2952.0712 1.5 0.5  0.17649 -0.015386 0.03571
33310.249 3001.2041 6.5 6.5 -0.28299  0.025498 0.21635
```

```
33310.249 3001.2041 6.5 5.5 -0.10066  0.009069 0.03365
33310.249 3001.2041 5.5 6.5 -0.29014  0.026142 0.03365
33310.249 3001.2041 5.5 5.5 -0.10781  0.009714 0.12787
33310.249 3001.2041 5.5 4.5  0.04648 -0.004188 0.05276
33310.249 3001.2041 4.5 5.5 -0.11386  0.010259 0.05276
33310.249 3001.2041 4.5 4.5  0.04043 -0.003643 0.06629
33310.249 3001.2041 4.5 3.5  0.16666 -0.015016 0.05952
33310.249 3001.2041 3.5 4.5  0.03548 -0.003197 0.05952
33310.249 3001.2041 3.5 3.5  0.16171 -0.014570 0.02721
33310.249 3001.2041 3.5 2.5  0.25989 -0.023416 0.05612
33310.249 3001.2041 2.5 3.5  0.15786 -0.014223 0.05612
33310.249 3001.2041 2.5 2.5  0.25604 -0.023069 0.00638
33310.249 3001.2041 2.5 1.5  0.32617 -0.029388 0.04464
33310.249 3001.2041 1.5 2.5  0.25329 -0.022822 0.04464
33310.249 3001.2041 1.5 1.5  0.32342 -0.029140 0.00000
33310.249 3001.2041 1.5 0.5  0.36549 -0.032931 0.02679
33310.249 3001.2041 0.5 1.5  0.32177 -0.028992 0.02679
33310.249 3001.2041 0.5 0.5  0.36384 -0.032783 0.00893
33138.271 3016.7801 5.5 6.5 -0.32394  0.029492 0.25000
33138.271 3016.7801 5.5 5.5 -0.14161  0.012892 0.04545
33138.271 3016.7801 5.5 4.5  0.01268 -0.001154 0.00455
33138.271 3016.7801 4.5 5.5 -0.11851  0.010789 0.16883
33138.271 3016.7801 4.5 4.5  0.03578 -0.003257 0.06926
33138.271 3016.7801 4.5 3.5  0.16201 -0.014749 0.01190
33138.271 3016.7801 3.5 4.5  0.05468 -0.004978 0.10476
33138.271 3016.7801 3.5 3.5  0.18091 -0.016470 0.07483
33138.271 3016.7801 3.5 2.5  0.27909 -0.025408 0.02041
33138.271 3016.7801 2.5 3.5  0.19561 -0.017808 0.05612
33138.271 3016.7801 2.5 2.5  0.29379 -0.026746 0.06531
33138.271 3016.7801 2.5 1.5  0.36392 -0.033131 0.02857
33138.271 3016.7801 1.5 2.5  0.30429 -0.027702 0.02143
33138.271 3016.7801 1.5 1.5  0.37442 -0.034087 0.04286
33138.271 3016.7801 1.5 0.5  0.41649 -0.037917 0.03571
32320.640 3093.1002 9.5 8.5 -0.27230  0.026060 0.19231
32320.640 3093.1002 8.5 8.5 -0.32170  0.030788 0.01188
32320.640 3093.1002 8.5 7.5 -0.13640  0.013054 0.16120
32320.640 3093.1002 7.5 8.5 -0.36590  0.035019 0.00036
32320.640 3093.1002 7.5 7.5 -0.18060  0.017284 0.01974
32320.640 3093.1002 7.5 6.5 -0.01710  0.001637 0.13374
32320.640 3093.1002 6.5 7.5 -0.21960  0.021017 0.00087
32320.640 3093.1002 6.5 6.5 -0.05610  0.005369 0.02400
32320.640 3093.1002 6.5 5.5  0.08560 -0.008192 0.10974
32320.640 3093.1002 5.5 6.5 -0.08990  0.008604 0.00134
32320.640 3093.1002 5.5 5.5  0.05180 -0.004957 0.02504
32320.640 3093.1002 5.5 4.5  0.17170 -0.016432 0.08900
32320.640 3093.1002 4.5 5.5  0.02320 -0.002220 0.00159
32320.640 3093.1002 4.5 4.5  0.14310 -0.013695 0.02318
32320.640 3093.1002 4.5 3.5  0.24120 -0.023084 0.07139
32320.640 3093.1002 3.5 4.5  0.11970 -0.011456 0.00146
32320.640 3093.1002 3.5 3.5  0.21780 -0.020844 0.01865
32320.640 3093.1002 3.5 2.5  0.29410 -0.028146 0.05682
32320.640 3093.1002 2.5 3.5  0.19960 -0.019102 0.00087
32320.640 3093.1002 2.5 2.5  0.27590 -0.026405 0.01136
32320.640 3093.1002 2.5 1.5  0.33040 -0.031620 0.04545
```

```
32224.793  3102.3005  8.5  7.5  -0.16975   0.016343  0.20455
32224.793  3102.3005  7.5  7.5  -0.22670   0.021826  0.01697
32224.793  3102.3005  7.5  6.5  -0.07295   0.007023  0.16485
32224.793  3102.3005  6.5  7.5  -0.27695   0.026663  0.00071
32224.793  3102.3005  6.5  6.5  -0.12320   0.011861  0.02785
32224.793  3102.3005  6.5  5.5   0.01005  -0.000968  0.13054
32224.793  3102.3005  5.5  6.5  -0.16675   0.016054  0.00175
32224.793  3102.3005  5.5  5.5  -0.03350   0.003225  0.03338
32224.793  3102.3005  5.5  4.5   0.07925  -0.007630  0.10124
32224.793  3102.3005  4.5  5.5  -0.07035   0.006773  0.00275
32224.793  3102.3005  4.5  4.5   0.04240  -0.004082  0.03428
32224.793  3102.3005  4.5  3.5   0.13465  -0.012963  0.07660
32224.793  3102.3005  3.5  4.5   0.01225  -0.001179  0.00337
32224.793  3102.3005  3.5  3.5   0.10450  -0.010061  0.03127
32224.793  3102.3005  3.5  2.5   0.17625  -0.016968  0.05628
32224.793  3102.3005  2.5  3.5   0.08105  -0.007803  0.00325
32224.793  3102.3005  2.5  2.5   0.15280  -0.014711  0.02494
32224.793  3102.3005  2.5  1.5   0.20405  -0.019645  0.04000
32224.793  3102.3005  1.5  2.5   0.13605  -0.013098  0.00212
32224.793  3102.3005  1.5  1.5   0.18730  -0.018032  0.01556
32224.793  3102.3005  1.5  0.5   0.21805  -0.020992  0.02778
32058.620  3118.3816  6.5  5.5   0.05670  -0.005515  0.25000
32058.620  3118.3816  5.5  5.5  -0.03690   0.003589  0.04545
32058.620  3118.3816  5.5  4.5   0.03735  -0.003633  0.16883
32058.620  3118.3816  4.5  5.5  -0.11610   0.011294  0.00455
32058.620  3118.3816  4.5  4.5  -0.04185   0.004071  0.06926
32058.620  3118.3816  4.5  3.5   0.01890  -0.001838  0.10476
32058.620  3118.3816  3.5  4.5  -0.10665   0.010374  0.01190
32058.620  3118.3816  3.5  3.5  -0.04590   0.004465  0.07483
32058.620  3118.3816  3.5  2.5   0.00135  -0.000131  0.05612
32058.620  3118.3816  2.5  3.5  -0.09630   0.009368  0.02041
32058.620  3118.3816  2.5  2.5  -0.04905   0.004771  0.06531
32058.620  3118.3816  2.5  1.5  -0.01530   0.001488  0.02143
32058.620  3118.3816  1.5  2.5  -0.08505   0.008273  0.02857
32058.620  3118.3816  1.5  1.5  -0.05130   0.004990  0.04286
32058.620  3118.3816  0.5  1.5  -0.07290   0.007091  0.03571
32030.216  3121.1470  8.5  8.5  -0.26425   0.025751  0.19051
32030.216  3121.1470  8.5  7.5  -0.07895   0.007694  0.01404
32030.216  3121.1470  7.5  8.5  -0.32120   0.031300  0.01404
32030.216  3121.1470  7.5  7.5  -0.13590   0.013243  0.14460
32030.216  3121.1470  7.5  6.5   0.02760  -0.002690  0.02318
32030.216  3121.1470  6.5  7.5  -0.18615   0.018140  0.02318
32030.216  3121.1470  6.5  6.5  -0.02265   0.002207  0.10794
32030.216  3121.1470  6.5  5.5   0.11905  -0.011601  0.02797
32030.216  3121.1470  5.5  6.5  -0.06620   0.006451  0.02797
32030.216  3121.1470  5.5  5.5   0.07550  -0.007357  0.07947
32030.216  3121.1470  5.5  4.5   0.19540  -0.019041  0.02893
32030.216  3121.1470  4.5  5.5   0.03865  -0.003766  0.02893
32030.216  3121.1470  4.5  4.5   0.15855  -0.015450  0.05820
32030.216  3121.1470  4.5  3.5   0.25665  -0.025010  0.02652
32030.216  3121.1470  3.5  4.5   0.12840  -0.012512  0.02652
32030.216  3121.1470  3.5  3.5   0.22650  -0.022072  0.04329
32030.216  3121.1470  3.5  2.5   0.30280  -0.029507  0.02110
32030.216  3121.1470  2.5  3.5   0.20305  -0.019787  0.02110
```

```
32030.216  3121.1470  2.5  2.5   0.27935  -0.027222  0.03435
32030.216  3121.1470  2.5  1.5   0.33385  -0.032533  0.01273
32030.216  3121.1470  1.5  2.5   0.26260  -0.025590  0.01273
32030.216  3121.1470  1.5  1.5   0.31710  -0.030900  0.03273
31987.803  3125.2856  5.5  4.5   0.22995  -0.022467  0.30000
31987.803  3125.2856  4.5  4.5   0.08695  -0.008496  0.09722
31987.803  3125.2856  4.5  3.5   0.02530  -0.002472  0.15278
31987.803  3125.2856  3.5  4.5  -0.03005   0.002936  0.01944
31987.803  3125.2856  3.5  3.5  -0.09170   0.008960  0.12698
31987.803  3125.2856  3.5  2.5  -0.13965   0.013645  0.05357
31987.803  3125.2856  2.5  3.5  -0.18270   0.017851  0.05357
31987.803  3125.2856  2.5  2.5  -0.23065   0.022536  0.09643
31987.803  3125.2856  1.5  2.5  -0.29565   0.028887  0.10000
31978.248  3126.2194  7.5  7.5  -0.16100   0.015740  0.20148
31978.248  3126.2194  7.5  6.5  -0.00725   0.000709  0.02074
31978.248  3126.2194  6.5  7.5  -0.22850   0.022339  0.02074
31978.248  3126.2194  6.5  6.5  -0.07475   0.007308  0.14005
31978.248  3126.2194  6.5  5.5   0.05850  -0.005719  0.03365
31978.248  3126.2194  5.5  6.5  -0.13325   0.013027  0.03365
31978.248  3126.2194  5.5  5.5   0.00000   0.000000  0.09324
31978.248  3126.2194  5.5  4.5   0.11275  -0.011023  0.03977
31978.248  3126.2194  4.5  5.5  -0.04950   0.004839  0.03977
31978.248  3126.2194  4.5  4.5   0.06325  -0.006184  0.05899
31978.248  3126.2194  4.5  3.5   0.15550  -0.015202  0.04012
31978.248  3126.2194  3.5  4.5   0.02275  -0.002224  0.04012
31978.248  3126.2194  3.5  3.5   0.11500  -0.011243  0.03527
31978.248  3126.2194  3.5  2.5   0.18675  -0.018257  0.03571
31978.248  3126.2194  2.5  3.5   0.08350  -0.008163  0.03571
31978.248  3126.2194  2.5  2.5   0.15525  -0.015178  0.02012
31978.248  3126.2194  2.5  1.5   0.20650  -0.020188  0.02750
31978.248  3126.2194  1.5  2.5   0.13275  -0.012978  0.02750
31978.248  3126.2194  1.5  1.5   0.18400  -0.017989  0.01185
31978.248  3126.2194  1.5  0.5   0.21475  -0.020995  0.01620
31978.248  3126.2194  0.5  1.5   0.17050  -0.016669  0.01620
31978.248  3126.2194  0.5  0.5   0.20125  -0.019675  0.01157
31936.869  3130.2701  6.5  6.5  -0.04200   0.004117  0.21635
31936.869  3130.2701  6.5  5.5   0.07760  -0.007606  0.03365
31936.869  3130.2701  5.5  6.5  -0.13560   0.013291  0.03365
31936.869  3130.2701  5.5  5.5  -0.01600   0.001568  0.12787
31936.869  3130.2701  5.5  4.5   0.08520  -0.008351  0.05276
31936.869  3130.2701  4.5  5.5  -0.09520   0.009331  0.05276
31936.869  3130.2701  4.5  4.5   0.00600  -0.000588  0.06629
31936.869  3130.2701  4.5  3.5   0.08880  -0.008704  0.05952
31936.869  3130.2701  3.5  4.5  -0.05880   0.005763  0.05952
31936.869  3130.2701  3.5  3.5   0.02400  -0.002352  0.02721
31936.869  3130.2701  3.5  2.5   0.08840  -0.008665  0.05612
31936.869  3130.2701  2.5  3.5  -0.02640   0.002588  0.05612
31936.869  3130.2701  2.5  2.5   0.03800  -0.003725  0.00638
31936.869  3130.2701  2.5  1.5   0.08400  -0.008233  0.04464
31936.869  3130.2701  1.5  2.5   0.00200  -0.000196  0.04464
31936.869  3130.2701  1.5  1.5   0.04800  -0.004705  0.00000
31936.869  3130.2701  1.5  0.5   0.07560  -0.007410  0.02679
31936.869  3130.2701  0.5  1.5   0.02640  -0.002588  0.02679
31936.869  3130.2701  0.5  0.5   0.05400  -0.005293  0.00893
```

```
31905.635  3133.3346  5.5  5.5   0.08750  -0.008593  0.23636
31905.635  3133.3346  5.5  4.5   0.16175  -0.015885  0.06364
31905.635  3133.3346  4.5  5.5  -0.05550   0.005451  0.06364
31905.635  3133.3346  4.5  4.5   0.01875  -0.001841  0.09470
31905.635  3133.3346  4.5  3.5   0.07950  -0.007808  0.09167
31905.635  3133.3346  3.5  4.5  -0.09825   0.009649  0.09167
31905.635  3133.3346  3.5  3.5  -0.03750   0.003683  0.01905
31905.635  3133.3346  3.5  2.5   0.00975  -0.000958  0.08929
31905.635  3133.3346  2.5  3.5  -0.12850   0.012620  0.08929
31905.635  3133.3346  2.5  2.5  -0.08125   0.007980  0.00071
31905.635  3133.3346  2.5  1.5  -0.04750   0.004665  0.06000
31905.635  3133.3346  1.5  2.5  -0.14625   0.014363  0.06000
31905.635  3133.3346  1.5  1.5  -0.11250   0.011049  0.04000
31777.439  3145.9755  6.5  7.5  -0.13580   0.013445  0.22222
31777.439  3145.9755  6.5  6.5   0.01795  -0.001777  0.02618
31777.439  3145.9755  6.5  5.5   0.15120  -0.014969  0.00160
31777.439  3145.9755  5.5  6.5  -0.07565   0.007490  0.16827
31777.439  3145.9755  5.5  5.5   0.05760  -0.005703  0.04196
31777.439  3145.9755  5.5  4.5   0.17035  -0.016865  0.00406
31777.439  3145.9755  4.5  5.5  -0.02160   0.002138  0.12311
31777.439  3145.9755  4.5  4.5   0.09115  -0.009024  0.04885
31777.439  3145.9755  4.5  3.5   0.18340  -0.018157  0.00661
31777.439  3145.9755  3.5  4.5   0.02635  -0.002609  0.08598
31777.439  3145.9755  3.5  3.5   0.11860  -0.011742  0.04837
31777.439  3145.9755  3.5  2.5   0.19035  -0.018845  0.00850
31777.439  3145.9755  2.5  3.5   0.06820  -0.006752  0.05612
31777.439  3145.9755  2.5  2.5   0.13995  -0.013856  0.04209
31777.439  3145.9755  2.5  1.5   0.19120  -0.018929  0.00893
31777.439  3145.9755  1.5  2.5   0.10395  -0.010291  0.03274
31777.439  3145.9755  1.5  1.5   0.15520  -0.015365  0.03175
31777.439  3145.9755  1.5  0.5   0.18595  -0.018410  0.00694
31777.439  3145.9755  0.5  1.5   0.13360  -0.013227  0.01488
31777.439  3145.9755  0.5  0.5   0.16435  -0.016271  0.02083
31555.205  3168.1325  6.5  5.5  -0.10803   0.010847  0.25000
31555.205  3168.1325  5.5  5.5  -0.17332   0.017402  0.04545
31555.205  3168.1325  5.5  4.5  -0.00557   0.000559  0.16883
31555.205  3168.1325  4.5  5.5  -0.22857   0.022949  0.00455
31555.205  3168.1325  4.5  4.5  -0.06082   0.006106  0.06926
31555.205  3168.1325  4.5  3.5   0.07643  -0.007674  0.10476
31555.205  3168.1325  3.5  4.5  -0.10602   0.010645  0.01190
31555.205  3168.1325  3.5  3.5   0.03123  -0.003136  0.07483
31555.205  3168.1325  3.5  2.5   0.13798  -0.013854  0.05612
31555.205  3168.1325  2.5  3.5  -0.00392   0.000394  0.02041
31555.205  3168.1325  2.5  2.5   0.10283  -0.010324  0.06531
31555.205  3168.1325  2.5  1.5   0.17908  -0.017980  0.02143
31555.205  3168.1325  1.5  2.5   0.07771  -0.007803  0.02857
31555.205  3168.1325  1.5  1.5   0.15396  -0.015458  0.04286
31555.205  3168.1325  0.5  1.5   0.13890  -0.013946  0.03571
31361.392  3187.7122  5.5  5.5  -0.07115   0.007233  0.23636
31361.392  3187.7122  5.5  4.5   0.09660  -0.009819  0.06364
31361.392  3187.7122  4.5  5.5  -0.18300   0.018601  0.06364
31361.392  3187.7122  4.5  4.5  -0.01525   0.001550  0.09470
31361.392  3187.7122  4.5  3.5   0.12200  -0.012401  0.09167
31361.392  3187.7122  3.5  4.5  -0.10676   0.010852  0.09167
```

```
31361.392 3187.7122 3.5 3.5   0.03049 -0.003100 0.01905
31361.392 3187.7122 3.5 2.5   0.13724 -0.013951 0.08929
31361.392 3187.7122 2.5 3.5  -0.04068  0.004135 0.08929
31361.392 3187.7122 2.5 2.5   0.06607 -0.006716 0.00071
31361.392 3187.7122 2.5 1.5   0.14232 -0.014467 0.06000
31361.392 3187.7122 1.5 2.5   0.01523 -0.001549 0.06000
31361.392 3187.7122 1.5 1.5   0.09148 -0.009299 0.04000
31353.517 3188.5129 6.5 6.5   0.04247 -0.004319 0.21635
31353.517 3188.5129 6.5 5.5   0.08147 -0.008285 0.03365
31353.517 3188.5129 5.5 6.5  -0.02282  0.002321 0.03365
31353.517 3188.5129 5.5 5.5   0.01618 -0.001645 0.12787
31353.517 3188.5129 5.5 4.5   0.04918 -0.005001 0.05276
31353.517 3188.5129 4.5 5.5  -0.03907  0.003973 0.05276
31353.517 3188.5129 4.5 4.5  -0.00607  0.000617 0.06629
31353.517 3188.5129 4.5 3.5   0.02093 -0.002129 0.05952
31353.517 3188.5129 3.5 4.5  -0.05127  0.005214 0.05952
31353.517 3188.5129 3.5 3.5  -0.02427  0.002468 0.02721
31353.517 3188.5129 3.5 2.5  -0.00327  0.000332 0.05612
31353.517 3188.5129 2.5 3.5  -0.05942  0.006043 0.05612
31353.517 3188.5129 2.5 2.5  -0.03842  0.003908 0.00638
31353.517 3188.5129 2.5 1.5  -0.02342  0.002382 0.04464
31353.517 3188.5129 1.5 2.5  -0.06354  0.006462 0.04464
31353.517 3188.5129 1.5 1.5  -0.04854  0.004936 0.00000
31353.517 3188.5129 1.5 0.5  -0.03954  0.004021 0.02679
31353.517 3188.5129 0.5 1.5  -0.06360  0.006468 0.02679
31353.517 3188.5129 0.5 0.5  -0.05460  0.005553 0.00893
31332.198 3190.6825 7.5 7.5   0.12383 -0.012611 0.20148
31332.198 3190.6825 7.5 6.5   0.10208 -0.010396 0.02074
31332.198 3190.6825 6.5 7.5   0.07924 -0.008070 0.02074
31332.198 3190.6825 6.5 6.5   0.05749 -0.005855 0.14005
31332.198 3190.6825 6.5 5.5   0.03864 -0.003935 0.03365
31332.198 3190.6825 5.5 6.5   0.01885 -0.001920 0.03365
31332.198 3190.6825 5.5 5.5   0.00000  0.000000 0.09324
31332.198 3190.6825 5.5 4.5  -0.01595  0.001624 0.03977
31332.198 3190.6825 4.5 5.5  -0.03270  0.003330 0.03977
31332.198 3190.6825 4.5 4.5  -0.04865  0.004954 0.05899
31332.198 3190.6825 4.5 3.5  -0.06170  0.006283 0.04012
31332.198 3190.6825 3.5 4.5  -0.07540  0.007679 0.04012
31332.198 3190.6825 3.5 3.5  -0.08845  0.009008 0.03527
31332.198 3190.6825 3.5 2.5  -0.09860  0.010041 0.03571
31332.198 3190.6825 2.5 3.5  -0.10926  0.011127 0.03571
31332.198 3190.6825 2.5 2.5  -0.11941  0.012160 0.02012
31332.198 3190.6825 2.5 1.5  -0.12666  0.012899 0.02750
31332.198 3190.6825 1.5 2.5  -0.13427  0.013674 0.02750
31332.198 3190.6825 1.5 1.5  -0.14152  0.014412 0.01185
31332.198 3190.6825 1.5 0.5  -0.14587  0.014855 0.01620
31332.198 3190.6825 0.5 1.5  -0.15044  0.015320 0.01620
31332.198 3190.6825 0.5 0.5  -0.15479  0.015763 0.01157
31159.704 3208.3461 5.5 6.5   0.07935 -0.008170 0.25000
31159.704 3208.3461 5.5 5.5   0.11835 -0.012186 0.04545
31159.704 3208.3461 5.5 4.5   0.15135 -0.015584 0.00455
31159.704 3208.3461 4.5 5.5   0.00650 -0.000670 0.16883
31159.704 3208.3461 4.5 4.5   0.03950 -0.004068 0.06926
31159.704 3208.3461 4.5 3.5   0.06650 -0.006848 0.01190
```

```
31159.704 3208.3461 3.5 4.5 -0.05201  0.005355 0.10476
31159.704 3208.3461 3.5 3.5 -0.02501  0.002575 0.07483
31159.704 3208.3461 3.5 2.5 -0.00401  0.000412 0.02041
31159.704 3208.3461 2.5 3.5 -0.09618  0.009903 0.05612
31159.704 3208.3461 2.5 2.5 -0.07518  0.007741 0.06531
31159.704 3208.3461 2.5 1.5 -0.06018  0.006196 0.02857
31159.704 3208.3461 1.5 2.5 -0.12602  0.012976 0.02143
31159.704 3208.3461 1.5 1.5 -0.11102  0.011431 0.04286
31159.704 3208.3461 1.5 0.5 -0.10202  0.010504 0.03571
31097.678 3214.7456 6.5 7.5  0.14607 -0.015100 0.22222
31097.678 3214.7456 6.5 6.5  0.12432 -0.012852 0.02618
31097.678 3214.7456 6.5 5.5  0.10547 -0.010903 0.00160
31097.678 3214.7456 5.5 6.5  0.05903 -0.006102 0.16827
31097.678 3214.7456 5.5 5.5  0.04018 -0.004154 0.04196
31097.678 3214.7456 5.5 4.5  0.02423 -0.002505 0.00406
31097.678 3214.7456 4.5 5.5 -0.01507  0.001558 0.12311
31097.678 3214.7456 4.5 4.5 -0.03102  0.003207 0.04885
31097.678 3214.7456 4.5 3.5 -0.04407  0.004556 0.00661
31097.678 3214.7456 3.5 4.5 -0.07622  0.007879 0.08598
31097.678 3214.7456 3.5 3.5 -0.08927  0.009228 0.04837
31097.678 3214.7456 3.5 2.5 -0.09942  0.010278 0.00850
31097.678 3214.7456 2.5 3.5 -0.12442  0.012863 0.05612
31097.678 3214.7456 2.5 2.5 -0.13457  0.013912 0.04209
31097.678 3214.7456 2.5 1.5 -0.14182  0.014662 0.00893
31097.678 3214.7456 1.5 2.5 -0.15969  0.016508 0.03274
31097.678 3214.7456 1.5 1.5 -0.16694  0.017258 0.03175
31097.678 3214.7456 1.5 0.5 -0.17129  0.017707 0.00694
31097.678 3214.7456 0.5 1.5 -0.18200  0.018815 0.01488
31097.678 3214.7456 0.5 0.5 -0.18635  0.019265 0.02083
30593.724 3267.7022 6.5 5.5 -0.03773  0.004030 0.25000
30593.724 3267.7022 5.5 5.5 -0.14654  0.015652 0.04545
30593.724 3267.7022 5.5 4.5  0.02121 -0.002266 0.16883
30593.724 3267.7022 4.5 5.5 -0.23861  0.025487 0.00455
30593.724 3267.7022 4.5 4.5 -0.07086  0.007569 0.06926
30593.724 3267.7022 4.5 3.5  0.06639 -0.007091 0.10476
30593.724 3267.7022 3.5 4.5 -0.14619  0.015615 0.01190
30593.724 3267.7022 3.5 3.5 -0.00894  0.000955 0.07483
30593.724 3267.7022 3.5 2.5  0.09781 -0.010447 0.05612
30593.724 3267.7022 2.5 3.5 -0.06753  0.007213 0.02041
30593.724 3267.7022 2.5 2.5  0.03922 -0.004189 0.06531
30593.724 3267.7022 2.5 1.5  0.11547 -0.012333 0.02143
30593.724 3267.7022 1.5 2.5 -0.00263  0.000281 0.02857
30593.724 3267.7022 1.5 1.5  0.07362 -0.007863 0.04286
30593.724 3267.7022 0.5 1.5  0.04851 -0.005181 0.03571
30561.737 3271.1225 7.5 6.5  0.07902 -0.008458 0.22222
30561.737 3271.1225 6.5 6.5  0.00294 -0.000314 0.02618
30561.737 3271.1225 6.5 5.5  0.04194 -0.004489 0.16827
30561.737 3271.1225 5.5 6.5 -0.06300  0.006743 0.00160
30561.737 3271.1225 5.5 5.5 -0.02400  0.002569 0.04196
30561.737 3271.1225 5.5 4.5  0.00900 -0.000963 0.12311
30561.737 3271.1225 4.5 5.5 -0.07979  0.008541 0.00406
30561.737 3271.1225 4.5 4.5 -0.04679  0.005008 0.04885
30561.737 3271.1225 4.5 3.5 -0.01979  0.002118 0.08598
30561.737 3271.1225 3.5 4.5 -0.09244  0.009895 0.00661
```

```
30561.737 3271.1225 3.5 3.5 -0.06544  0.007005 0.04837
30561.737 3271.1225 3.5 2.5 -0.04444  0.004757 0.05612
30561.737 3271.1225 2.5 3.5 -0.10094  0.010805 0.00850
30561.737 3271.1225 2.5 2.5 -0.07994  0.008557 0.04209
30561.737 3271.1225 2.5 1.5 -0.06494  0.006951 0.03274
30561.737 3271.1225 1.5 2.5 -0.10530  0.011271 0.00893
30561.737 3271.1225 1.5 1.5 -0.09030  0.009666 0.03175
30561.737 3271.1225 1.5 0.5 -0.08130  0.008703 0.01488
30561.737 3271.1225 0.5 1.5 -0.10552  0.011295 0.00694
30561.737 3271.1225 0.5 0.5 -0.09652  0.010331 0.02083
30515.075 3276.1247 8.5 7.5  0.16523 -0.017740 0.20455
30515.075 3276.1247 7.5 7.5  0.10470 -0.011241 0.01697
30515.075 3276.1247 7.5 6.5  0.08295 -0.008906 0.16485
30515.075 3276.1247 6.5 7.5  0.05128 -0.005506 0.00071
30515.075 3276.1247 6.5 6.5  0.02953 -0.003171 0.02785
30515.075 3276.1247 6.5 5.5  0.01068 -0.001147 0.13054
30515.075 3276.1247 5.5 6.5 -0.01676  0.001799 0.00175
30515.075 3276.1247 5.5 5.5 -0.03561  0.003823 0.03338
30515.075 3276.1247 5.5 4.5 -0.05156  0.005536 0.10124
30515.075 3276.1247 4.5 5.5 -0.07478  0.008029 0.00275
30515.075 3276.1247 4.5 4.5 -0.09073  0.009741 0.03428
30515.075 3276.1247 4.5 3.5 -0.10378  0.011142 0.07660
30515.075 3276.1247 3.5 4.5 -0.12278  0.013182 0.00337
30515.075 3276.1247 3.5 3.5 -0.13583  0.014583 0.03127
30515.075 3276.1247 3.5 2.5 -0.14598  0.015673 0.05628
30515.075 3276.1247 2.5 3.5 -0.16076  0.017260 0.00325
30515.075 3276.1247 2.5 2.5 -0.17091  0.018349 0.02494
30515.075 3276.1247 2.5 1.5 -0.17816  0.019128 0.04000
30515.075 3276.1247 1.5 2.5 -0.18871  0.020261 0.00212
30515.075 3276.1247 1.5 1.5 -0.19596  0.021039 0.01556
30515.075 3276.1247 1.5 0.5 -0.20031  0.021506 0.02778
30305.898 3298.7378 7.5 7.5  0.18262 -0.019878 0.20148
30305.898 3298.7378 7.5 6.5  0.16087 -0.017510 0.02074
30305.898 3298.7378 6.5 7.5  0.10654 -0.011597 0.02074
30305.898 3298.7378 6.5 6.5  0.08479 -0.009229 0.14005
30305.898 3298.7378 6.5 5.5  0.06594 -0.007177 0.03365
30305.898 3298.7378 5.5 6.5  0.01885 -0.002052 0.03365
30305.898 3298.7378 5.5 5.5  0.00000  0.000000 0.09324
30305.898 3298.7378 5.5 4.5 -0.01595  0.001736 0.03977
30305.898 3298.7378 4.5 5.5 -0.05579  0.006073 0.03977
30305.898 3298.7378 4.5 4.5 -0.07174  0.007809 0.05899
30305.898 3298.7378 4.5 3.5 -0.08479  0.009230 0.04012
30305.898 3298.7378 3.5 4.5 -0.11739  0.012778 0.04012
30305.898 3298.7378 3.5 3.5 -0.13044  0.014199 0.03527
30305.898 3298.7378 3.5 2.5 -0.14059  0.015304 0.03571
30305.898 3298.7378 2.5 3.5 -0.16594  0.018063 0.03571
30305.898 3298.7378 2.5 2.5 -0.17609  0.019168 0.02012
30305.898 3298.7378 2.5 1.5 -0.18334  0.019957 0.02750
30305.898 3298.7378 1.5 2.5 -0.20145  0.021929 0.02750
30305.898 3298.7378 1.5 1.5 -0.20870  0.022718 0.01185
30305.898 3298.7378 1.5 0.5 -0.21305  0.023191 0.01620
30305.898 3298.7378 0.5 1.5 -0.22392  0.024374 0.01620
30305.898 3298.7378 0.5 0.5 -0.22827  0.024848 0.01157
28422.776 3517.2994 6.5 7.5  0.06685 -0.008273 0.22222
```

```
28422.776 3517.2994 6.5 6.5   0.04510 -0.005581 0.02618
28422.776 3517.2994 6.5 5.5   0.02625 -0.003249 0.00160
28422.776 3517.2994 5.5 6.5   0.02885 -0.003570 0.16827
28422.776 3517.2994 5.5 5.5   0.01000 -0.001238 0.04196
28422.776 3517.2994 5.5 4.5  -0.00595  0.000736 0.00406
28422.776 3517.2994 4.5 5.5  -0.00375  0.000464 0.12311
28422.776 3517.2994 4.5 4.5  -0.01970  0.002438 0.04885
28422.776 3517.2994 4.5 3.5  -0.03275  0.004053 0.00661
28422.776 3517.2994 3.5 4.5  -0.03095  0.003830 0.08598
28422.776 3517.2994 3.5 3.5  -0.04400  0.005445 0.04837
28422.776 3517.2994 3.5 2.5  -0.05415  0.006701 0.00850
28422.776 3517.2994 2.5 3.5  -0.05275  0.006528 0.05612
28422.776 3517.2994 2.5 2.5  -0.06290  0.007784 0.04209
28422.776 3517.2994 2.5 1.5  -0.07015  0.008681 0.00893
28422.776 3517.2994 1.5 2.5  -0.06915  0.008558 0.03274
28422.776 3517.2994 1.5 1.5  -0.07640  0.009455 0.03175
28422.776 3517.2994 1.5 0.5  -0.08075  0.009993 0.00694
28422.776 3517.2994 0.5 1.5  -0.08015  0.009919 0.01488
28422.776 3517.2994 0.5 0.5  -0.08450  0.010457 0.02083
28362.971 3524.7160 6.5 6.5  -0.00688  0.000856 0.21635
28362.971 3524.7160 6.5 5.5   0.03212 -0.003991 0.03365
28362.971 3524.7160 5.5 6.5  -0.04162  0.005173 0.03365
28362.971 3524.7160 5.5 5.5  -0.00262  0.000326 0.12787
28362.971 3524.7160 5.5 4.5   0.03038 -0.003775 0.05276
28362.971 3524.7160 4.5 5.5  -0.03202  0.003979 0.05276
28362.971 3524.7160 4.5 4.5   0.00098 -0.000122 0.06629
28362.971 3524.7160 4.5 3.5   0.02798 -0.003478 0.05952
28362.971 3524.7160 3.5 4.5  -0.02307  0.002867 0.05952
28362.971 3524.7160 3.5 3.5   0.00393 -0.000489 0.02721
28362.971 3524.7160 3.5 2.5   0.02493 -0.003099 0.05612
28362.971 3524.7160 2.5 3.5  -0.01477  0.001836 0.05612
28362.971 3524.7160 2.5 2.5   0.00623 -0.000774 0.00638
28362.971 3524.7160 2.5 1.5   0.02123 -0.002638 0.04464
28362.971 3524.7160 1.5 2.5  -0.00713  0.000886 0.04464
28362.971 3524.7160 1.5 1.5   0.00787 -0.000978 0.00000
28362.971 3524.7160 1.5 0.5   0.01687 -0.002096 0.02679
28362.971 3524.7160 0.5 1.5  -0.00015  0.000019 0.02679
28362.971 3524.7160 0.5 0.5   0.00885 -0.001100 0.00893
28314.324 3530.7720 4.5 5.5  -0.11690  0.014578 0.30000
28314.324 3530.7720 4.5 4.5   0.05085 -0.006341 0.09722
28314.324 3530.7720 4.5 3.5   0.18810 -0.023456 0.01944
28314.324 3530.7720 3.5 4.5  -0.07335  0.009147 0.15278
28314.324 3530.7720 3.5 3.5   0.06390 -0.007969 0.12698
28314.324 3530.7720 3.5 2.5   0.17065 -0.021281 0.05357
28314.324 3530.7720 2.5 3.5  -0.03269  0.004077 0.05357
28314.324 3530.7720 2.5 2.5   0.07406 -0.009235 0.09643
28314.324 3530.7720 2.5 1.5   0.15031 -0.018743 0.10000
28199.129 3545.1959 5.5 6.5   0.00713 -0.000897 0.25000
28199.129 3545.1959 5.5 5.5   0.04613 -0.005800 0.04545
28199.129 3545.1959 5.5 4.5   0.07913 -0.009949 0.00455
28199.129 3545.1959 4.5 5.5  -0.00897  0.001128 0.16883
28199.129 3545.1959 4.5 4.5   0.02403 -0.003021 0.06926
28199.129 3545.1959 4.5 3.5   0.05103 -0.006416 0.01190
28199.129 3545.1959 3.5 4.5  -0.02106  0.002647 0.10476
```

```
28199.129 3545.1959 3.5 3.5  0.00594 -0.000747 0.07483
28199.129 3545.1959 3.5 2.5  0.02694 -0.003387 0.02041
28199.129 3545.1959 2.5 3.5 -0.02913  0.003662 0.05612
28199.129 3545.1959 2.5 2.5 -0.00813  0.001022 0.06531
28199.129 3545.1959 2.5 1.5  0.00687 -0.000864 0.02857
28199.129 3545.1959 1.5 2.5 -0.03317  0.004171 0.02143
28199.129 3545.1959 1.5 1.5 -0.01817  0.002285 0.04286
28199.129 3545.1959 1.5 0.5 -0.00917  0.001153 0.03571
28033.222 3566.1777 5.5 5.5 -0.15757  0.020046 0.23636
28033.222 3566.1777 5.5 4.5  0.01018 -0.001295 0.06364
28033.222 3566.1777 4.5 5.5 -0.20152  0.025636 0.06364
28033.222 3566.1777 4.5 4.5 -0.03377  0.004296 0.09470
28033.222 3566.1777 4.5 3.5  0.10348 -0.013165 0.09167
28033.222 3566.1777 3.5 4.5 -0.06972  0.008869 0.09167
28033.222 3566.1777 3.5 3.5  0.06753 -0.008591 0.01905
28033.222 3566.1777 3.5 2.5  0.17428 -0.022171 0.08929
28033.222 3566.1777 2.5 3.5  0.03957 -0.005034 0.08929
28033.222 3566.1777 2.5 2.5  0.14632 -0.018614 0.00071
28033.222 3566.1777 2.5 1.5  0.22257 -0.028314 0.06000
28033.222 3566.1777 1.5 2.5  0.12634 -0.016073 0.06000
28033.222 3566.1777 1.5 1.5  0.20259 -0.025773 0.04000
27849.075 3589.7591 4.5 5.5 -0.12993  0.016748 0.30000
27849.075 3589.7591 4.5 4.5  0.03782 -0.004875 0.09722
27849.075 3589.7591 4.5 3.5  0.17507 -0.022567 0.01944
27849.075 3589.7591 3.5 4.5 -0.06963  0.008975 0.15278
27849.075 3589.7591 3.5 3.5  0.06762 -0.008717 0.12698
27849.075 3589.7591 3.5 2.5  0.17437 -0.022477 0.05357
27849.075 3589.7591 2.5 3.5 -0.01595  0.002056 0.05357
27849.075 3589.7591 2.5 2.5  0.09080 -0.011704 0.09643
27849.075 3589.7591 2.5 1.5  0.16705 -0.021533 0.10000
27831.534 3592.0216 5.5 6.5 -0.00707  0.000913 0.25000
27831.534 3592.0216 5.5 5.5  0.03193 -0.004121 0.04545
27831.534 3592.0216 5.5 4.5  0.06493 -0.008380 0.00455
27831.534 3592.0216 4.5 5.5 -0.01202  0.001551 0.16883
27831.534 3592.0216 4.5 4.5  0.02098 -0.002708 0.06926
27831.534 3592.0216 4.5 3.5  0.04798 -0.006193 0.01190
27831.534 3592.0216 3.5 4.5 -0.01497  0.001932 0.10476
27831.534 3592.0216 3.5 3.5  0.01203 -0.001553 0.07483
27831.534 3592.0216 3.5 2.5  0.03303 -0.004263 0.02041
27831.534 3592.0216 2.5 3.5 -0.01593  0.002056 0.05612
27831.534 3592.0216 2.5 2.5  0.00507 -0.000654 0.06531
27831.534 3592.0216 2.5 1.5  0.02007 -0.002590 0.02857
27831.534 3592.0216 1.5 2.5 -0.01491  0.001924 0.02143
27831.534 3592.0216 1.5 1.5  0.00009 -0.000012 0.04286
27831.534 3592.0216 1.5 0.5  0.00909 -0.001174 0.03571
27821.377 3593.3330 6.5 7.5  0.09205 -0.011889 0.22222
27821.377 3593.3330 6.5 6.5  0.07030 -0.009080 0.02618
27821.377 3593.3330 6.5 5.5  0.05145 -0.006645 0.00160
27821.377 3593.3330 5.5 6.5  0.03845 -0.004966 0.16827
27821.377 3593.3330 5.5 5.5  0.01960 -0.002532 0.04196
27821.377 3593.3330 5.5 4.5  0.00365 -0.000471 0.00406
27821.377 3593.3330 4.5 5.5 -0.00735  0.000949 0.12311
27821.377 3593.3330 4.5 4.5 -0.02330  0.003009 0.04885
27821.377 3593.3330 4.5 3.5 -0.03635  0.004695 0.00661
```

```
27821.377 3593.3330 3.5 4.5 -0.04535  0.005857 0.08598
27821.377 3593.3330 3.5 3.5 -0.05840  0.007543 0.04837
27821.377 3593.3330 3.5 2.5 -0.06855  0.008854 0.00850
27821.377 3593.3330 2.5 3.5 -0.07555  0.009758 0.05612
27821.377 3593.3330 2.5 2.5 -0.08570  0.011069 0.04209
27821.377 3593.3330 2.5 1.5 -0.09295  0.012006 0.00893
27821.377 3593.3330 1.5 2.5 -0.09795  0.012651 0.03274
27821.377 3593.3330 1.5 1.5 -0.10520  0.013588 0.03175
27821.377 3593.3330 1.5 0.5 -0.10955  0.014150 0.00694
27821.377 3593.3330 0.5 1.5 -0.11255  0.014537 0.01488
27821.377 3593.3330 0.5 0.5 -0.11690  0.015099 0.02083
26821.148 3727.3412 7.5 7.5  0.00319 -0.000443 0.20148
26821.148 3727.3412 7.5 6.5  0.04607 -0.006402 0.02074
26821.148 3727.3412 6.5 7.5 -0.04140  0.005753 0.02074
26821.148 3727.3412 6.5 6.5  0.00148 -0.000206 0.14005
26821.148 3727.3412 6.5 5.5  0.03864 -0.005370 0.03365
26821.148 3727.3412 5.5 6.5 -0.03716  0.005165 0.03365
26821.148 3727.3412 5.5 5.5  0.00000  0.000000 0.09324
26821.148 3727.3412 5.5 4.5  0.03145 -0.004370 0.03977
26821.148 3727.3412 4.5 5.5 -0.03270  0.004544 0.03977
26821.148 3727.3412 4.5 4.5 -0.00125  0.000174 0.05899
26821.148 3727.3412 4.5 3.5  0.02447 -0.003401 0.04012
26821.148 3727.3412 3.5 4.5 -0.02801  0.003892 0.04012
26821.148 3727.3412 3.5 3.5 -0.00228  0.000317 0.03527
26821.148 3727.3412 3.5 2.5  0.01773 -0.002464 0.03571
26821.148 3727.3412 2.5 3.5 -0.02309  0.003208 0.03571
26821.148 3727.3412 2.5 2.5 -0.00308  0.000427 0.02012
26821.148 3727.3412 2.5 1.5  0.01122 -0.001559 0.02750
26821.148 3727.3412 1.5 2.5 -0.01794  0.002493 0.02750
26821.148 3727.3412 1.5 1.5 -0.00365  0.000507 0.01185
26821.148 3727.3412 1.5 0.5  0.00493 -0.000685 0.01620
26821.148 3727.3412 0.5 1.5 -0.01256  0.001746 0.01620
26821.148 3727.3412 0.5 0.5 -0.00399  0.000554 0.01157
26652.922 3750.8677 6.5 6.5  0.01759 -0.002475 0.21635
26652.922 3750.8677 6.5 5.5  0.07199 -0.010131 0.03365
26652.922 3750.8677 5.5 6.5 -0.04770  0.006713 0.03365
26652.922 3750.8677 5.5 5.5  0.00670 -0.000943 0.12787
26652.922 3750.8677 5.5 4.5  0.05273 -0.007421 0.05276
26652.922 3750.8677 4.5 5.5 -0.04854  0.006832 0.05276
26652.922 3750.8677 4.5 4.5 -0.00251  0.000354 0.06629
26652.922 3750.8677 4.5 3.5  0.03515 -0.004947 0.05952
26652.922 3750.8677 3.5 4.5 -0.04771  0.006715 0.05952
26652.922 3750.8677 3.5 3.5 -0.01005  0.001415 0.02721
26652.922 3750.8677 3.5 2.5  0.01924 -0.002708 0.05612
26652.922 3750.8677 2.5 3.5 -0.04521  0.006362 0.05612
26652.922 3750.8677 2.5 2.5 -0.01591  0.002240 0.00638
26652.922 3750.8677 2.5 1.5  0.00501 -0.000705 0.04464
26652.922 3750.8677 1.5 2.5 -0.04103  0.005774 0.04464
26652.922 3750.8677 1.5 1.5 -0.02010  0.002829 0.00000
26652.922 3750.8677 1.5 0.5 -0.00755  0.001062 0.02679
26652.922 3750.8677 0.5 1.5 -0.03517  0.004949 0.02679
26652.922 3750.8677 0.5 0.5 -0.02261  0.003183 0.00893
26586.628 3760.2208 6.5 7.5  0.02543 -0.003596 0.22222
26586.628 3760.2208 6.5 6.5  0.06831 -0.009661 0.02618
```

```
26586.628 3760.2208 6.5 5.5   0.10547 -0.014917 0.00160
26586.628 3760.2208 5.5 6.5   0.00302 -0.000427 0.16827
26586.628 3760.2208 5.5 5.5   0.04018 -0.005683 0.04196
26586.628 3760.2208 5.5 4.5   0.07162 -0.010130 0.00406
26586.628 3760.2208 4.5 5.5  -0.01507  0.002131 0.12311
26586.628 3760.2208 4.5 4.5   0.01638 -0.002316 0.04885
26586.628 3760.2208 4.5 3.5   0.04211 -0.005955 0.00661
26586.628 3760.2208 3.5 4.5  -0.02882  0.004077 0.08598
26586.628 3760.2208 3.5 3.5  -0.00309  0.000438 0.04837
26586.628 3760.2208 3.5 2.5   0.01692 -0.002392 0.00850
26586.628 3760.2208 2.5 3.5  -0.03825  0.005410 0.05612
26586.628 3760.2208 2.5 2.5  -0.01824  0.002580 0.04209
26586.628 3760.2208 2.5 1.5  -0.00395  0.000558 0.00893
26586.628 3760.2208 1.5 2.5  -0.04335  0.006132 0.03274
26586.628 3760.2208 1.5 1.5  -0.02906  0.004110 0.03175
26586.628 3760.2208 1.5 0.5  -0.02048  0.002897 0.00694
26586.628 3760.2208 0.5 1.5  -0.04413  0.006241 0.01488
26586.628 3760.2208 0.5 0.5  -0.03555  0.005028 0.02083
25733.819 3884.8361 6.5 6.5   0.05297 -0.007996 0.21635
25733.819 3884.8361 6.5 5.5   0.08547 -0.012903 0.03365
25733.819 3884.8361 5.5 6.5  -0.01232  0.001860 0.03365
25733.819 3884.8361 5.5 5.5   0.02018 -0.003046 0.12787
25733.819 3884.8361 5.5 4.5   0.04768 -0.007198 0.05276
25733.819 3884.8361 4.5 5.5  -0.03507  0.005294 0.05276
25733.819 3884.8361 4.5 4.5  -0.00757  0.001142 0.06629
25733.819 3884.8361 4.5 3.5   0.01493 -0.002254 0.05952
25733.819 3884.8361 3.5 4.5  -0.05277  0.007966 0.05952
25733.819 3884.8361 3.5 3.5  -0.03027  0.004569 0.02721
25733.819 3884.8361 3.5 2.5  -0.01277  0.001927 0.05612
25733.819 3884.8361 2.5 3.5  -0.06542  0.009877 0.05612
25733.819 3884.8361 2.5 2.5  -0.04792  0.007235 0.00638
25733.819 3884.8361 2.5 1.5  -0.03542  0.005348 0.04464
25733.819 3884.8361 1.5 2.5  -0.07304  0.011026 0.04464
25733.819 3884.8361 1.5 1.5  -0.06054  0.009139 0.00000
25733.819 3884.8361 1.5 0.5  -0.05304  0.008007 0.02679
25733.819 3884.8361 0.5 1.5  -0.07560  0.011413 0.02679
25733.819 3884.8361 0.5 0.5  -0.06810  0.010281 0.00893
25526.395 3916.4045 5.5 4.5   0.07013 -0.010761 0.30000
25526.395 3916.4045 4.5 4.5   0.01503 -0.002306 0.09722
25526.395 3916.4045 4.5 3.5   0.01503 -0.002306 0.15278
25526.395 3916.4045 3.5 4.5  -0.03006  0.004612 0.01944
25526.395 3916.4045 3.5 3.5  -0.03006  0.004612 0.12698
25526.395 3916.4045 3.5 2.5  -0.03006  0.004612 0.05357
25526.395 3916.4045 2.5 3.5  -0.06513  0.009992 0.05357
25526.395 3916.4045 2.5 2.5  -0.06513  0.009992 0.09643
25526.395 3916.4045 1.5 2.5  -0.09017  0.013835 0.10000
25296.759 3951.9572 6.5 5.5   0.05612 -0.008767 0.25000
25296.759 3951.9572 5.5 5.5   0.02138 -0.003340 0.04545
25296.759 3951.9572 5.5 4.5   0.02138 -0.003340 0.16883
25296.759 3951.9572 4.5 5.5  -0.00802  0.001252 0.00455
25296.759 3951.9572 4.5 4.5  -0.00802  0.001252 0.06926
25296.759 3951.9572 4.5 3.5  -0.00802  0.001252 0.10476
25296.759 3951.9572 3.5 4.5  -0.03207  0.005010 0.01190
25296.759 3951.9572 3.5 3.5  -0.03207  0.005010 0.07483
```

```
25296.759 3951.9572 3.5 2.5 -0.03207  0.005010 0.05612
25296.759 3951.9572 2.5 3.5 -0.05077  0.007932 0.02041
25296.759 3951.9572 2.5 2.5 -0.05077  0.007932 0.06531
25296.759 3951.9572 2.5 1.5 -0.05077  0.007932 0.02143
25296.759 3951.9572 1.5 2.5 -0.06413  0.010019 0.02857
25296.759 3951.9572 1.5 1.5 -0.06413  0.010019 0.04286
25296.759 3951.9572 0.5 1.5 -0.07215  0.011272 0.03571
24974.653 4002.9279 4.5 4.5  0.08357 -0.013395 0.25463
24974.653 4002.9279 4.5 3.5  0.08357 -0.013395 0.16204
24974.653 4002.9279 3.5 4.5 -0.02388  0.003827 0.16204
24974.653 4002.9279 3.5 3.5 -0.02388  0.003827 0.01058
24974.653 4002.9279 3.5 2.5 -0.02388  0.003827 0.16071
24974.653 4002.9279 2.5 3.5 -0.10745  0.017223 0.16071
24974.653 4002.9279 2.5 2.5 -0.10745  0.017223 0.08929
24957.357 4005.7021 8.5 8.5 -0.12993  0.020855 0.19051
24957.357 4005.7021 8.5 7.5 -0.00629  0.001009 0.01404
24957.357 4005.7021 7.5 8.5 -0.19047  0.030572 0.01404
24957.357 4005.7021 7.5 7.5 -0.06682  0.010726 0.14460
24957.357 4005.7021 7.5 6.5  0.04228 -0.006786 0.02318
24957.357 4005.7021 6.5 7.5 -0.12024  0.019299 0.02318
24957.357 4005.7021 6.5 6.5 -0.01114  0.001788 0.10794
24957.357 4005.7021 6.5 5.5  0.08342 -0.013389 0.02797
24957.357 4005.7021 5.5 6.5 -0.05743  0.009218 0.02797
24957.357 4005.7021 5.5 5.5  0.03712 -0.005959 0.07947
24957.357 4005.7021 5.5 4.5  0.11713 -0.018800 0.02893
24957.357 4005.7021 4.5 5.5 -0.00205  0.000328 0.02893
24957.357 4005.7021 4.5 4.5  0.07796 -0.012513 0.05820
24957.357 4005.7021 4.5 3.5  0.14342 -0.023020 0.02652
24957.357 4005.7021 3.5 4.5  0.04591 -0.007369 0.02652
24957.357 4005.7021 3.5 3.5  0.11137 -0.017876 0.04329
24957.357 4005.7021 3.5 2.5  0.16229 -0.026048 0.02110
24957.357 4005.7021 2.5 3.5  0.08645 -0.013875 0.02110
24957.357 4005.7021 2.5 2.5  0.13736 -0.022047 0.03435
24957.357 4005.7021 2.5 1.5  0.17373 -0.027884 0.01273
24957.357 4005.7021 1.5 2.5  0.11955 -0.019189 0.01273
24957.357 4005.7021 1.5 1.5  0.15592 -0.025026 0.03273
24942.039 4008.1622 7.5 6.5  0.08952 -0.014385 0.22222
24942.039 4008.1622 6.5 6.5  0.01344 -0.002159 0.02618
24942.039 4008.1622 6.5 5.5  0.04594 -0.007382 0.16827
24942.039 4008.1622 5.5 6.5 -0.05250  0.008437 0.00160
24942.039 4008.1622 5.5 5.5 -0.02000  0.003214 0.04196
24942.039 4008.1622 5.5 4.5  0.00750 -0.001205 0.12311
24942.039 4008.1622 4.5 5.5 -0.07579  0.012180 0.00406
24942.039 4008.1622 4.5 4.5 -0.04829  0.007761 0.04885
24942.039 4008.1622 4.5 3.5 -0.02579  0.004145 0.08598
24942.039 4008.1622 3.5 4.5 -0.09394  0.015096 0.00661
24942.039 4008.1622 3.5 3.5 -0.07144  0.011481 0.04837
24942.039 4008.1622 3.5 2.5 -0.05394  0.008668 0.05612
24942.039 4008.1622 2.5 3.5 -0.10694  0.017186 0.00850
24942.039 4008.1622 2.5 2.5 -0.08944  0.014374 0.04209
24942.039 4008.1622 2.5 1.5 -0.07694  0.012365 0.03274
24942.039 4008.1622 1.5 2.5 -0.11480  0.018449 0.00893
24942.039 4008.1622 1.5 1.5 -0.10230  0.016441 0.03175
24942.039 4008.1622 1.5 0.5 -0.09480  0.015235 0.01488
```

```
24942.039 4008.1622 0.5 1.5 -0.11752  0.018886 0.00694
24942.039 4008.1622 0.5 0.5 -0.11002  0.017681 0.02083
21901.696 4564.5771 6.5 5.5 -0.00850  0.001771 0.25000
21901.696 4564.5771 5.5 5.5 -0.07379  0.015378 0.04545
21901.696 4564.5771 5.5 4.5  0.01576 -0.003284 0.16883
21901.696 4564.5771 4.5 5.5 -0.12903  0.026892 0.00455
21901.696 4564.5771 4.5 4.5 -0.03949  0.008230 0.06926
21901.696 4564.5771 4.5 3.5  0.03377 -0.007039 0.10476
21901.696 4564.5771 3.5 4.5 -0.08469  0.017650 0.01190
21901.696 4564.5771 3.5 3.5 -0.01143  0.002381 0.07483
21901.696 4564.5771 3.5 2.5  0.04556 -0.009495 0.05612
21901.696 4564.5771 2.5 3.5 -0.04658  0.009708 0.02041
21901.696 4564.5771 2.5 2.5  0.01040 -0.002168 0.06531
21901.696 4564.5771 2.5 1.5  0.05110 -0.010650 0.02143
21901.696 4564.5771 1.5 2.5 -0.01471  0.003066 0.02857
21901.696 4564.5771 1.5 1.5  0.02599 -0.005417 0.04286
21901.696 4564.5771 0.5 1.5  0.01092 -0.002277 0.03571
```

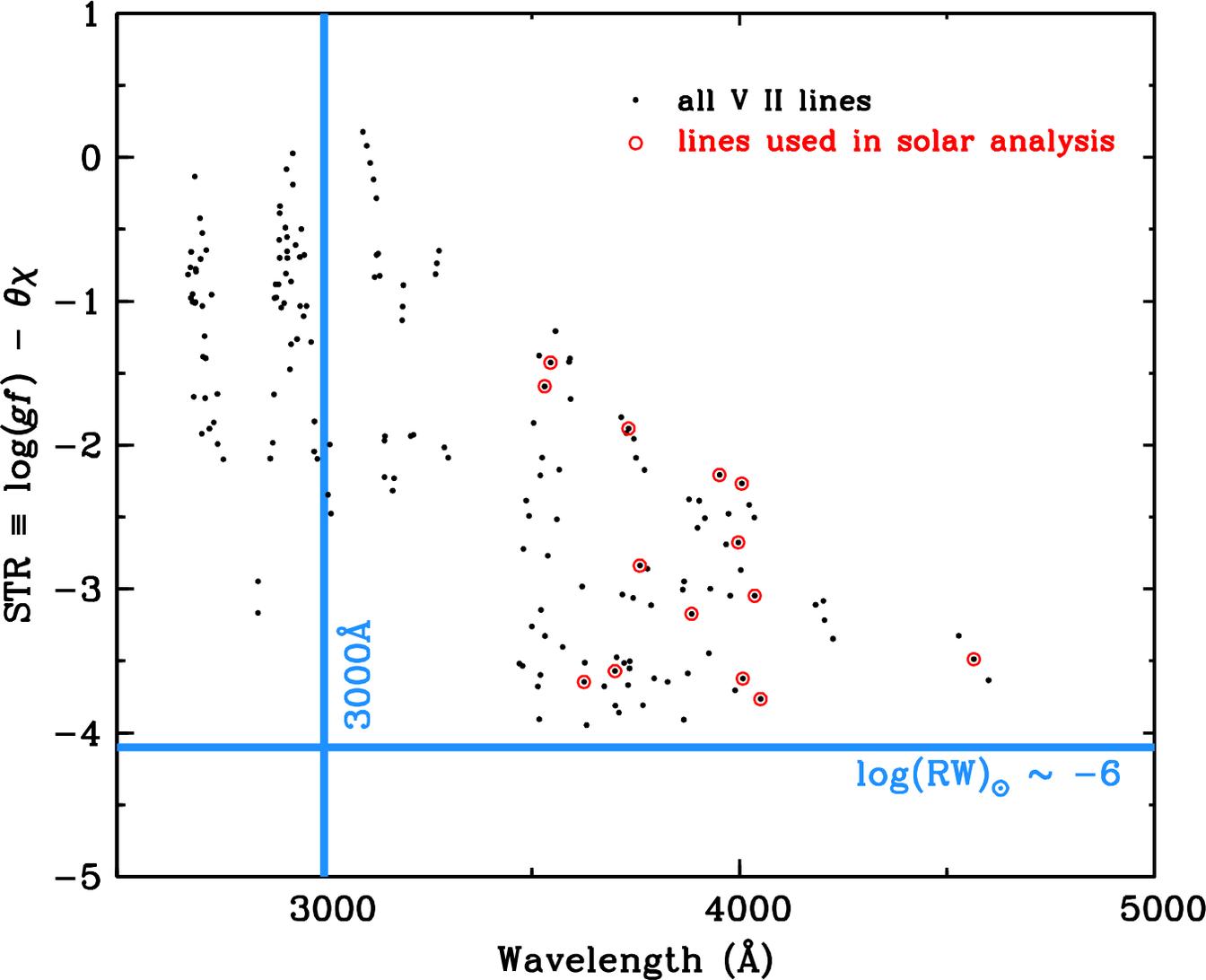

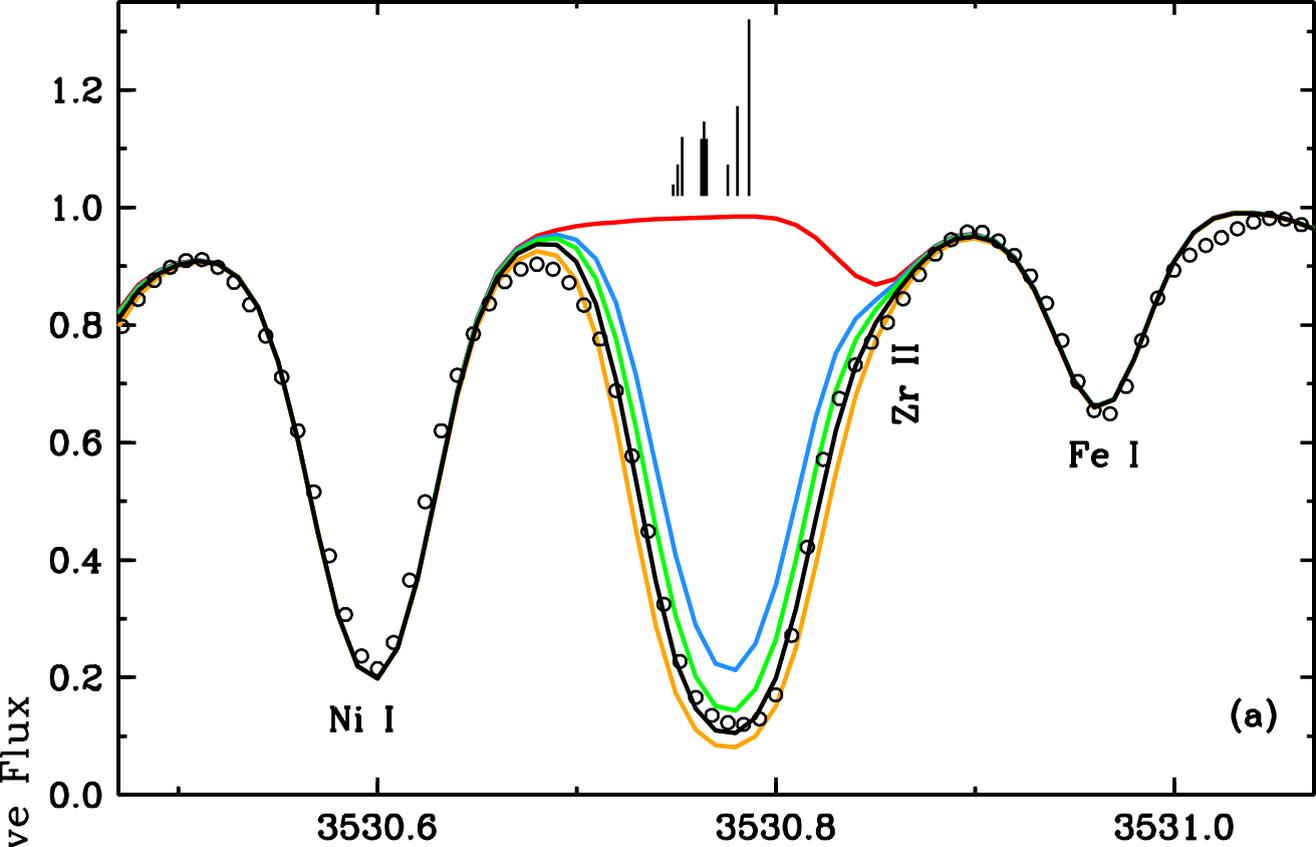
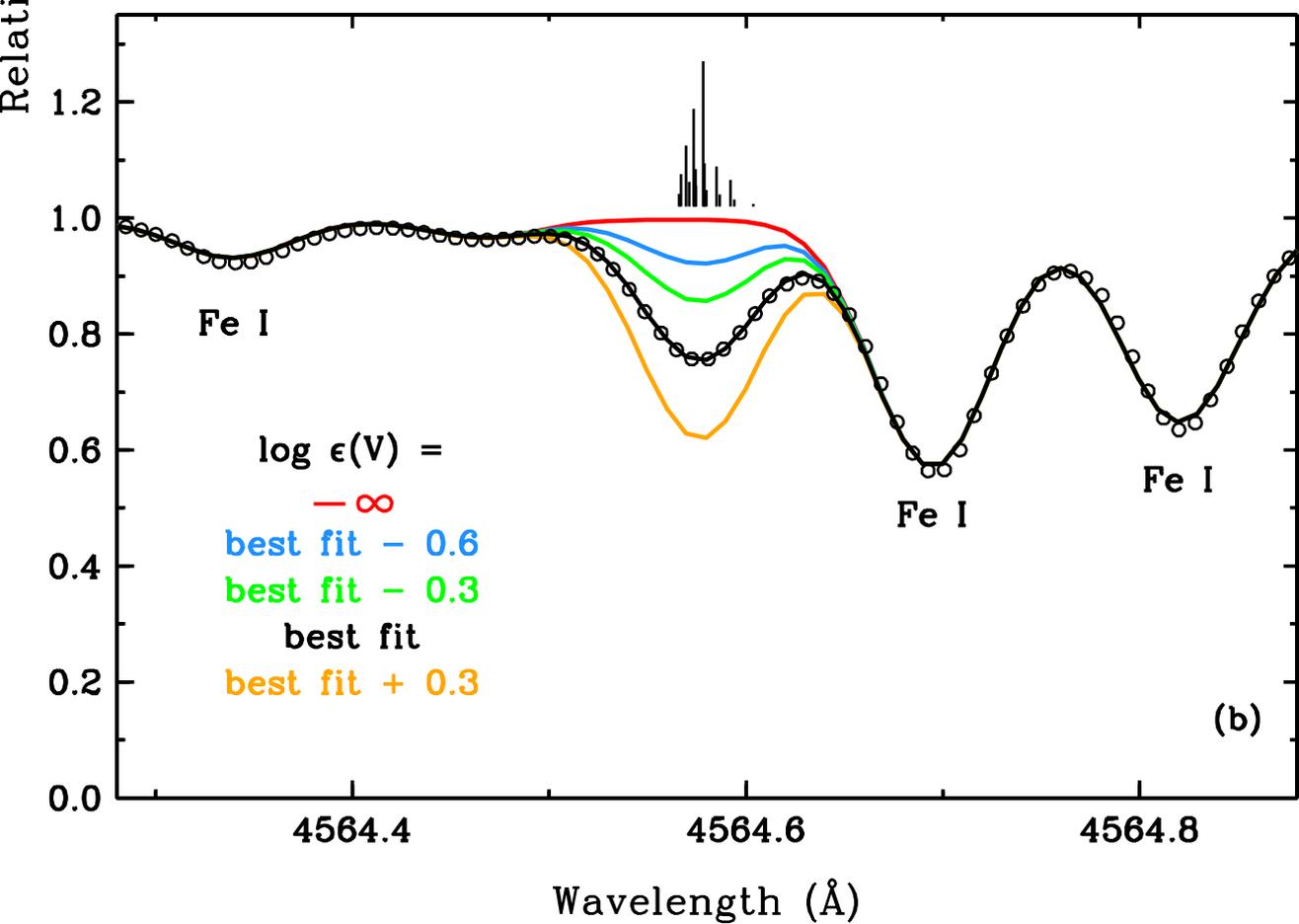

Table 6. Lines of ionized vanadium in the solar photosphere.

| Wavelength in air (Å) | Excitation energy (eV) | $\log_{10}(gf)$ | $\log_{10}(\varepsilon)$ | HFS[a] |
|---|---|---|---|---|
| 3530.772 | 1.070 | -0.53 | 3.95 | yes |
| 3545.196 | 1.095 | -0.32 | 4.00 | yes |
| 3625.611 | 2.375 | -1.22 | 3.90 | no |
| 3700.125 | 2.489 | -1.11 | 3.90 | no |
| 3732.748 | 1.564 | -0.32 | 3.83 | no |
| 3760.221 | 1.686 | -1.15 | 3.97 | yes |
| 3884.836 | 1.792 | -1.38 | 3.91 | yes |
| 3866.722 | 1.427 | -1.52 | 3.95 | no |
| 3951.957 | 1.475 | -0.73 | 3.95 | yes |
| 3997.110 | 1.475 | -1.20 | 3.98 | no |
| 4005.702 | 1.816 | -0.45 | 3.95 | yes |
| 4008.162 | 1.792 | -1.83 | 3.97 | yes |
| 4036.764 | 1.475 | -1.57 | 4.00 | yes |
| 4051.045 | 1.803 | -1.97 | 4.03 | no |
| 4564.577 | 2.266 | -1.22 | 4.03 | yes |

[a]This column denotes whether or not hyperfine substructure has been included in the abundance calculation.

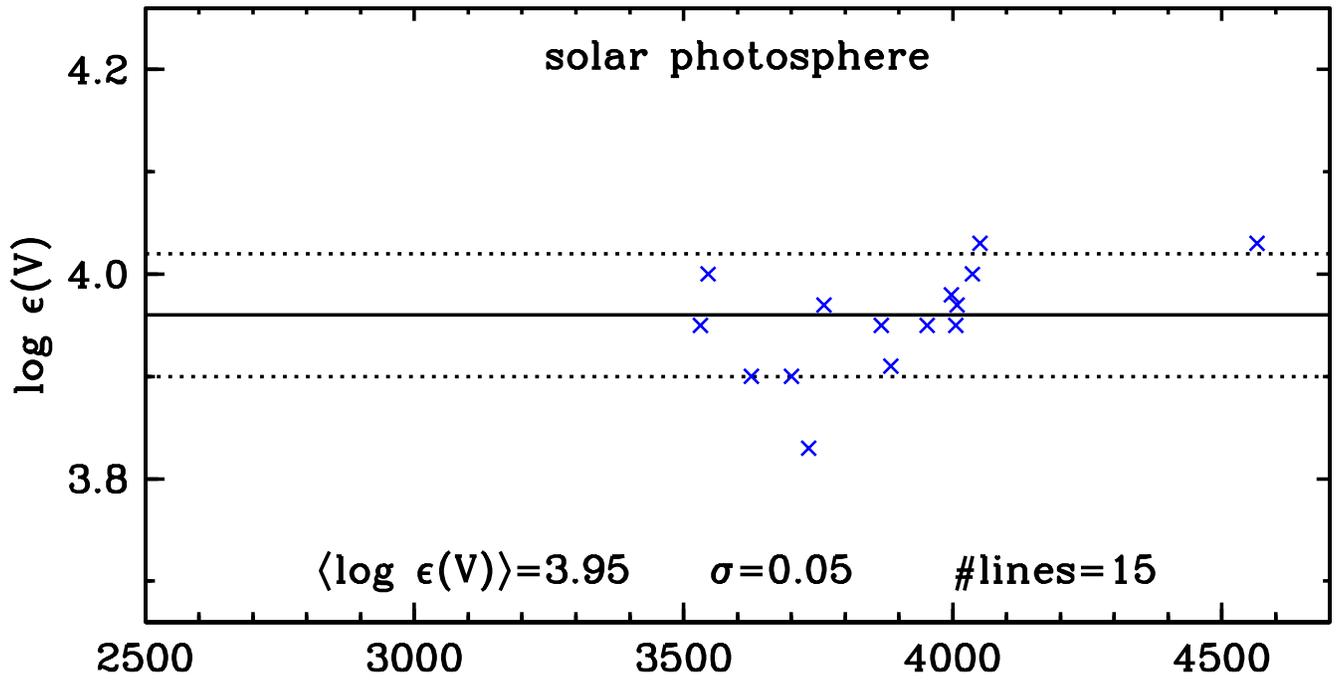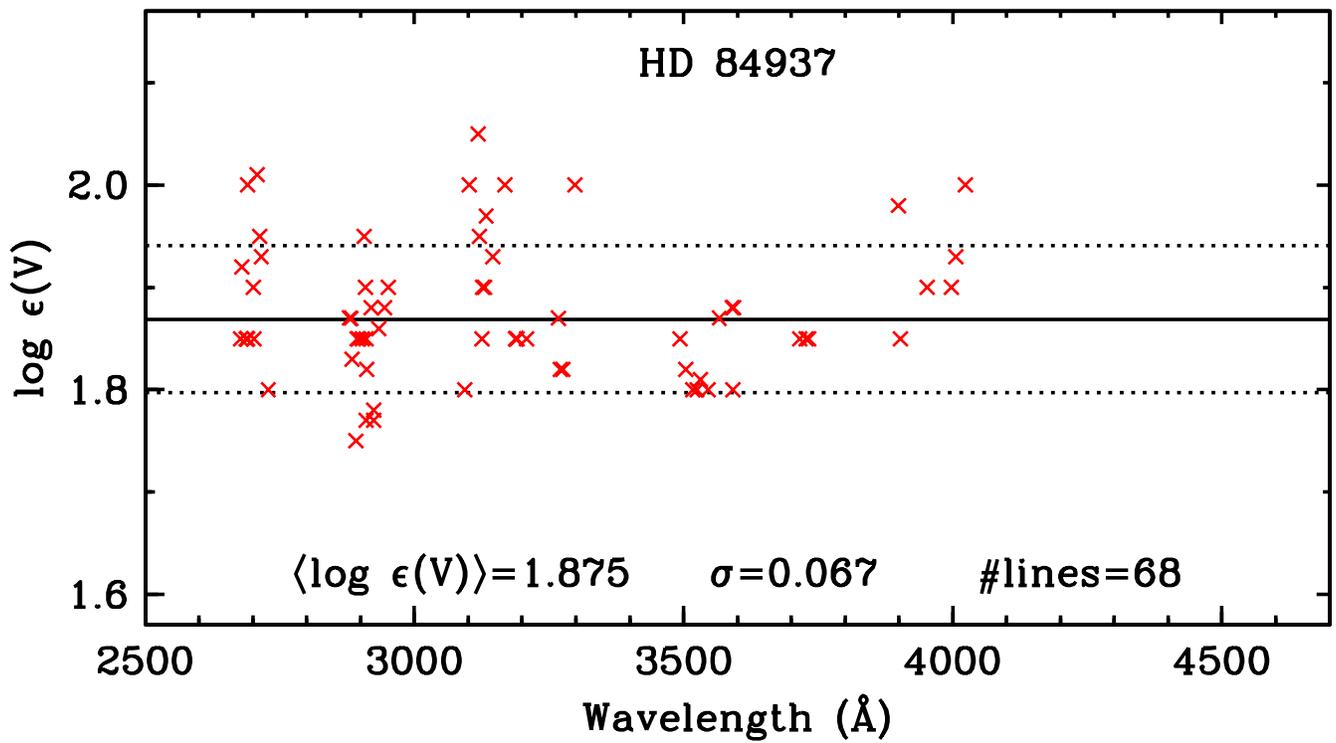

Table 7. Lines of ionized vanadium in HD 84937.

| Wavelength in air (Å) | Excitation energy (eV) | $\log_{10}(gf)$ | $\log_{10}(\varepsilon)$ | HFS[a] |
|---|---|---|---|---|
| 2677.796 | 0.004 | -0.76 | 1.85 | no |
| 2679.316 | 0.026 | -0.63 | 1.92 | yes |
| 2687.951 | 0.042 | -0.08 | 1.85 | yes |
| 2688.708 | 0.042 | -0.98 | 1.85 | yes |
| 2690.241 | 0.026 | -0.76 | 1.93 | no |

Note—Table 7 is available in its entirety via the link to the machine-readable version online.

[a]This column denotes whether or not hyperfine substructure has been included in the abundance calculation.

```
Title: Improved V II log(gf) Values, Hyperfine Structure Constants, and
       Abundance Determinations in the Photospheres of the Sun and
       Metal-poor Star HD 84937
Authors: Wood M.P., Lawler J.E., Den Hartog E.A., Sneden C., & Cowan J.J.
Table: Lines of ionized vanadium in HD 84937.
================================================================================
Byte-by-byte Description of file: Table7mr.txt
--------------------------------------------------------------------------------
   Bytes Format Units  Label    Explanations
--------------------------------------------------------------------------------
   1-  8 F8.3   0.1nm  WaveAir  Air Wavelength in Angstroms
  10- 14 F5.3   eV     EP       Lower level excitation energy
  16- 20 F5.2   ---    log(gf)  Log of degeneracy times oscillator
                                 strength
  22- 25 F4.2   ---    log(eps) Log of abundance epsilon
  27- 29 A3     ---    HFS      Hyperfine structure used in abundance
                                 calculation
--------------------------------------------------------------------------------

2677.796 0.004 -0.76 1.85 no
2679.316 0.026 -0.63 1.92 yes
2687.951 0.042 -0.08 1.85 yes
2688.708 0.042 -0.98 1.85 yes
2690.241 0.026 -0.76 1.93 no
2690.782 0.013 -0.77 2.00 no
2700.928 0.042 -0.37 1.90 yes
2702.177 0.026 -0.69 1.85 yes
2707.860 0.004 -1.39 2.01 yes
2713.044 0.013 -1.67 1.95 yes
2715.655 0.013 -0.63 1.93 yes
2728.637 0.004 -0.95 1.80 yes
2880.028 0.348 -0.64 1.87 yes
2882.499 0.333 -0.55 1.87 no
2884.783 0.323 -0.64 1.83 no
2891.640 0.333 -0.24 1.75 no
2893.317 0.368  0.03 1.85 yes
2896.206 0.333 -0.72 1.85 yes
2903.075 0.323 -0.70 1.85 yes
2906.458 0.348 -0.15 1.95 yes
2908.817 0.392  0.31 1.90 yes
2910.019 0.333 -0.22 1.85 yes
2910.386 0.323 -0.34 1.77 yes
2911.063 0.348 -0.35 1.82 yes
2919.993 0.368 -0.92 1.88 no
2920.384 0.333 -0.53 1.78 no
2924.019 0.392  0.42 1.78 yes
2924.641 0.368  0.18 1.77 yes
2934.401 0.323 -0.95 1.86 yes
2944.571 0.368 -0.13 1.88 yes
2952.071 0.348 -0.33 1.90 yes
3093.100 0.392  0.57 1.80 yes
3102.301 0.368  0.45 2.00 yes
3118.382 0.333  0.18 2.05 yes
3121.147 0.392 -0.46 1.95 yes
```

```
3125.286 0.323   0.04 1.85 yes
3126.219 0.368  -0.31 1.90 yes
3130.270 0.348  -0.32 1.90 yes
3133.335 0.333  -0.50 1.97 yes
3145.976 0.368  -1.57 1.93 yes
3168.133 1.070  -1.12 2.00 yes
3187.712 1.070  -0.07 1.85 yes
3188.513 1.095   0.06 1.85 yes
3190.683 1.127   0.24 1.85 yes
3208.346 1.095  -0.85 1.85 yes
3267.702 1.070   0.25 1.87 yes
3271.123 1.095   0.36 1.82 yes
3276.125 1.127   0.48 1.82 yes
3298.738 1.127  -0.97 2.00 yes
3493.162 1.070  -1.42 1.85 no
3504.436 1.095  -0.77 1.82 no
3517.299 1.127  -0.24 1.80 yes
3524.716 1.095  -0.99 1.80 yes
3530.772 1.070  -0.53 1.81 yes
3545.196 1.095  -0.32 1.80 yes
3566.178 1.070  -1.08 1.87 yes
3589.759 1.070  -0.35 1.88 yes
3592.022 1.095  -0.30 1.80 yes
3593.333 1.127  -0.55 1.88 yes
3715.464 1.574  -0.22 1.85 no
3727.341 1.686  -0.23 1.85 yes
3732.748 1.564  -0.32 1.85 no
3899.128 1.803  -0.77 1.98 no
3903.253 1.475  -0.91 1.85 no
3951.957 1.475  -0.73 1.90 yes
3997.110 1.475  -1.20 1.90 no
4005.702 1.816  -0.45 1.93 yes
4023.377 1.803  -0.61 2.00 no
```

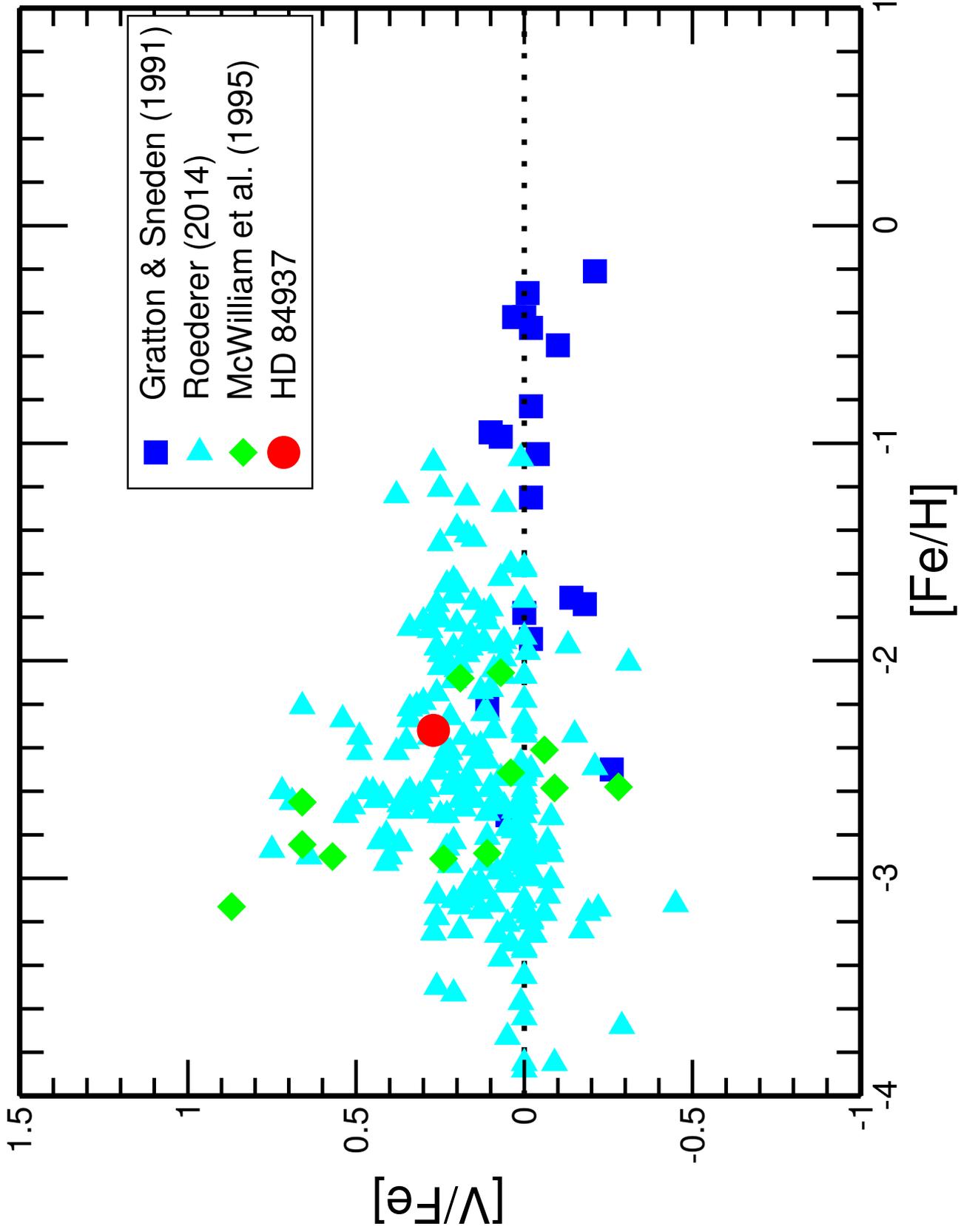